\documentclass[11pt,a4paper]{article}
\usepackage[top=1.25in, bottom=1.25in, left=.7in, right=.7in]{geometry}
\usepackage{setspace}		

\usepackage{amssymb}
\usepackage{amscd,amsfonts,amsmath,amsthm,bbm,bm,latexsym,mathrsfs}
\usepackage{epsfig,graphics,graphicx,natbib,subfigure,longtable}
\usepackage{eurosym}

\usepackage[figuresright]{rotating}
\usepackage{bookmark}
\usepackage{booktabs}
\usepackage[usenames]{color,colortbl}

\newtheorem{remark}{Remark}[section]

\theoremstyle{remark}

\theoremstyle{plain}

\newtheorem{exa}{Example}

\usepackage{lineno}
\usepackage{ulem}

\begin{document}
	
	\newcommand{\lp}{\left(}
	\newcommand{\rp}{\right)}
	\newcommand{\lsp}{\left[}
	\newcommand{\rsp}{\right]}
	\newcommand{\lop}{\left]}
	\newcommand{\rop}{\right[}
	\newcommand{\lbr}{\left\{}
	\newcommand{\rbr}{\right\}}
	\newcommand{\lang}{\langle}
	\newcommand{\rang}{\rangle}

	\newcommand{\A}{\mathcal{A}}
	\newcommand{\Se}{\mathcal{S}}
	\newcommand{\Q}{\mathcal{Q}}
	\newcommand{\C}{\mathcal{C}}
	
	\newcommand{\DA}{\mathcal{D}_{\A}}
	\newcommand{\DS}{\mathcal{D}_{\Se}}
	\newcommand{\DC}{\mathcal{D}_{\C}}

	\newcommand{\AB}{\mathcal{A}_{B}^{\alpha,\beta}}
	\newcommand{\SB}{\Se_{B}}
	\newcommand{\QB}{\Q_{B}}
	\newcommand{\CB}{\C_{B}}
	\newcommand{\CEV}{\C_{EV}}
	
	\newcommand{\DSB}{\mathcal{D}_{\SB}}
	\newcommand{\DQB}{\mathcal{D}_{\QB}}
	\newcommand{\DCB}{\mathcal{D}_{\CB}}
	
	\newcommand{\Pic}{\mathcal{P}}
	
	\newcommand{\Ext}{\textnormal{Ext}}
	
	\newcommand{\Dex}{\Gamma}
			
		\title{A Multivariate Dependence Analysis for Electricity Prices, Demand and Renewable Energy Sources}
		
		\author{Fabrizio Durante\thanks{Dipartimento di Scienze dell'Economia,  Universit\`a del Salento, Lecce, Italy,  \color{blue}\texttt{fabrizio.durante@unisalento.it}} \and Angelica Gianfreda\thanks{Department of Economics ``Marco Biagi'',  University of Modena and Reggio Emilia, Modena, Italy, Energy Markets Group, London Business School, UK, \color{blue}\texttt{angelica.gianfreda@unimore.it}} \and
Francesco Ravazzolo\thanks{Department of Data Science and Analytics,  BI Norwegian Business School, Norway and Faculty of Economics and Management,  Free University of Bozen-Bolzano, Italy and RCEA, \color{blue}\texttt{francesco.ravazzolo@bi.no}} \and
		Luca Rossini\thanks{Department of Economics, Management and Quantitative Methods,  University of Milan, Italy, \color{blue}\texttt{luca.rossini@unimi.it}}
 }
		
	\date{\today}
\maketitle	
		
\abstract{\noindent This paper examines the dependence between electricity prices, demand, and renewable energy sources by means of a multivariate copula model {while studying Germany, the widest studied market in Europe}. The inter-dependencies are investigated in-depth and monitored over time, with particular emphasis on the tail behavior. To this end, suitable tail dependence measures are introduced to take into account a multivariate extreme scenario appropriately identified {through the} Kendall's distribution function. The empirical evidence demonstrates a strong association between electricity prices, renewable energy sources, and demand within a day and over the studied years. Hence, this analysis provides guidance for further and different incentives for promoting green energy generation while considering the time-varying dependencies of the involved variables.
	
		\noindent \textbf{Keywords:}  			Copula; Electricity; Kendall distribution; Solar and Wind Power; Tail Dependence.
		}

	\section{Introduction}
	\label{sec_Intro}
	In recent years, {the electricity generation from} renewable energy sources (RES) has increased in importance in the economies of all countries, especially in Europe, due to {stringent} regulations {to reduce} carbon emissions and to {provide incentives for investments in clean technologies}. However, the interrelationships between RES and demand, and their combined effect on electricity prices {have been under-investigated and} there are still few works {focusing on this multivariate dependence}. These relations are particularly important since RES 
	can reduce the demand for electricity if weather conditions allow. Indeed, it has been largely proved that wind generation reduces the mean (and the skewness) of {the distribution of}  electricity price {while increasing the} price variability. In contrast, there is no clear understanding of the effect of solar power generation, especially regarding its interactions with demand, and eventually with wind power generation.
	{Therefore,} this paper aims at  exploring these interdependencies {in details}. 
	
	To this aim, a new database is compiled using hourly electricity prices determined on the day-ahead German market together with predictions for both RES and demand. This allows to consider the dependence between these variables and the effects of their different combinations across all 24 hours and across a sample of years, going from 2011 (a year in which RES were at their early introduction) to 2019.
	Note that Germany is the largest European {electricity} market {for traded volume and production (see} \cite{QREC2018}). Moreover, it is a leading country for the total wind power capacity per inhabitant (jointly with Denmark) and solar PV capacity per inhabitant (recently flanked by Italy and Spain).\footnote{Indeed, the RES share of total power capacity increased from $24\%$ to $44\%$ from 2010 to 2015 in the major European countries.} Therefore, studying the German market allows us to understand the dependence structure among prices, demand, and RES, which could provide useful guidance for policymakers. In particular, the uncovered multivariate dependence structure could {display important effects due to the} increasing RES penetration and could provide support for further investments to reduce carbon emissions.
	
	Here, the dependence among prices, demand, and RES is investigated by using a copula approach. This method is appropriate since it allows for a careful description of the multivariate stochastic behavior and for an accurate analysis of different types of association and tail dependence. This is particularly important since, for example, situations in which high wind generation is coupled with high demand levels, together with high solar production, may represent co-movements in extreme behavior that are not easily detected with other methodologies. In particular, copulas allow to proceed in two steps: first, individual variables are modelled according to their features{;} and, then, the dependencies between price, demand, wind, and solar generation are described with {a} great{er} flexibility. 
	
	Several papers have applied copula models {for modelling} energy markets.
	\cite{Grothe2011} adopt copulas to evaluate investment decisions regarding the placement of wind turbines {with respect to wind speed} in order to reduce output fluctuations and stabilize the supply.  \cite{Denault2009} {use} copulas to model and investigate the complementarity between hydro and wind, {aiming at reducing} the risk of shortages in water inflows. Multivariate copulas are {instead} considered {to inspect the integration of wind energy in the European grid; see}  \cite{Hagspiel2012}.  \cite{Haghi2010} {implement a multivariate non-normal copula model for studying} the behavior of wind speed, solar radiation, and load profiles of a network. 
	
	Moreover,  copulas {have been used} for the relationships between electricity prices observed over different regions, {or to depict the relationships} between prices and fundamental variables.  For example, robust partial correlations are estimated between changes in electricity prices in the connected zones of New York state in \cite{Dupuis2017}. {In addition,} \cite{IgTr2016} examine the dependence structure of electricity spot prices across Australian regional markets.
	Several regime-switching AR--GARCH copulas are proposed in \cite{PircalabuBenth2017} to study the pairwise behavior of electricity prices over interconnected European markets (Germany, France, Netherlands, Belgium, and Western Denmark). In particular, the skewed t distribution is considered because it describes the marginal dynamics better than the normal distribution and can also capture the pair-wise tail dependence.
	
	Regarding the study of the dependence between electricity prices and/or renewable energy sources, the literature has focused largely on bivariate models, mainly by considering 
	prices and wind generation. 
	For instance, the dependence between wind power production and electricity prices is examined in \cite{Ketterer2014}, \cite{Pircalabu2017}, \cite{Rintamaki2017} and  \cite{Jonsson2010}.  
	\cite{Elberg2015} develop stochastic simulation model able to capture the full spatial dependence structure of wind power by using copula models incorporating a{lso} demand and supply {information}.
	
	Regarding solar power, \cite{Paraschiv2014} show that it decreases price volatility and {more recently,} \cite{Gianfreda2018a} show that both wind and solar power reduce mean electricity prices, but increase their volatility. More importantly, they provide new insights regarding the negative effect of wind on the skewness of price distributions, hence suggesting {to} control for the behavior of the tails. 
	
	The{refore,} this paper extends the recent literature on the multivariate dependence of electricity prices by providing {first} new methodological tools for {the} joint tail behavior and {then new} empirical {results based on}  a novel dataset.
	
	Tri- and quadrivariate copulas {are used} in order to capture the dependence (at an hourly level) between the stochastic variables that are different in their nature, namely, electricity prices, forecasted electricity demand and forecasted wind, together with the newly included forecasted solar PV generation. Note that in a previous study, \cite{Liebl2013} consider the dependence between electricity prices and demand by means of functional factor models, {without including} solar and wind power. 
	
	Second,  to explore the dependence structure and demonstrate the importance of considering all possible interaction effects, {two analyses are implemented: a global and static one, over the full sample of studied years; and a dynamic inspection, using an approach of rolling windows.}
	
	Third,  coefficients {for the multivariate tail dependence} are proposed in order to detect possible joint tail dependencies. Following \cite{SalDeMDur11},  \cite{NapSpi09} and also \cite{Beretal18SERRA}, who introduced these indices based on the concept of Kendall extreme scenario in the {analysis of} (environmental) risks, we consider these novel  measures for the inspection of extreme scenarios that market operators, analysts and policymakers may be forced to face.
	
	Specifically, following a copula-based ARMA-GARCH model for multivariate time series, 
	we describe the relationships between prices, demand, and RES; and, those among RES and demand. Additionally, and if necessary, one could also detect relations across solar and wind power production. 
	In particular, 
	we focus on the relationship between (i) demand and prices, (ii) prices and wind, or (iii) demand and solar. In case (i), one should expect a positive dependence, as demand increases (even if `corrected' or reduced by solar generation), prices should increase as well. An inverse relation is instead expected in case (ii), i.e., prices should decrease as wind increases (and solar is considered an additional supply factor reducing the demand). In the latter case (iii), again a negative dependence is expected, since when solar PV increases, demand is expected to be reduced. Our results confirm these expectations, indicating a strong negative dependence between electricity prices and RES variables during the day; and the evidence is identical using different copula models. 
		
	The paper is organized as follows. Section \ref{sec_Data} describes the German market and the dataset employed. Section \ref{sec_Cop} briefly recalls the notion of copula, the methodology, and the estimation procedure. Here,  the coefficients {for the multivariate tail dependence} are also introduced. Section \ref{sec_Copula_Insample} is devoted to both the global and time-varying analyses on the dependence parameter of tri- or quadrivariate dimensional copulas. Finally, conclusions are presented in Section \ref{sec_Conclusions}.

	%
	%
	\section{Data description}
	\label{sec_Data}
	
	This empirical study relies on a new hourly dataset consisting of German electricity prices, forecasted demand, forecasted wind, and forecasted solar PV generation from January 1, 2011, to December 31, 2019. Electricity prices are quoted in \euro/MWh on a daily basis. They have been pre-processed for time-clock changes, that is the 25th hour in October has been excluded, whereas the missing 24th hour in March has been interpolated. Hence, there are no missing observations.
	
	The hourly auction prices in Germany are determined on the day-ahead market before noon, and then, in practice, they are forward prices for delivery during the predetermined hours on the following day. These prices have been collected directly from the German power market, \textit{European Energy Exchange} (EEX). In addition, by considering the day-ahead determination of prices, the forecasted values for demand, wind, and solar PV generation have been used, as provided by Thomson Reuters with an hourly frequency. Specifically, the forecasts used in this analysis are those obtained by the European Centre for Medium-Range Weather Forecast (ECMWF), which result from the running of the operational model at midnight (technically, the model is said to \textit{run at hour $00$}). This represents the latest information available to market operators before they submit their bids/offers, because this model updates from $05.40$ a.m. to $06.55$ a.m.
	
	It is important to emphasize that other data sources are commonly used in similar research about forecasted consumption, wind, and solar generation. Specifically, researchers collect this information from the official websites of the transmission system operators (TSOs) of the market under investigation and, then, they additionally provide these data to the European network of the TSOs for electricity (ENTSOE)\footnote{For further details see \texttt{www.entose.org} and its transparency platform at \texttt{https://transparency.entsoe.eu}}. As far as the consumption forecasts are concerned, the transparency data, provided as a \textit{day-ahead forecast of the total load}, are published (hence, publicly available) per time unit (currently having a quarter hourly frequency) either \textit{at the latest two hours before the gate closure time of the day-ahead market} or at 12:00 (in local time) \textit{at the latest when the gate closure time does not apply}. This represents the \textit{publication deadline} for ENTSOE (as named on the website) and refers to data available to market operators at (the latest) 10 a.m., whereas the data used in this analysis is published immediately after the update and is already accessible at 8 a.m. when traders start to run their forecasting models to construct a portfolio of 24 hourly prices representing their bidding strategy submitted on the day-ahead market before noon.
	
	More importantly, the relevance and novelty of the database used in this research are highlighted when considering the public availability of RES forecasts. Indeed, ENTSOE publishes also \textit{day-ahead} forecasted values of electricity generated by wind and solar photo-voltaic plants but only by 6.00 p.m. (in Brussels time). Recently, ENTSOE started to provide additional \textit{current} and \textit{intraday} forecasts representing the last current update and the most recent intraday forecasts, respectively, at 8.00 a.m. for all 24 hours of the day of delivery, which are not expected to be \textit{regularly updated after 8.00}.
	However, at the time of writing this paper, the field of current forecasts was still empty, whereas the field for \textit{intraday} forecasts was available for wind offshore only from 01/01/2018, whereas those for wind onshore and solar were available only from the 26th  February 2018 (hence the lenght of the series is too short for historical dynamic analyses). Instead, the database used here contains RES forecasts produced by early hours in the mornings and consistently from 2011, thus representing an extremely important source of information for detecting dependencies and comparing their historical evolution.
	
	Regarding the details of the ECMWF forecasts, and as far as demand forecasts are concerned, weather forecasts (accounting for temperature, precipitation, pressure, wind speeds, and cloud cover or radiation) are used in the models,  whereas the forecasts for wind generation make use of wind speed and installed capacity. Finally, PV installations, solar radiation, and installed capacity (because of the predominance of photovoltaic plants over solar thermal ones) are used to generate forecasts for solar power generation.
	
\begin{figure}[h!]
		\centering
		\begin{tabular}{cc}
			\includegraphics[width=6cm]{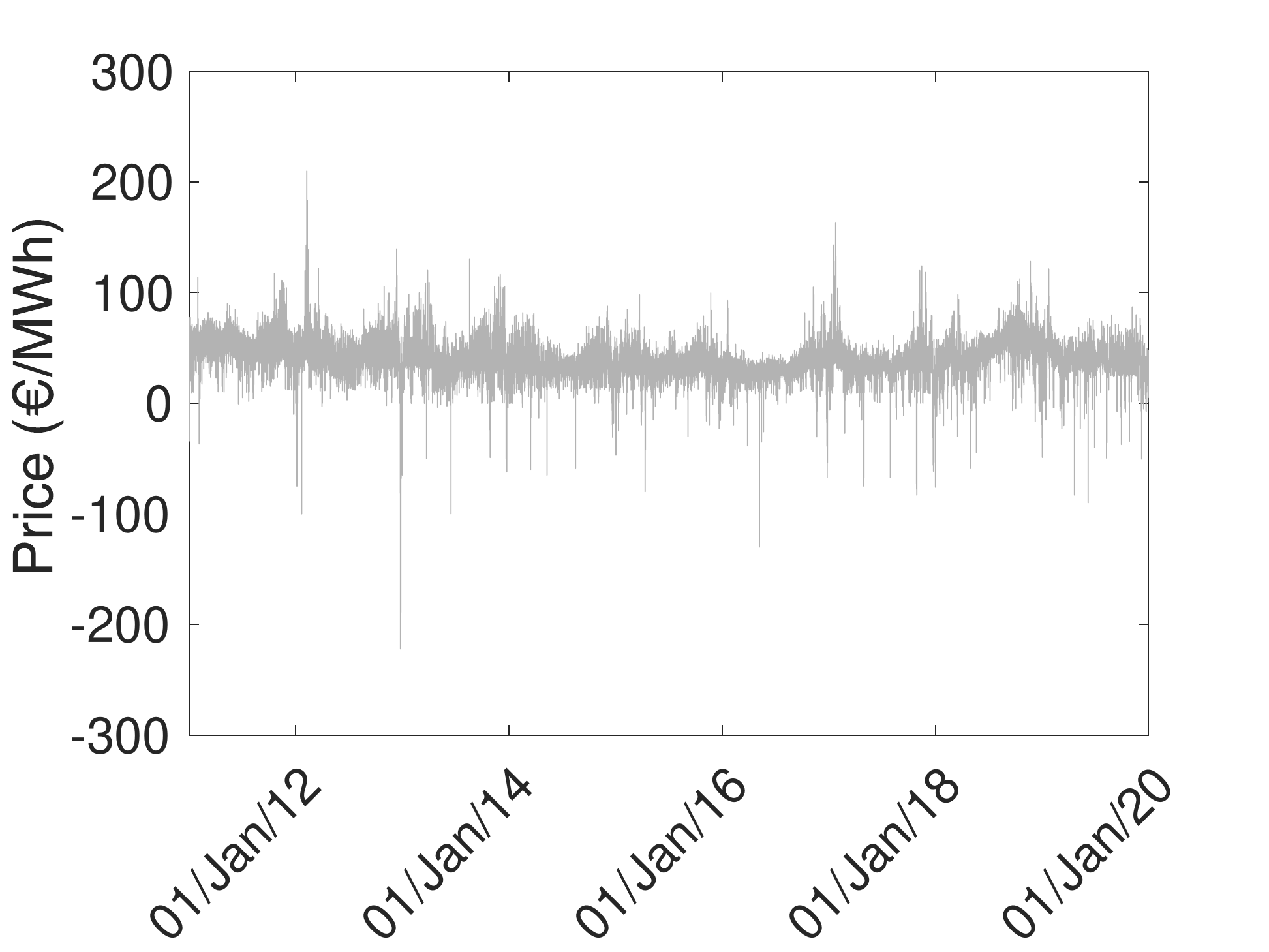}&
			\includegraphics[width=6cm]{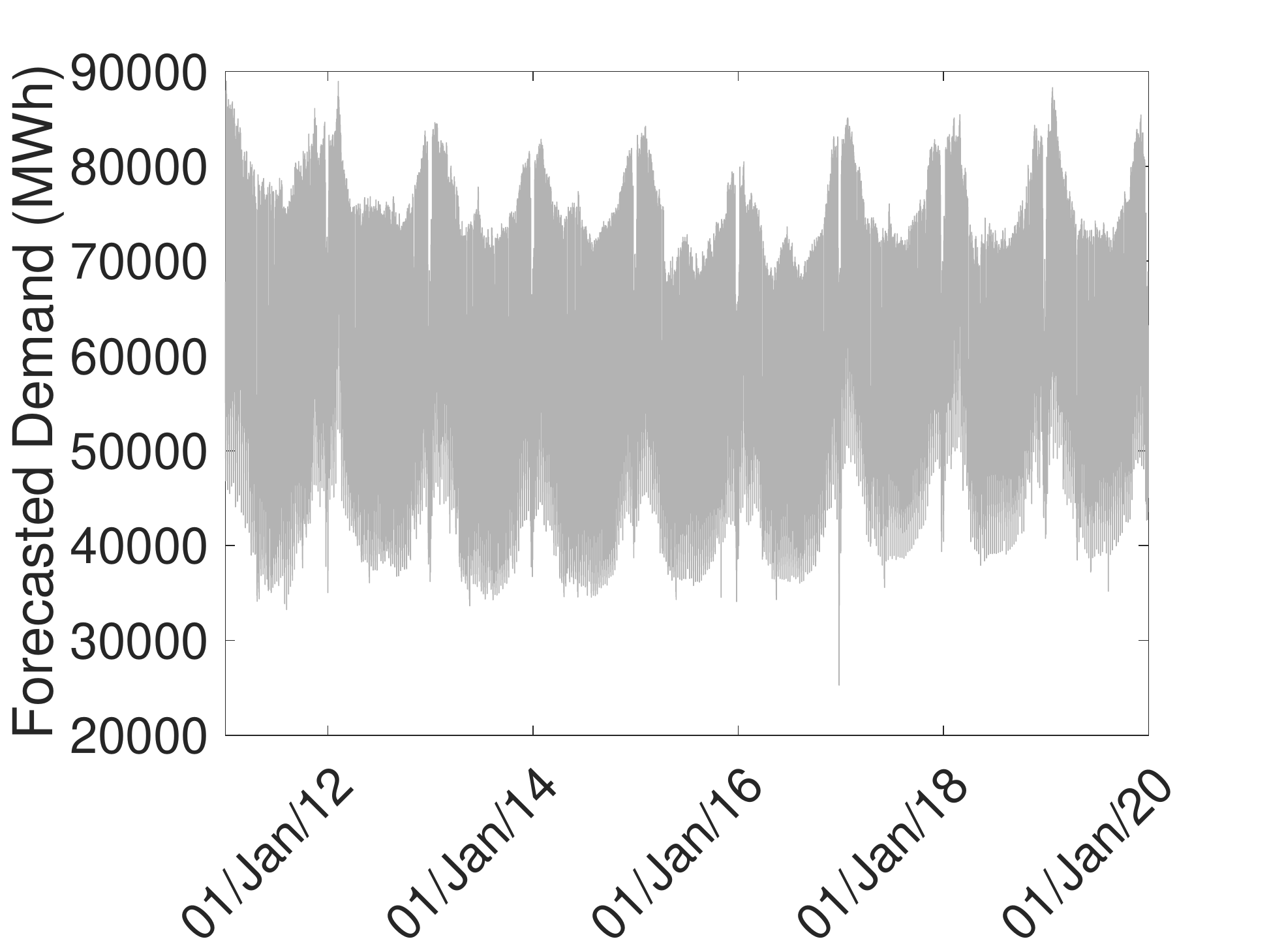}\\
			(a)&(b)\\
			\smallskip
			\includegraphics[width=6cm]{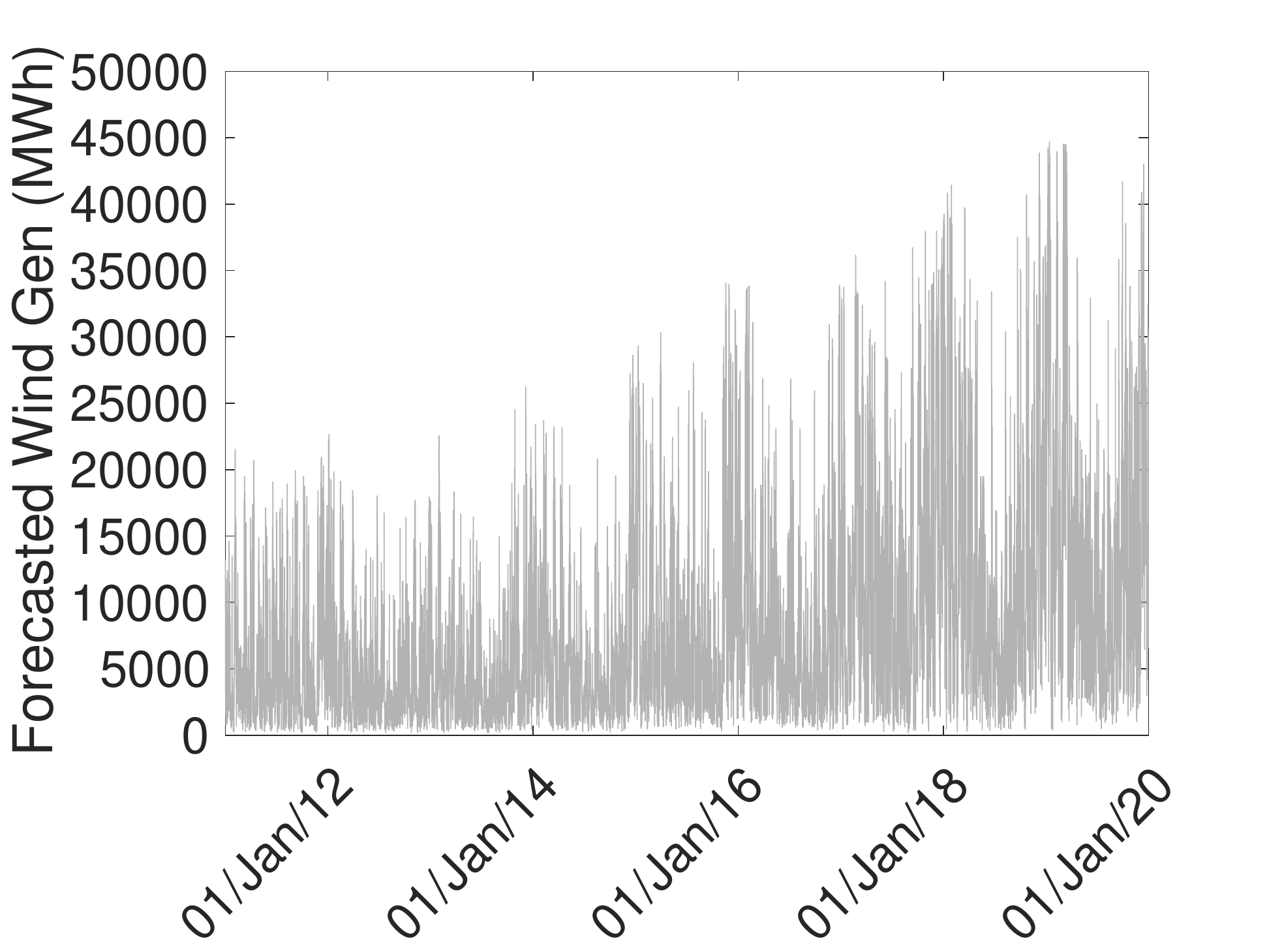}&
			\includegraphics[width=6cm]{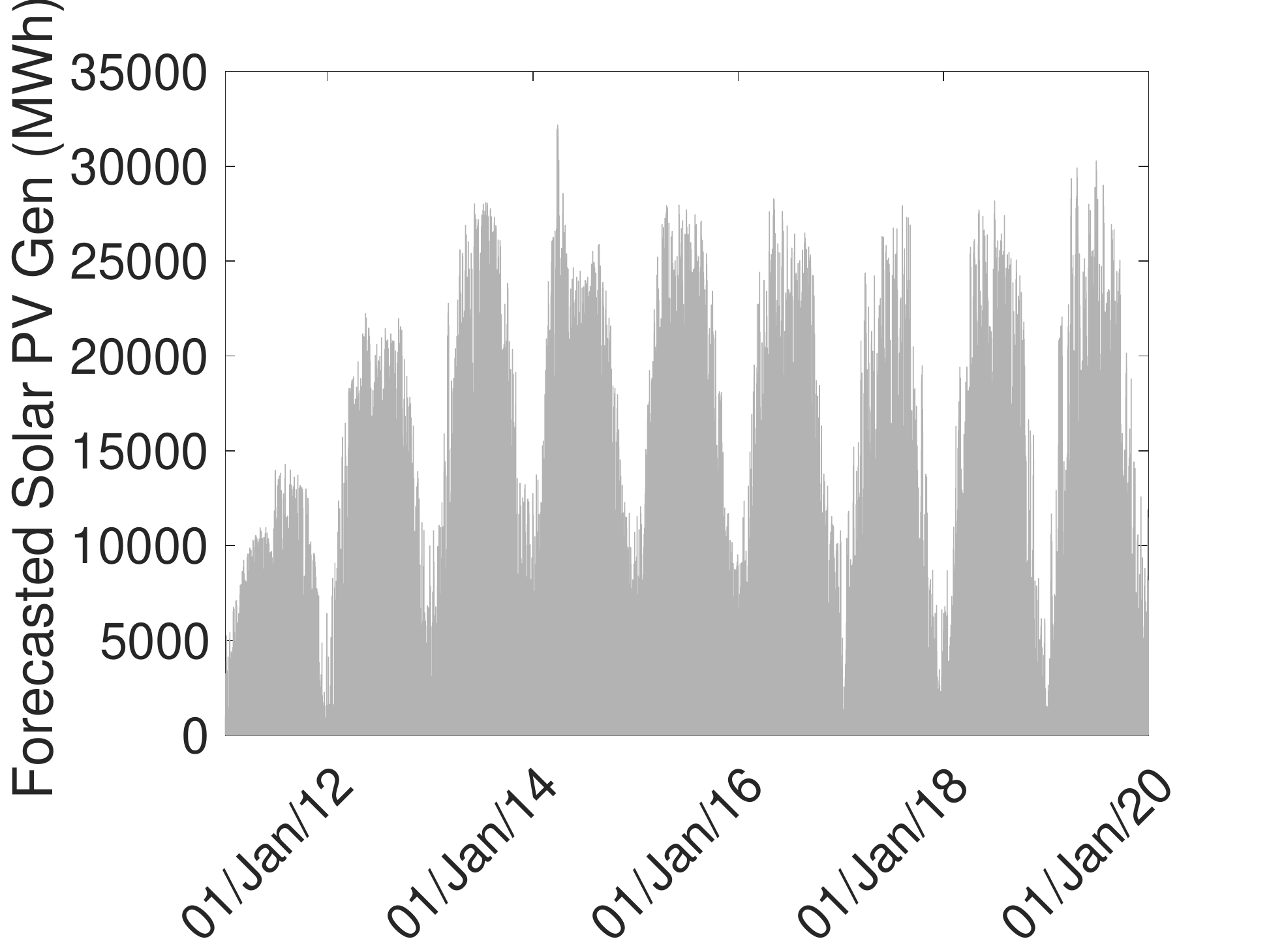}\\
			(c) & (d) \\
		\end{tabular}
		\caption{\footnotesize{Hourly Time Series for Electricity Day-ahead Prices (panel a), Forecasted Demand (panel b), Forecasted Wind Generation (panel c) and Forecasted Solar PV Generation (panel d) observed in Germany from 01/01/2011 to 31/12/2019.}}
		\label{Plot_Germany}
	\end{figure}

	Figure \ref{Plot_Germany} shows the dynamics of all time series. The hourly electricity prices in panel (a) show ``downside'' spikes together with mean-reversion and seasonality, especially in the last years of the sample, when negative prices also reduced their occurrences. The behavior of the forecasted demand series is shown in panel (b), with peaks during winters and lows during summers, representing the typical calendar seasonality. The forecasted wind generation is depicted in panel (c), and it shows high variability due to weather conditions, together with a sharp increasing trend due to new investments in additional wind capacity. Finally, the forecasted solar generation is shown in panel (d), where strong seasonal patterns are again visible through the calendar year.
	
\begin{figure}[h!]
		\centering
		\begin{tabular}{ccc}
 			\includegraphics[width=5cm]{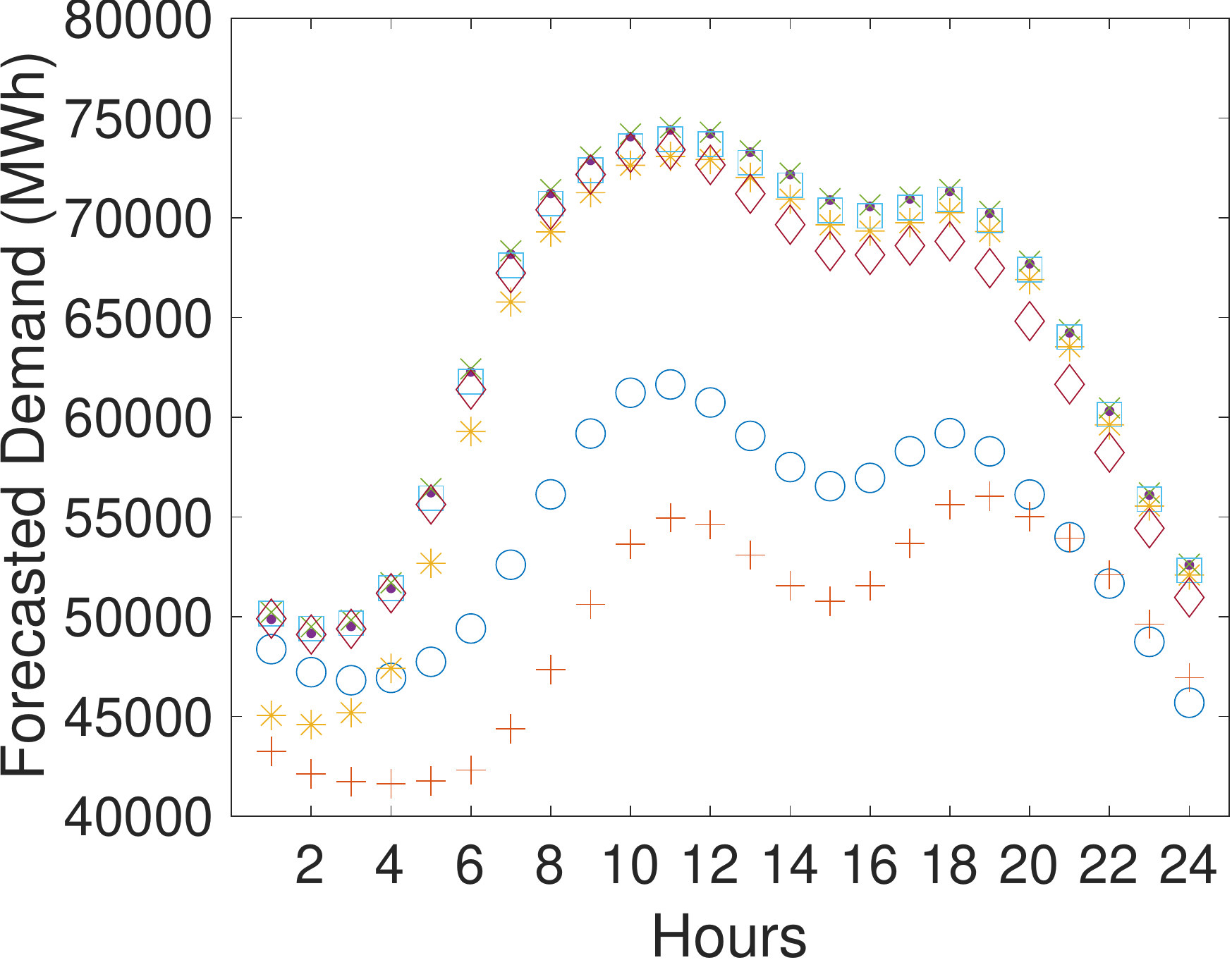}&
			\includegraphics[width=5cm]{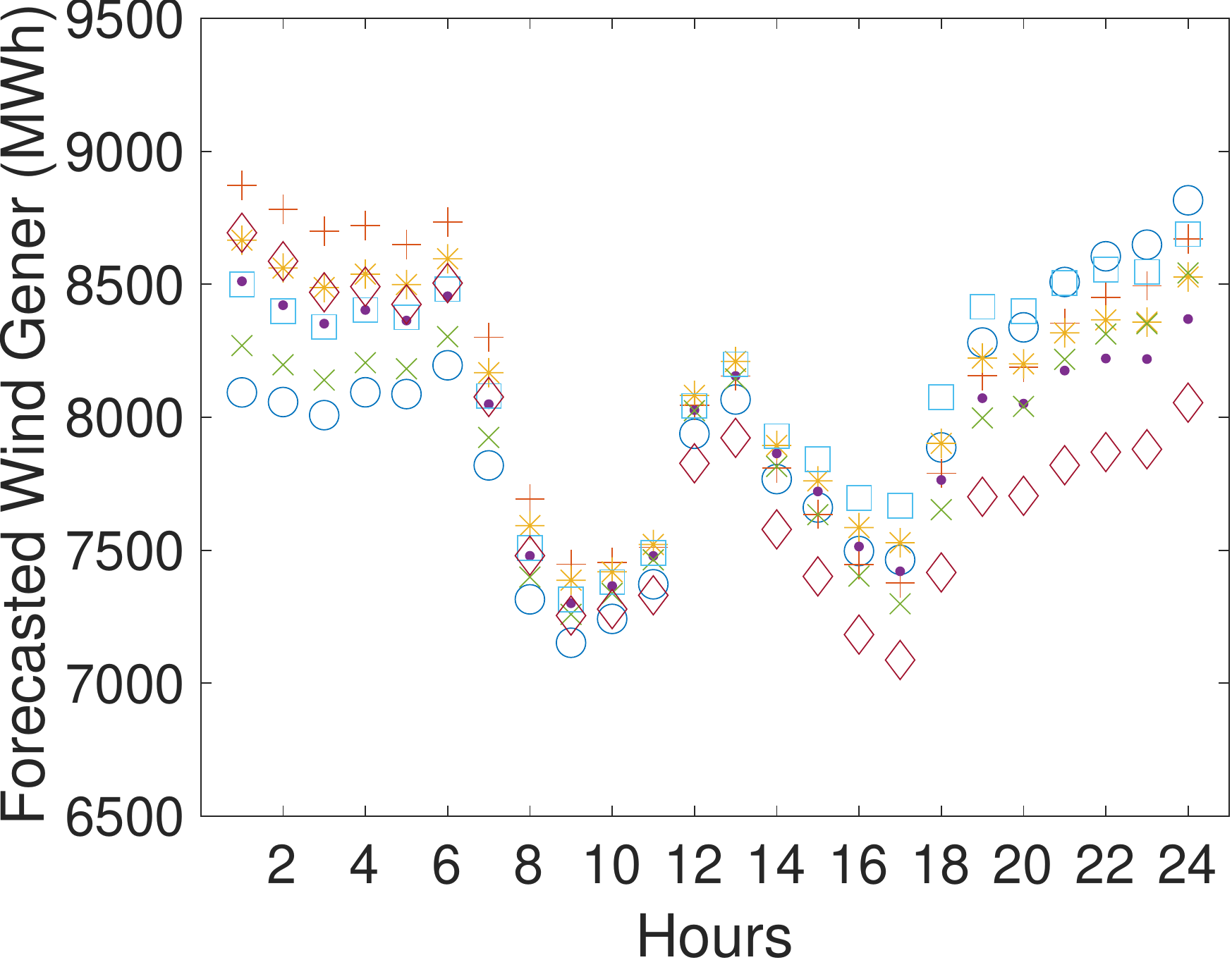} &
			\includegraphics[width=5cm]{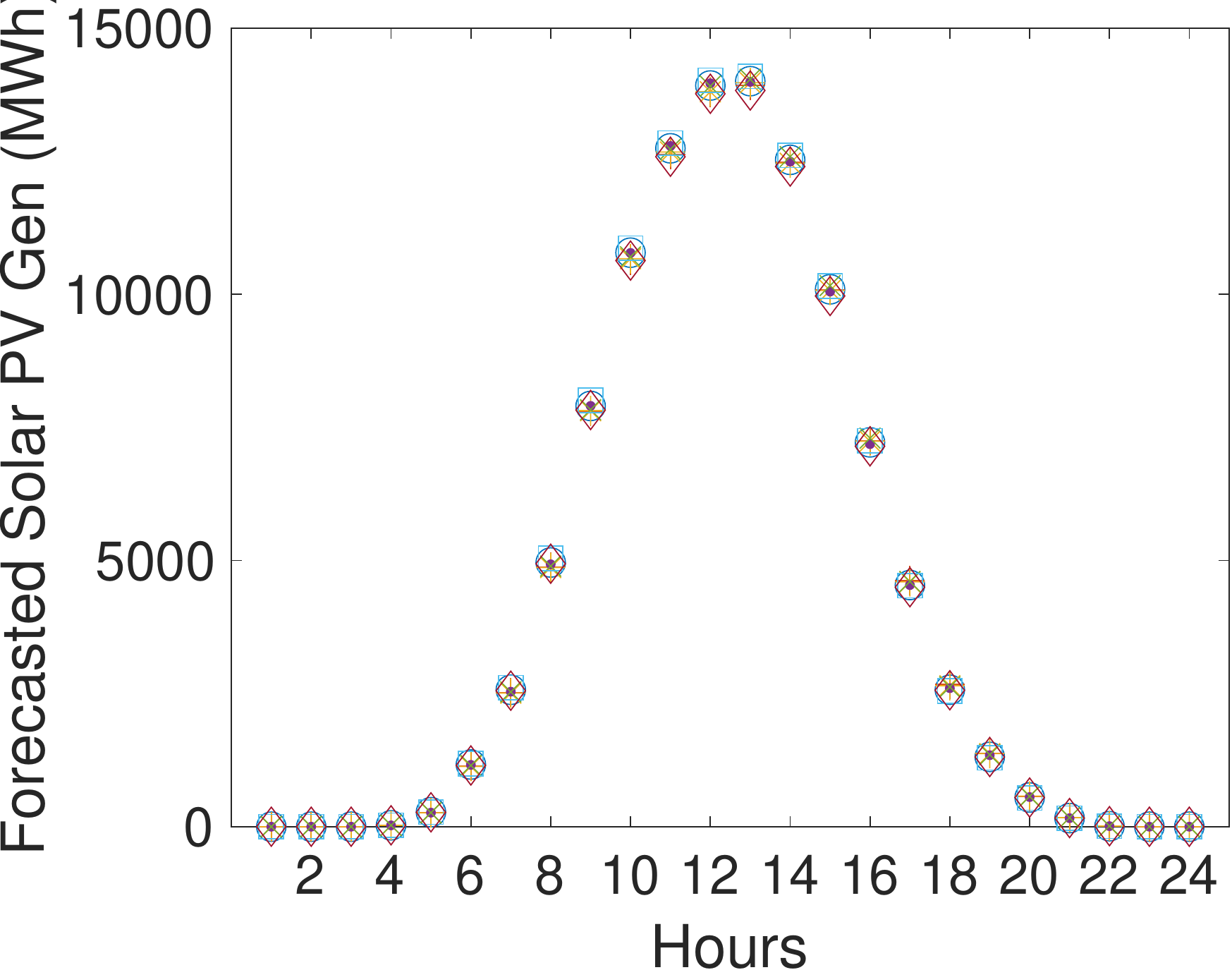} \\
			\includegraphics[width=5cm]{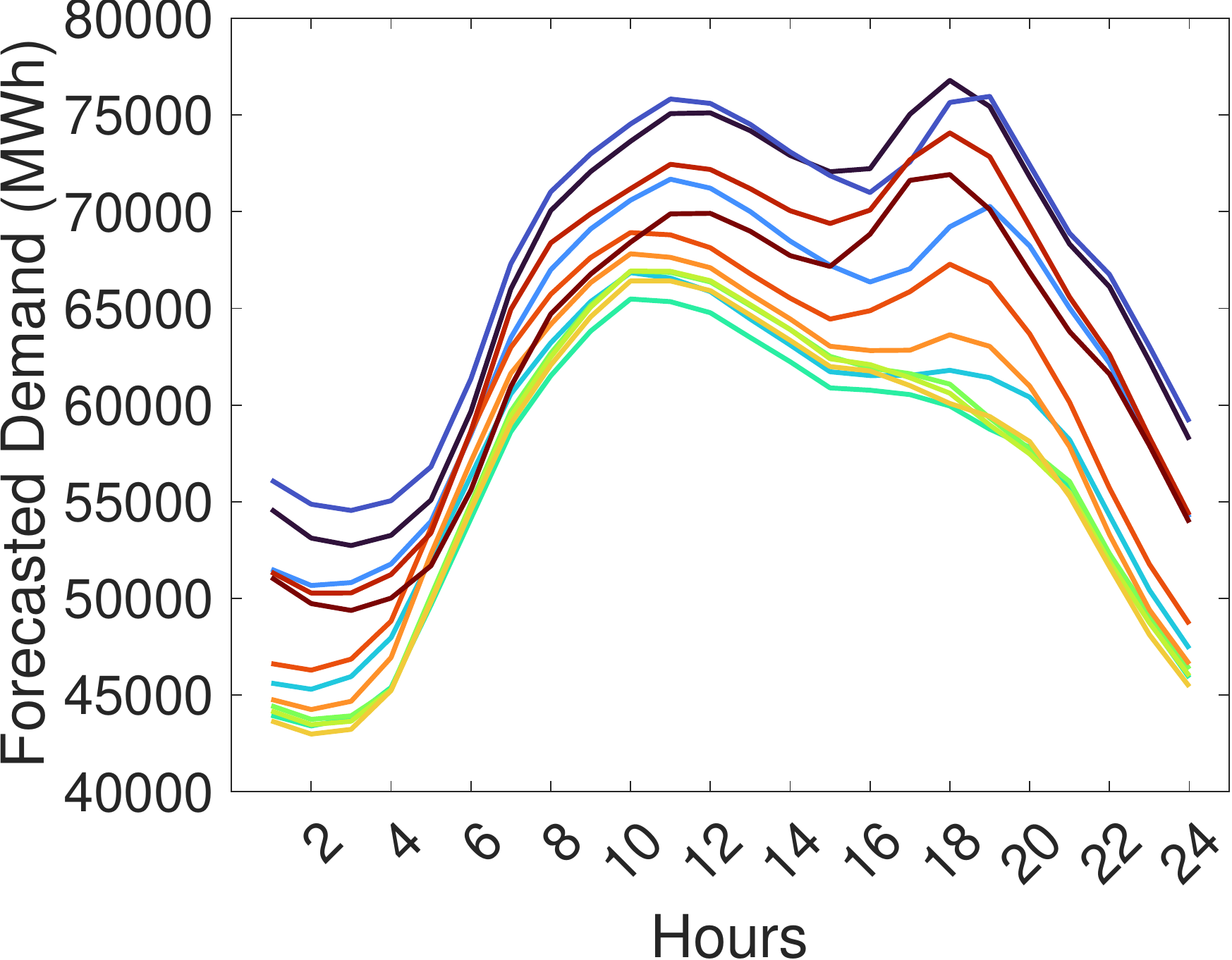}&
			\includegraphics[width=5cm]{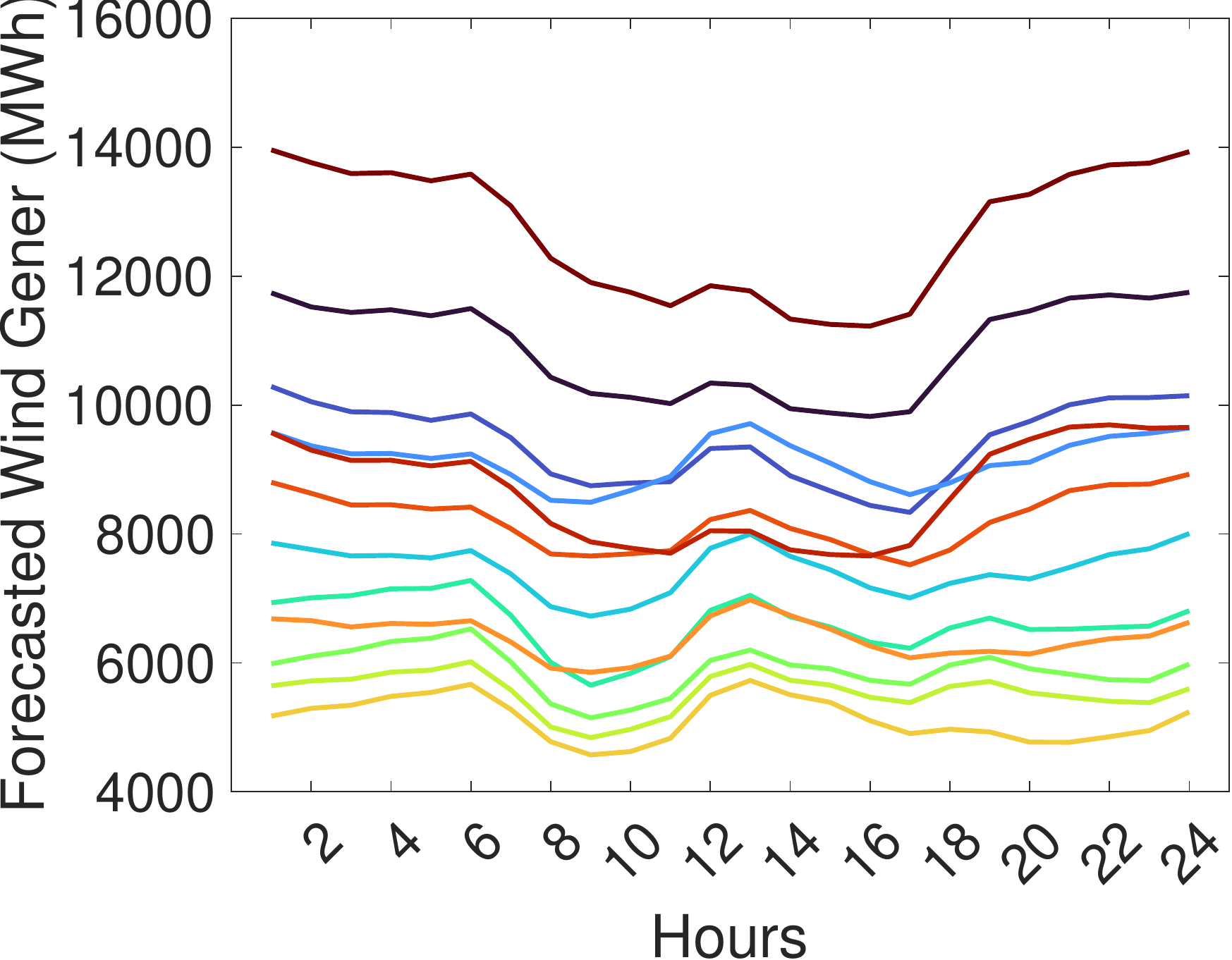} &
			\includegraphics[width=5cm]{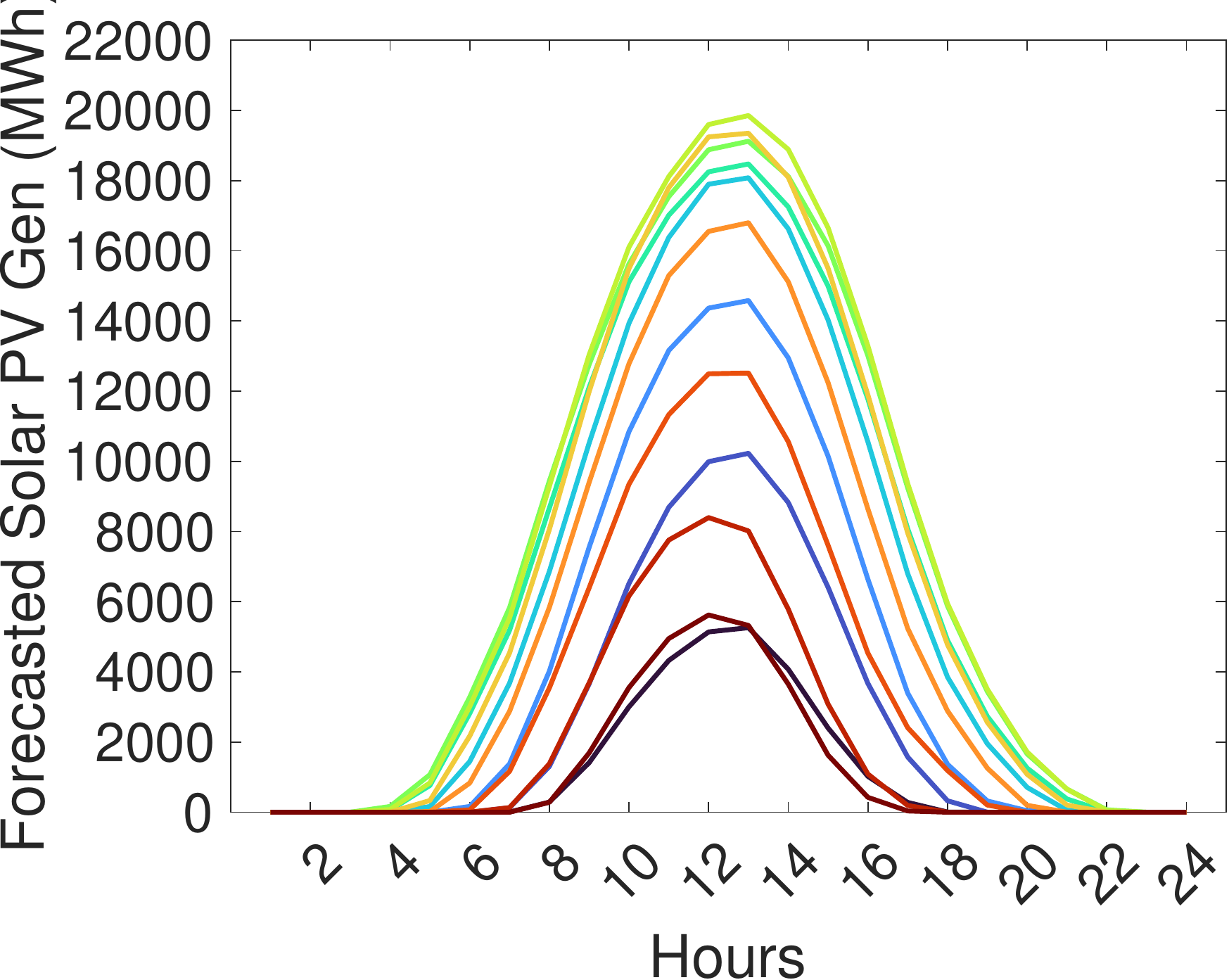} \\
		\end{tabular}
		\caption{\footnotesize{Intra-daily profiles for the different days of the week (top row) and for different months (bottom row) for German Forecasted Demand (left column), Forecasted Wind Generation (middle column) and Forecasted Solar PV Generation (right column). [Sat ($\circ$), Sun ($+$), Mon ($\star$), Tue ($\bullet$), Wed ($\times$), Thu ($\square$), Fri ($\diamond$)]. [Jan (black),  Feb (dark blue), Mar (blue),  Apr (light blue),  May (green),  Jun (light green),  Jul (yellow),  Aug (light orange),  Sep (orange),  Oct (red),  Nov (dark red),  Dec (brown).]}}
		\label{intra_daily}
	\end{figure}

	Interestingly, the panels in Figure \ref{intra_daily} show the profiles for demand, wind and solar generation forecasted over 24 hours, across days of the week and months of the years. These clearly support the importance of modelling weekly and monthly seasonality before undertaking further analysis. In addition, these emphasize the different intra-daily dynamics of demand and RES, which influence the multivariate dependence. In fact,  higher demand is available during peak periods (from hour 8 to hour 20), similarly for solar power, with its peaks around noon, whereas wind is higher during off-peak periods (that is in early mornings and late afternoons) but lower during peak hours.
	
	%
	%
	
	\section{{Methodology}}
	\label{sec_Cop}
	
	The main purpose of the paper is to develop a joint stochastic model that characterizes the marginal behavior of electricity prices, demand, and renewable energy sources by capturing the related dependence structure. 
	To this end, we exploit the advantages of {the} copula methodology, which 
	has been used for economic and financial applications in a number of works (e.g., \cite{Cherubinietal2012}, \cite{Czado19} and \cite{MaiScherer2014}, references therein). Specifically,  an $n-$dimensional copula is a distribution function supported on the unit cube $[0,1]^n$ with a uniform marginal distribution. As well-known, an $n-$dimensional joint distribution function can be decomposed into its $n$ univariate marginal distributions and an $n-$dimensional copula, which is unique when the marginal distributions are continuous. For more details, see also \cite{Durante2016} and \cite{Nelsen2006}. 
	
	Specifically, in view of Sklar's theorem, given an $n$-dimensional distribution function $F$ with marginals $F_j$, for $j=1,\dots,n$, a copula $C:[0,1]^n \to [0,1]$ exists that satisfies
	\begin{equation}
		F(\textbf{y})=C(F_1(y_1),\dots,F_n(y_n)) \notag
	\end{equation} 
	for every $\textbf{y}=(y_1,\dots,y_n) \in \mathbb{R}^n$. If $F$ is continuous, then the copula is uniquely determined by 
	\begin{equation*}
		C(\textbf{u})=F(F_1^{-1}(u_1),\dots,F_n^{-1}(u_n)), \quad \textbf{u} \in [0,1]^n, \label{Sklar}
	\end{equation*} 
	where $F_1^{-1},\dots,F_n^{-1}$ are the quantile functions of $F_1,\dots,F_n$, respectively. In particular, for an absolutely continuous $F$, its density $f$ can be decomposed in the form
	$$
	f(\mathbf y)=c(F_1(y_1),\dots,F_n(y_n))\prod_{i=1}^nf_i(y_i),
	$$
	where $c$ and $f_1,\dots,f_n$ are the density of the copula and of the marginals, respectively.
	
	Here, a version of Sklar's Theorem, adapted to the case of a time series, is considered. Specifically, let $y_{h,i,t}$ be the value of variable $i$ at hour $h$ and on day $t$. To simplify the notation in what follows, the subscript $h$ is suppressed and whenever $y_{i,t}$ is considered, it refers to the electricity price or the renewable energy sources for some given hour of the day ($h=1,\ldots, 24$). 
	Moreover, $\mathbf{Y}_t = (Y_{1,t},\dots,Y_{n,t})$ denotes the random vector for the different variables and for $t=1,\ldots,T$.
	
	Following \cite{Patton2012}, the conditional information generated by past observations of the variables is considered, called $\mathcal{F}_{t-1}^{h}$, for each hour $h$. For simplicity, hereafter $\mathcal{F}_{t-1}$ denotes the information set containing past observations. If we let $F(\cdot|\mathcal{F}_{t-1})$ be the multivariate conditional distribution function of the random vector $\mathbf{Y}_t$ with conditional marginal distribution functions $(F_1(\cdot|\mathcal{F}_{t-1}),\dots,F_n(\cdot|\mathcal{F}_{t-1}))$, then a multi-dimensional conditional copula $C(\cdot|\mathcal{F}_{t-1})$ exists such that 
	\begin{equation*}
		F((y_{1,t},\dots,y_{n,t})|\mathcal{F}_{t-1}) = C\left(F_1(y_{1,t}|\mathcal{F}_{t-1}),\dots, F_n(y_{n,t}|\mathcal{F}_{t-1})|\mathcal{F}_{t-1}\right). \label{Cond_Sklar}
	\end{equation*} 
	Moreover, if the marginal distribution functions are continuous, then the copula is unique. On the opposite side, given the conditional marginal distributions, a copula can be used to link the variables to form a conditional joint distribution with the specified margins.
	
	Furthermore, the \textit{pseudo-observations} are defined as follows: 
	\begin{equation*}
		u_{i,t} = F_i(y_{i,t}|\mathcal{F}_{t-1}), \quad \mbox{ for } i=1,\dots,n \label{PIT_U}
	\end{equation*} 
	and we denote $\textbf{u}_{t} = (u_{1,t},\dots,u_{n,t})$. If the marginal models are correctly specified, then $u_{i,t}$ is uniformly distributed on $(0,1)$ and the conditional copula can be estimated from $\textbf{u}_t|\mathcal{F}_{t-1}$.
	
	As emphasized in \cite{FerWeg12} and \cite{Patton2012}, note that here the same information set is used in each of the marginals and for the copula, then the resulting function is a joint (conditional) distribution function. 
	However, empirically, we can assume that  $F_i(y_{i,t}|\mathcal{F}_{t-1})=F_i(y_{i,t}|\mathcal{F}_{t-1}^i)$ for $i=1,\dots,n$, i.e.  each variable depends on its own past information $\mathcal{F}_{t-1}^i$ but not directly on the past information of any other variable. 
	
	\subsection{The marginal models}

	To find proper marginal distribution models, we consider the four target variables (electricity prices, forecasted demand, wind, and solar PV) separately. Then, following \cite{Patton2012, Pircalabu2017}, AR-GARCH copula models are considered for each hour of the day.
	
	The modelling procedure can be divided into two steps. In the first step, AR-GARCH models are applied to the individual series of prices, demand, and renewable energies, and in addition a deseasonalization is implemented by using dummy variables, for months of the year and weekends. In the second step, the dependence among innovations is studied by applying the copula models proposed in the literature. These two steps are described in what follows. Initially, the AR-GARCH marginals are considered to model the conditional mean and the conditional variance of every single marginal variable. In particular, the AR($p$)-GARCH(1,1) model for the marginal distributions is defined as
	\begin{align*}
		y_{i,t} &= \sum_{j=1}^p \phi_{i,j} y_{i,t-j} + \sum_{k=1}^K \psi_{k} d_{k,t} + \varepsilon_{i,t}, \notag \\
		\varepsilon_{i,t} &= \sigma_{i,t} \eta_{i,t} \, \, \text{ for } \, \,i =1,\dots,n, \label{AR-GARCH}\\
		\sigma_{i,t}^2 &= \omega_i + \alpha_{i} \varepsilon_{i,t-1}^2 + \beta_{i} \sigma_{i,t-1}^2,  \notag
	\end{align*} 
	where $d_{k,t}$ are the dummy variables representing the twelve months of the year plus Saturdays and Sundays, hence $K=14$. Moreover, the parameters $\omega_i,\alpha_i,\beta_i$ follow the usual restrictions for GARCH models, that is $\omega_i>0$ and $\alpha_i + \beta_i<1$ (e.g., \cite{Nelson1992}).
	
	In our empirical application, the total number of variables $n$ is equal to $4$ (which are the electricity prices, forecasted demand, forecasted wind, and forecasted solar PV generation). Following \cite{Gianfreda2018} for the choice of the lag parameters of the different AR-GARCH models, $p=3$ is assumed for the electricity prices; including only the first, the second, and the seventh lag of the hourly prices, hence with a slight abuse of notation. On the other hand, $p=1$ is considered for demand and renewable energy sources, since forecasted variables are used.
	
	Using the AR($p$)-GARCH(1,1) representation described above, the residuals $\eta_{i,t}$ can be represented as follows: 
	\begin{equation*}
		\eta_{i,t}| \mathcal{F}_{t-1} \sim F_{i} \quad \text{ for } i=1,2,\dots,n \, \mbox{ and } \forall t,
	\end{equation*} 
	where $F_i$ comes from the Gaussian distribution. It can be observed that the AR-GARCH models, when properly fitted to univariate time series, produce innovation processes $(\eta_{1,t},\dots,\eta_{n,t})$ that can be considered as serially independent (see \cite{Remillard2017}). Recalling the previous description, the variables of interest are modelled separately for each hour of the day by using four AR($p$)-GARCH(1,1) models with different lags. The following part describes the copula model employed for the residuals. 
	
	\subsection{The copula model}
	
	After having modelled individually the four different variables, their possible dependence is described, 
	at one specific hour of a day, by means of a multivariate copula that capture the relationships among the residuals of the estimated univariate time series. In particular, here we use \textit{vine copulas}. 
	
	Introduced by \cite{BedCoo02} and \cite{Joe96}, vine copulas are built using a cascade of bivariate copulas, called \textit{pair copulas}. This cascade is identified using a set of nested trees called a regular vine tree sequence or regular vine (in short, \textit{R--vine}), which allows to organize and illustrate the needed pairs of variables and their corresponding sets of conditioning variables (see \cite{Aas16} and \cite{Czado19}). In particular, examples of (simplified) regular vine copulas are: (a) multivariate Gaussian copulas, where the pair copulas are bivariate Gaussian copulas with dependence parameter given by the corresponding partial correlation; (b) multivariate Student $t$ copulas with $\nu$ degrees of freedom. 
	
	The estimation procedure for R-vine copulas requires a vine tree structure and the associated bivariate copula families with corresponding parameters. 
	For the selection of vine tree structures, we follow the sequential top-down approach proposed by \cite{Disetal13}. It starts with the tree level one and finds the maximum spanning tree, where each edge has a predefined weight, e.g., the absolute value of the empirical Kendall's $\tau$ between the nodes forming the edge. Then, from a set of bivariate copula families, we select the optimal pair copula families using the Akaike Information Criterion (AIC). For these latter steps, we benefit from the estimation and simulation procedures implemented in \cite{rvinecopulib,VineCopula}. More details about R-vines and related inference procedures are given by \cite{Czado19,Joe2015}.  For a historical account about their use, see \cite{GenSch19}.
	
	\subsection{Modelling tail dependencies}\label{sec:tail}
	
	Different copula types can accommodate flexible dependence patterns in the multivariate case. However, classical families may have some limitations. For instance, the multivariate Gaussian copula does not accommodate any tail dependence and has been criticized after the financial crisis in 2008 (see \cite{PucSch18}). On the other hand, the multivariate Student's $t$ copula does not capture any asymmetry in the tails.
	
	To accommodate a great variety of dependence structures in higher dimensions and overcome the issues of the multivariate elliptical and Archimedean copulas, vine copulas have been used in this analysis. As emphasized by \cite{Joeetal10}, these copulas allow a variety of joint tail behavior of the related distributions.
	
	In order to quantify the degree of dependence in the tails, the so-called \textit{tail dependence coefficients} can be used (see \citealp{DurFerPap15}). Let us recall that, given continuous random variables $X$ and $Y$ defined on the same probability space $(\Omega,\mathcal{F},\mathbb{P})$ with distribution functions $F_X$ and $F_Y$, respectively, the \textit{lower tail dependence coefficient $\lambda_L$} of $(X,Y)$ is defined by
	\[
	\lambda_L(X,Y)=\lim_{t\to 0^+}\mathbb{P}\left( Y\le F_{Y}^{(-1)}(t)\mid X\le F_{X}^{(-1)}(t)\right);
	\]
	and the \textit{upper tail dependence coefficient $\lambda_U$} of $(X,Y)$ is defined by
	\[
	\lambda_U(X,Y)=\lim_{t\to 1^-}\mathbb{P}\left(Y> F_{Y}^{(-1)}(t) \, \mid\, X> F_{X}^{(-1)}(t)\right);
	\]
	provided that the above limits exist.  Here, given a random variable $X$ with distribution function $F$, the quantile function associated with $X$ is given by $F^{(-1)}(t)=\inf\{x\in \mathbb{R}\colon F(x)\ge t\}$.
	
	The upper tail dependence coefficient indicates the asymptotic limit of the probability that one random variable exceeds a high quantile, given that the other variable exceeds a high quantile. A similar interpretation holds for the case of the lower tail dependence coefficient.
	As known (see, for instance, \cite{Durante2016}), tail dependence coefficients only depend on the copula $C$ of $(X,Y)$ in view of the formulas:
	\[
	\lambda_L=\lim_{t \to 0^+}\dfrac{C(t,t)}{t} \quad\mbox{and}\quad
	\lambda_U=\lim_{t \to 1^-}\dfrac{1-2t+C(t,t)}{1-t}.
	\]
	Clearly, both coefficients take values in $[0,1]$. In particular, $X$ and $Y$ are said to be asymptotically independent in the lower (respectively, upper) tail when $\lambda_L(X,Y)=0$ (respectively, $\lambda_U(X,Y)=0)$.
	
	Now, let us assume to have a multidimensional random vector and there is an interest in the tendency of some of the components to achieve extreme values simultaneously, that is taking extremely small or extremely large values. Given that there are more than two components, it is not obvious how to define a tail dependence index, and several contributions attempting to provide a solution appeared in the literature (see \cite{Duretal16,Jaw09,Gijetal20}). 
	Here, in order to describe the tail dependence in a multivariate setting, we propose two novel tail dependence coefficients, inspired by the recent studies in conditional value-at-risk in \cite{Beretal18SERRA}.
	
	Specifically, in order to quantify how high (respectively, small) values of one variable, say $Y$, are influenced by two or more variables, say $X_1,\dots,X_l$, we proceed as follows. First, we select a given threshold level for the variable $Y$, corresponding to its $\beta$ quantile. Second, we select a suitable region $B\subseteq \mathbb{R}^l$ such that $\mathbb{P}((X_1,\dots,X_l)\in B)=\alpha$. Such a region $B$ collects all the realizations of the vector $(X_1,\dots,X_l)$ that are judged to be extreme (i.e. either very small or very high). Finally, given a copula-based model for the random vector $(X_1,\dots,X_l,Y)$ we calculate, for a suitable $\beta\in (0,0.5)$,
	
	\begin{equation}\label{eq:qL}
		q_L(\alpha,\beta)=\mathbb{P}(Y\le F_Y^{-1}(\beta)|(X_1,\dots,X_l)\in B_L);
	\end{equation}
	to take into account negative tails, and
	\begin{equation}\label{eq:qU}
		q_U(\alpha,\beta)=\mathbb{P}(Y\ge F_Y^{-1}(1-\beta)|(X_1,\dots,X_l)\in B_U);
	\end{equation}
	instead for positive tails, for some sets $B_L$ and $B_U$ in $\mathbb{R}^l$.
	
\begin{figure}[ht]
		\centering
		\begin{tabular}{lr}
			\includegraphics[scale=0.45]{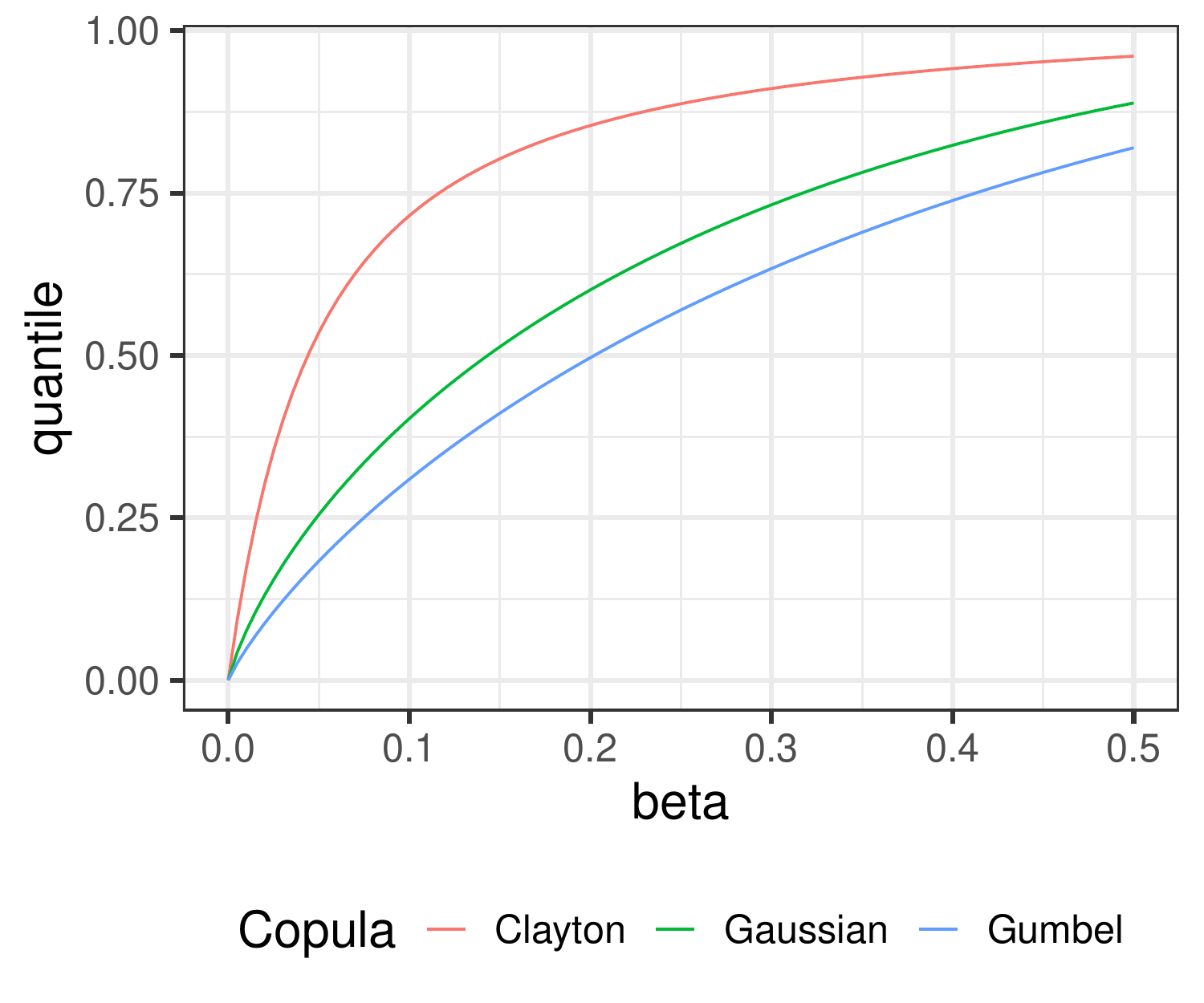}
			&
			\includegraphics[scale=0.45]{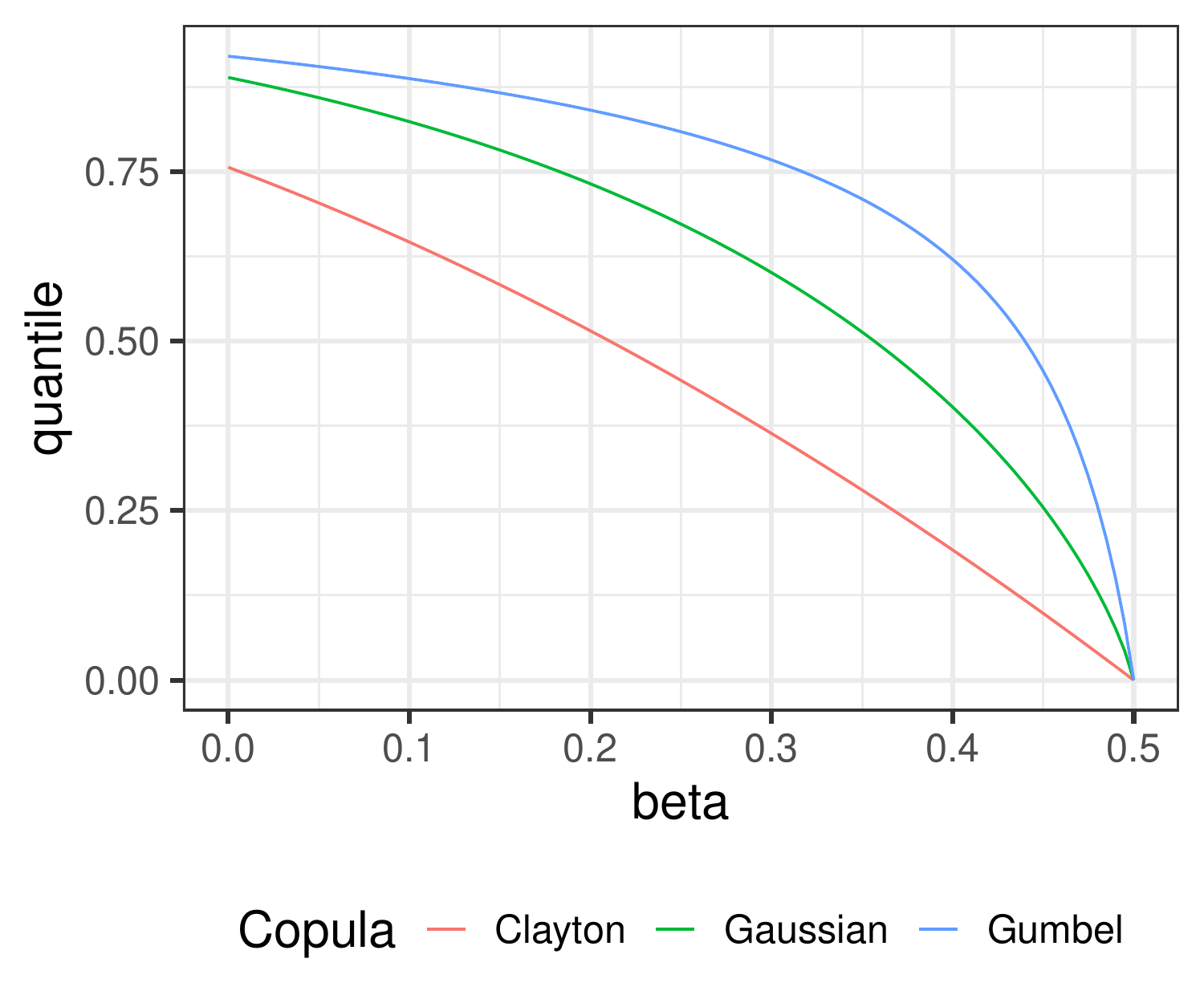}
		\end{tabular}
		\caption{\footnotesize{Graphs of the functions $q_L$ in \eqref{eq:QL2} and $q_U$ in \eqref{eq:QU2} for $\alpha=0.05$ and different copula models with the same Spearman's correlation equal to $0.5$}}
		\label{Fig_QL}
	\end{figure}

	\begin{exa}
		Consider, for instance, $l=1$, i.e. we are interested in the random pair $(X,Y)$. Then it is natural to select $B_L=(-\infty,F_X^{-1}(\alpha)]$, and $B_U=[F_X^{-1}(1-\alpha),+\infty)$ for $\alpha\in (0,0.5)$. Then
		\begin{equation}\label{eq:QL2}
			q_L(\beta)
			=\mathbb{P}(Y\le F_Y^{-1}(\beta)|X\le F_X^{-1}(\alpha))
			=\dfrac{C(\alpha,\beta)}{\alpha},
		\end{equation}
		where $C$ is the copula of $(X,Y)$  and $\beta\in (0,0.5)$. For $\alpha=\beta$, $q_L(\beta)$ defines the tail concentration function used in \cite{DurFerPap15,Patton2012}. Analogously, it holds
		\begin{equation}\label{eq:QU2}
			q_U(\beta)
			=\mathbb{P}(Y\ge F_Y^{-1}(1-\beta)|X\ge F_X^{-1}(1-\alpha))
			=\dfrac{\overline{C}(1-\alpha,1-\beta)}{\alpha},
		\end{equation}
		where $\widehat{C}$ is the survival copula associated with $C$, given by $\widehat{C}(x,y)=x+y-1+C(1-x,1-y)$ for every $(x,y)\in [0,1]^2$. For $\alpha=0.05$ we show in Figure \ref{Fig_QL} the graphs of $q_L(\beta)$ and $q_U(\beta)$ for three families of copulas with the same Spearman's correlation equal to $0.5$, namely Gaussian copula (that is symmetric in the tail), Gumbel copula (that has an upper tail dependence coefficient different from $0$, while it shows asymptotic independence in the lower tail), Clayton copula (that has a lower tail dependence coefficient different from $0$, while it shows asymptotic independence in the upper tail).
	\end{exa}
	
	The selection of the region $B$ is crucial and depends on the application. As in \cite{DiB14}, \cite{Cobetal18} and \cite{SalDeMDur11}, the notion of Kendall scenario is used.
	Specifically, let $L_{F}(t)$ denote the $t$-level curve of the joint distribution function $F$ of $(X_1,\dots,X_l)$. Thus, we set
	\begin{itemize}
		\item $B_L=\bigcup_{0\le t\le t_L^\alpha} L_{F}(t)
		=\{\mathbf{x}\in \mathbb{R}^l\colon F(\mathbf{x})\le t_L^\alpha\}$,
		\item $B_U=\bigcup_{t_U^\alpha\le t\le 1} L_{F}(t)
		=\{\mathbf{x}\in \mathbb{R}^l\colon F(\mathbf{x})\ge t_U^\alpha\}$,
	\end{itemize}
	where $t_L^\alpha$ and $t_U^\alpha$ are suitable values so that the probability that $(X_1,\dots,X_l)$ belongs to $B_L$ (respectively, $B_U$) is equal to $\alpha$.
	
	Now, let $\alpha,\beta\in (0,0.5)$. If we denote by $F_\mathbf{X}$ the distribution function of the random vector $(X_1,\dots,X_l)$, then it holds
	\begin{align*}
		q_L^K(\alpha,\beta)
		&=\mathbb{P}(Y\le F_Y^{-1}(\beta)\mid \mathbf{X}\in B_L)\\
		&=\dfrac{\mathbb{P}(F_{\mathbf{X}}(\mathbf{X})\le t_L^{\alpha},F_Y(Y)\le\beta))}{\mathbb{P}(\mathbf{X}\in B_L)}
		=\dfrac{D(K_{\mathbf{X}}(t_L^{\alpha}),\beta)}{\alpha},
	\end{align*}
	where $K_{\mathbf{X}}$ is the distribution function of $F_{\mathbf{X}}(\mathbf{X})$, known as \textit{Kendall function} associated with $\mathbf{X}$ (see \cite{Durante2016,Hofetal19,NapSpi09}). Moreover, $D$ is the copula associated with the random pair $(F_{\mathbf{X}}(\mathbf{X}),F_Y(Y))$. Note that $F_Y(Y)$ is uniformly distributed on $[0,1]$, and it is known as the \textit{probability integral transform} of $Y$. However,  $F_{\mathbf{X}}(\mathbf{X})$ is not uniformly distributed on $[0,1]$ and it can be considered as a multivariate probability integral transform.
	
	Analogously, let $\alpha,\beta\in (0,0.5)$. It holds
	\begin{align*}
		q_U^K(\alpha,\beta)
		&=\mathbb{P}(Y\ge F_Y^{-1}(1-\beta)\mid \mathbf{X}\in B_U)\\
		&=\dfrac{\mathbb{P}(F_{\mathbf{X}}(\mathbf{X})\ge t_U^\alpha,F_Y(Y)\ge 1-\beta)}{\mathbb{P}(\mathbf{X}\in B_U)}
		=\dfrac{\widehat{D}(1-K_{\mathbf{X}}(t_U^\alpha),1-\beta)}{\alpha},
	\end{align*}
	where $K_{\mathbf{X}}$ is the distribution function of $F_{\mathbf{X}}(\mathbf{X})$, and $\widehat{D}$ is the survival copula of $(F_{\mathbf{X}}(\mathbf{X}),F_Y(Y))$.
	
	\begin{remark}
		Note that, under independence of $Y$ and $\mathbf{X}$, it holds that
		$$
		q_L^K(\alpha,\beta)=\beta,\qquad q_U^K(\alpha,\beta)=1-\beta.
		$$
		Therefore, the ratio $q_L^K(\alpha,\beta)/\beta$ (respectively,
		$q_U^K(\alpha,\beta)/(1-\beta)$) quantifies the relative effect on the tail of $Y$ provided by an extreme scenario related to $\mathbf{X}$.
	\end{remark}
	
	Now, analogously to bivariate tail dependence coefficients, we can introduce the following multivariate versions of tail dependence coefficients
	\begin{equation}\label{eq:MTDC}
		\lambda_L^K(Y\mid \mathbf{X})=\lim_{\alpha\to 0^+}q_L^K(\alpha,\alpha),\qquad
		\lambda_U^K(Y\mid \mathbf{X})=\lim_{\alpha\to 0^+}q_U^K(\alpha,\alpha),
	\end{equation}
	provided that the above limits exist and are finite.
	Here, the suffix $K$ is helpful to remind that the conditional event is obtained from the Kendall distribution.
	
	Operationally, the coefficients defined in Eq.~\eqref{eq:MTDC} are classical (bivariate) tail dependence coefficients between $Y$ and $F_{\mathbf{X}}(\mathbf{X})$, that is an aggregation of $\mathbf{X}$ via the collapsing function $F_{\mathbf{X}}$ (see \cite{Hofetal19}). Then, their estimation depends on the bivariate copula of $(F_{\mathbf{X}}(\mathbf{X}),Y)$ and it can be obtained by implementing standard techniques like those described in \cite{SchSch07}.
	
	%
	\section{{Empirical Results}}
	\label{sec_Copula_Insample}
	
	Given the high penetration of wind and solar power in Germany, it is interesting to consider a joint model for electricity prices, demand, and RES to capture the dependence effects. Specifically, first it is considered the relationships between prices, demand, and wind, since solar is only available during midday hours (that is from hour 8 to hour 16). Then, we focus on a more interesting quadrivariate copula model to account for the possible interactions between prices, wind, and demand, while also considering solar power. To understand the dependence structure,  
	vine copula models are used to account for possible different behaviors in the tails. 
	In what follows, a global analysis of the whole dataset is presented, and subsequently, a rolling window approach is considered to depict the time-varying correlations. 
	
	\subsection{Global analysis over {the full sample of years} 2011-2019}
	
	Using the entire sample of the full nine years, the joint dependence structure between electricity prices, forecasted demand, and RES is investigated through vine copula models.
	
\begin{figure}[h!]
		\begin{center}
			\includegraphics[scale=0.50,angle=270]{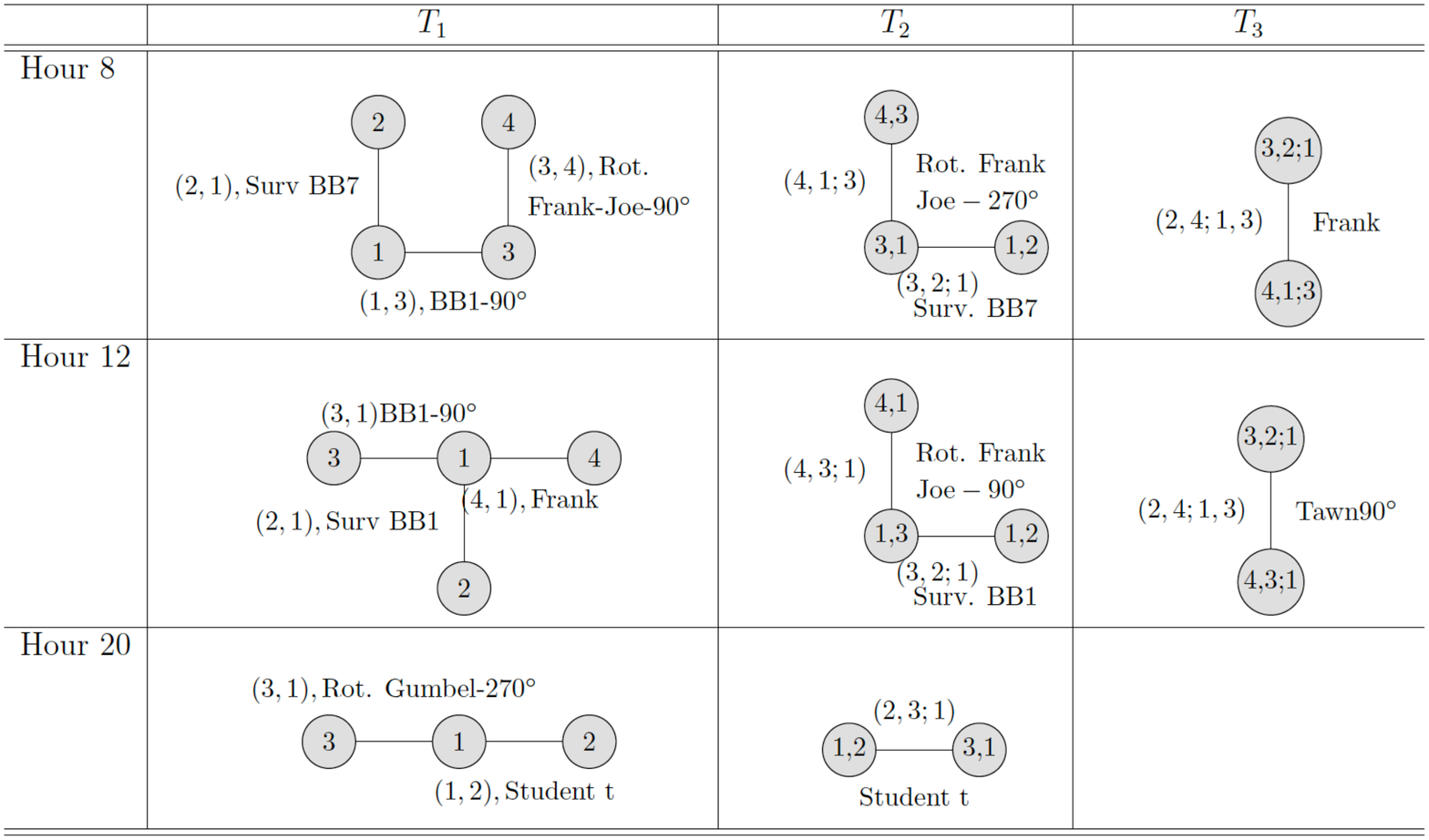}
		\end{center}
		\caption{\footnotesize{Tree Structure of a Vine Copula Model for Hours 8 to 12 and 20. Note that $1$ stands for the Electricity Prices; $2$ for the Forecasted Demand; $3$ for the Forecasted Wind and $4$ for the Forecasted Solar PV. From $17$ to midnight and from midnight to $7$, we run a trivariate copula (without Forecasted Solar PV).}} \label{VineTree}
	\end{figure}

	First, the R-vine copula is estimated for each hour, i.e., $h=1,\dots,24$,  by determining the tree structure and the involved pair-copulas. An example is visualized in Figure \ref{VineTree}, where results for hours $8$, $12$, and $20$ are presented. Other results are omitted but are available upon request.
	
\begin{figure}[ht]
		\centering
		\begin{tabular}{lr}
			\includegraphics[scale=0.48]{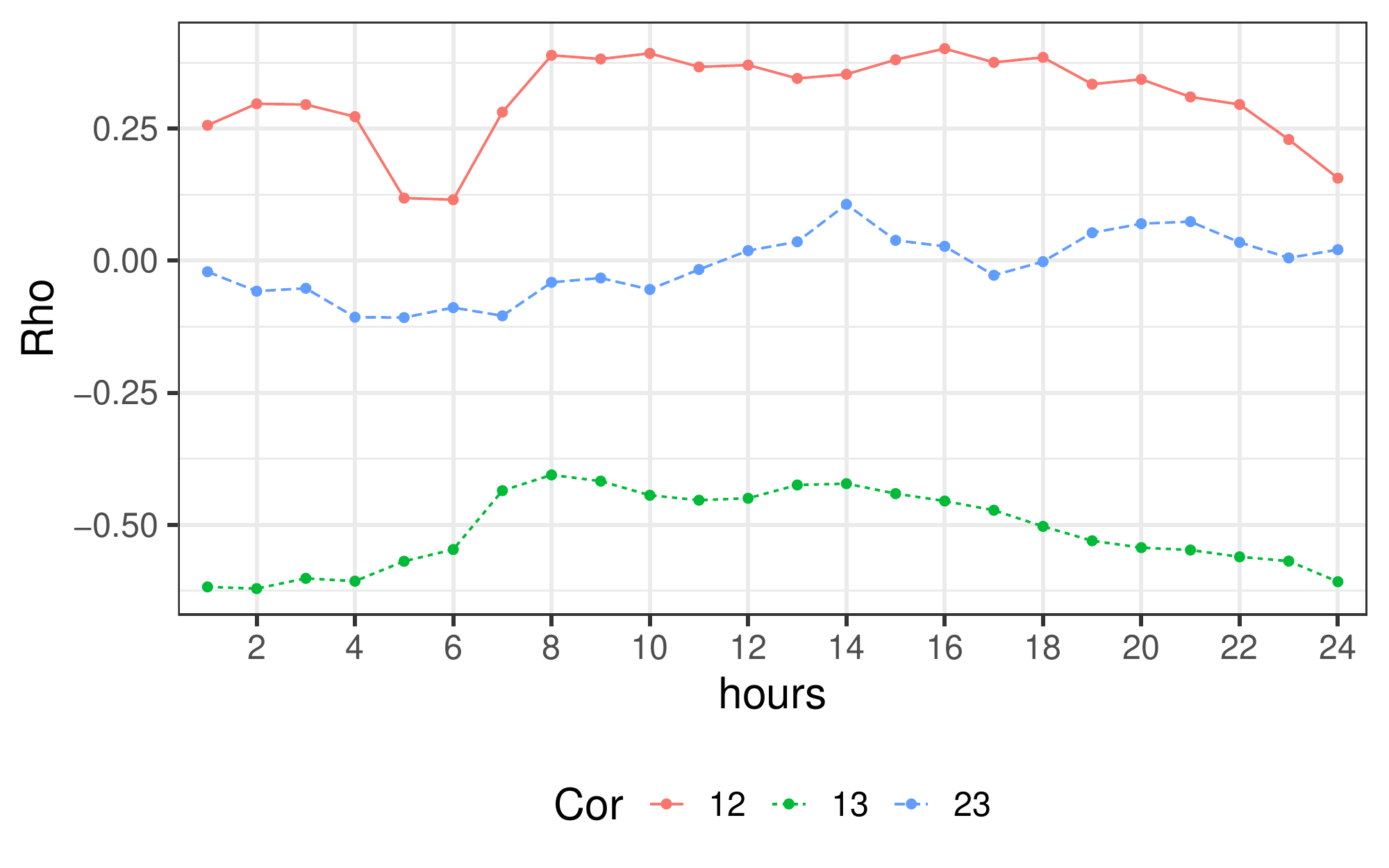}
			&
			\includegraphics[scale=0.48]{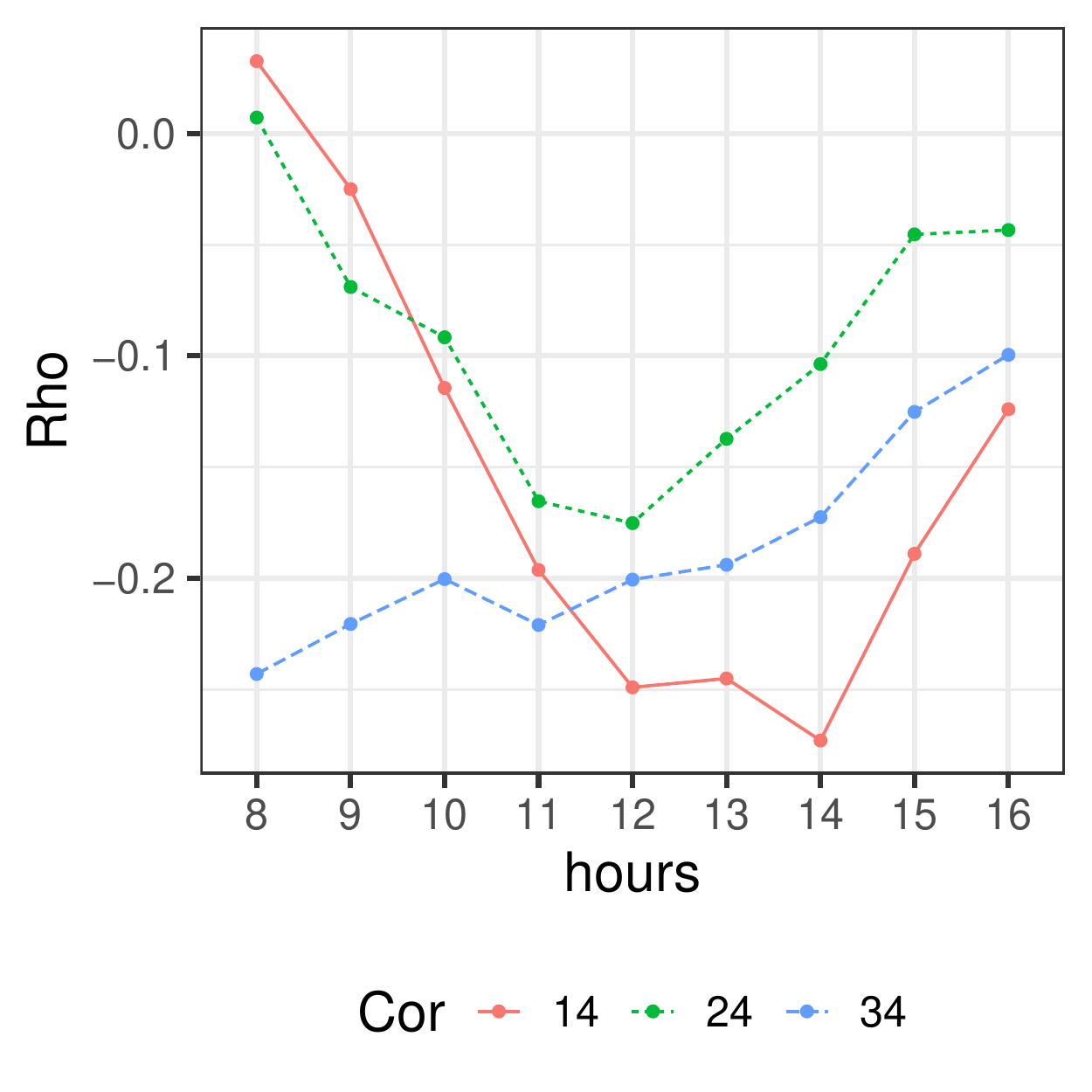}
		\end{tabular}
		\caption{\footnotesize{Pairwise Spearman's correlations induced by the R-Vine Copula model specification over the 24 Hours between Electricity Prices (1), Forecasted Demand (2) and Forecasted Wind (3) on the left; and, among Forecasted Solar PV (4) and the other variables on the right.}}
		\label{Fig_Correl_Insample}
	\end{figure}

	Then, the induced pairwise (Spearman's) correlation is computed for each hour and presented in Figure \ref{Fig_Correl_Insample}. Note that the link between the forecasted solar power generation and the other variables is included only from hour $8$ to hour $16$. In particular, Figure \ref{Fig_Correl_Insample} shows a positive dependence between the electricity prices and the forecasted demand during the entire $24$ hours. The correlation falls during the early morning (i.e. from $5$ to $6$ approximately at $0.2$), and in the late evening after $20$); hence, confirming the known fact that prices follow the intra-day dynamics of demand, being higher during peak hours and lower in off-peak hours. When the relation between electricity prices and forecasted wind generation is instead considered, a negative correlation is detected, recalling the reverse dynamics of the intra-daily wind profile. Indeed, the negative correlation is larger when wind generation is high (during early or late hours) and it diminishes, keeping its sign, when wind generation decreases (during peak hours, as shown on the left side of Figure \ref{intra_daily}). As expected, the correlation between forecasted demand and wind fluctuates around zero and indeed this is not of concern for this analysis since both variables are influenced by weather conditions.
	
	The most interesting results concern the dependence of electricity prices on solar power (see right side of  Figure \ref{intra_daily}). Similar to wind, a reverse situation to the intra-daily profile observed for the forecasted solar generation can be detected, with the correlation becoming progressively more negative when solar generation increases over the central hours. More specifically, the hours between $8$ and $16$ show a correlation found to be mostly negative, apart from the first hours when the sun is weakly shining (that is at $8$ and $9$ in the morning). This confirms that the increasing forecasted solar PV production leads to a decrease in electricity prices. In particular, the lowest negative value of $-0.25$ is observed at midday.
	
	Moving forward and considering the less investigated dependencies between demand and solar, we do empirically observe a negative correlation between the forecasted demand and solar PV production, with a major impact around noon, recalling again the intra-day dynamics of solar PV generation.  In this way, an increase in the forecasted PV production at noon leads to a decrease in the forecasted demand.
	Finally, the correlation between forecasted wind and solar PV is also considered. And, in this case, interestingly, the correlation is found to be negative across all considered central hours. Results regarding the negative correlation between demand and solar are in agreement with the common practice of thinking of solar power as \textit{negative demand}. 
	
	Overall, these results confirm the well-known \textit{merit order effect}, according to which RES (wind and solar) decrease the electricity prices because they enter the supply curve before the other generation sources and, consequently, they shift the supply curve towards the right, thus decreasing the equilibrium price. However, the results presented in this specific analysis do refer to correlations when considering a multivariate dependence model, that is when all possible interactions between involved variables are considered. More explicitly, correlations are studied when prices interact with individual RES and when RES interact with demand as well. The same results may not occur, for instance, when simple pair-wise correlations or regression models are considered, in which the marginal effects of RES are hypothesized \textit{ceteris paribus}.
	
\begin{figure}[h!]
		\centering
		\begin{tabular}{cc}
			\includegraphics[scale=0.48]{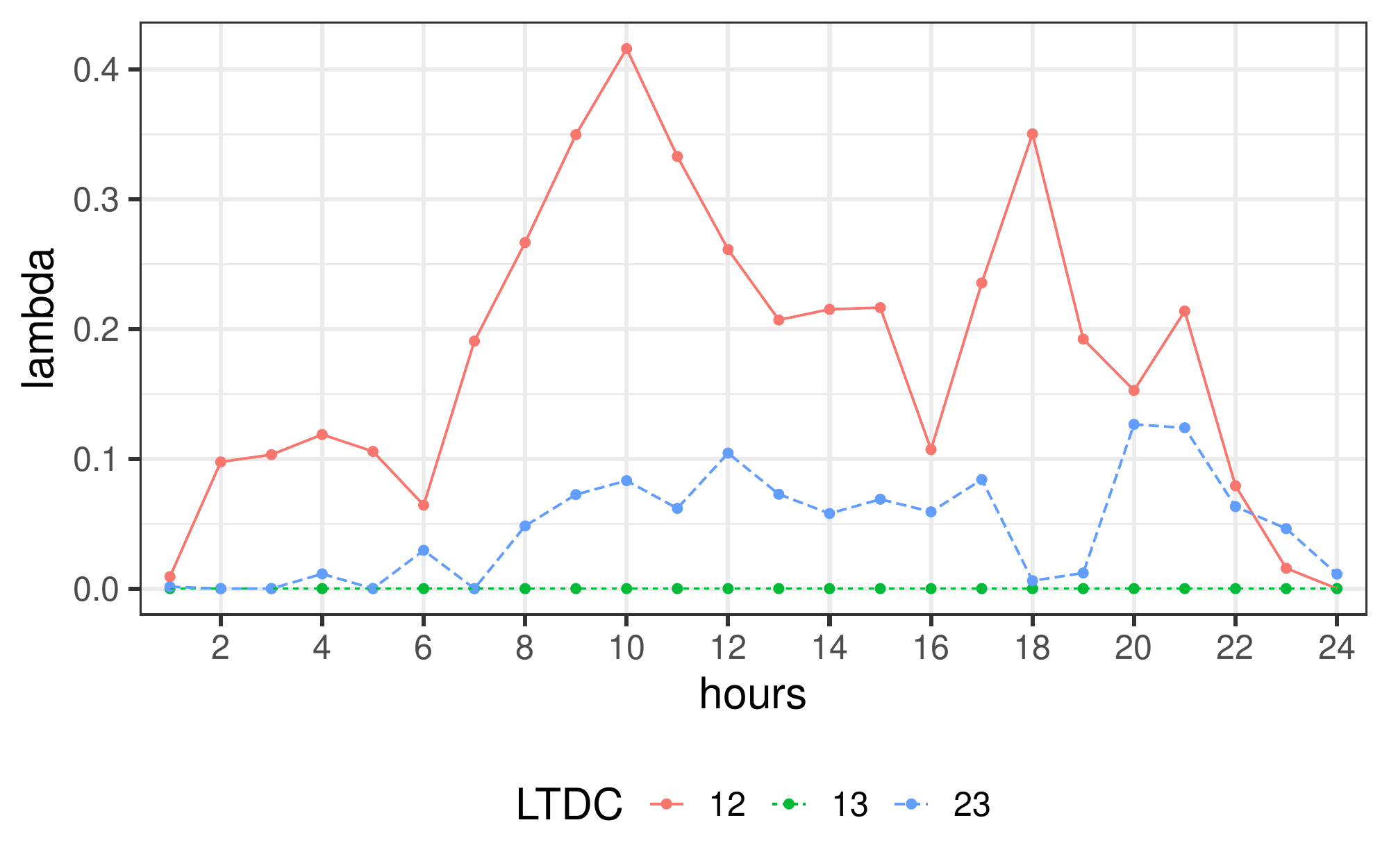}
			&
			\includegraphics[scale=0.48]{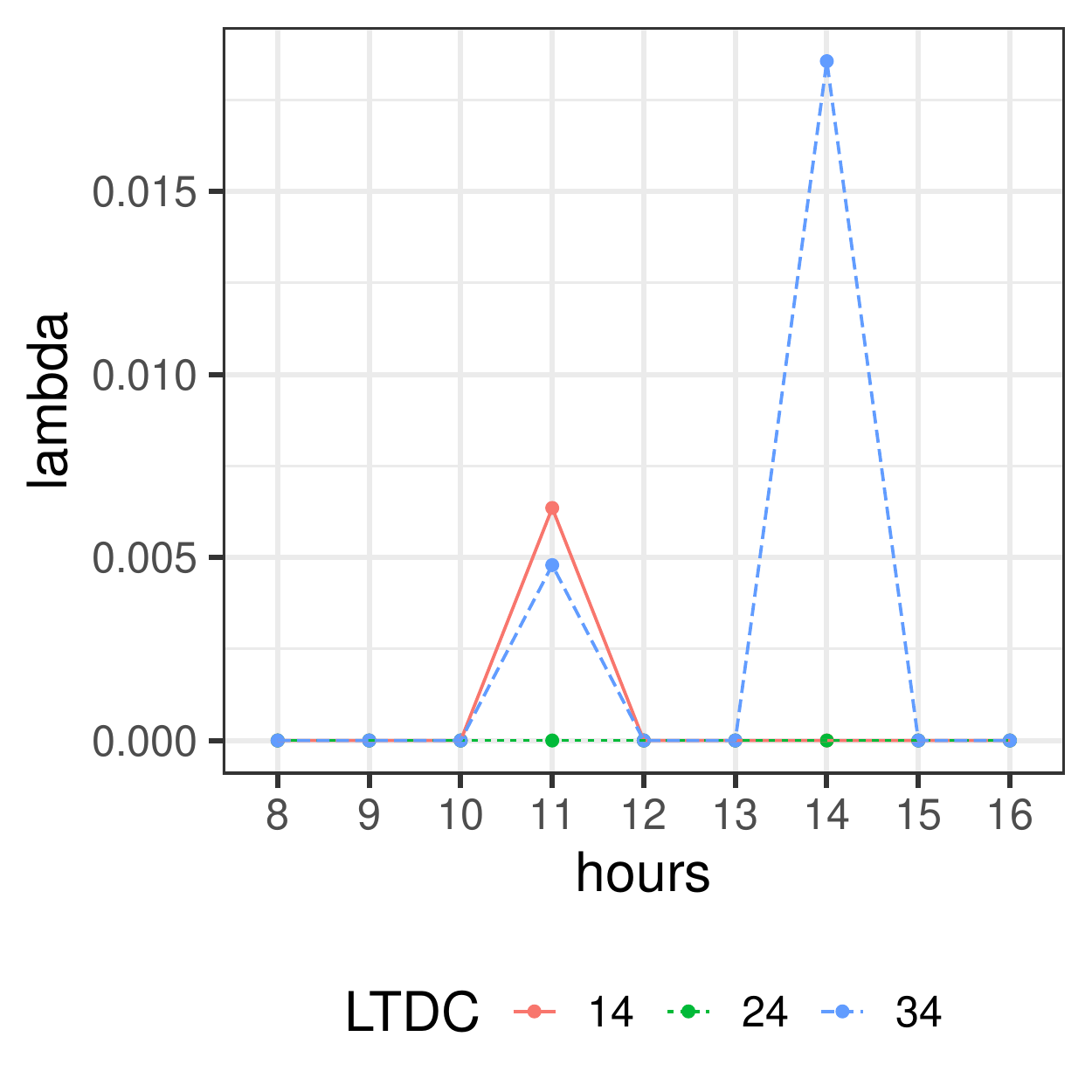}\\
			(a) & (b)\\
			
			\medskip\\
			
			\includegraphics[scale=0.48]{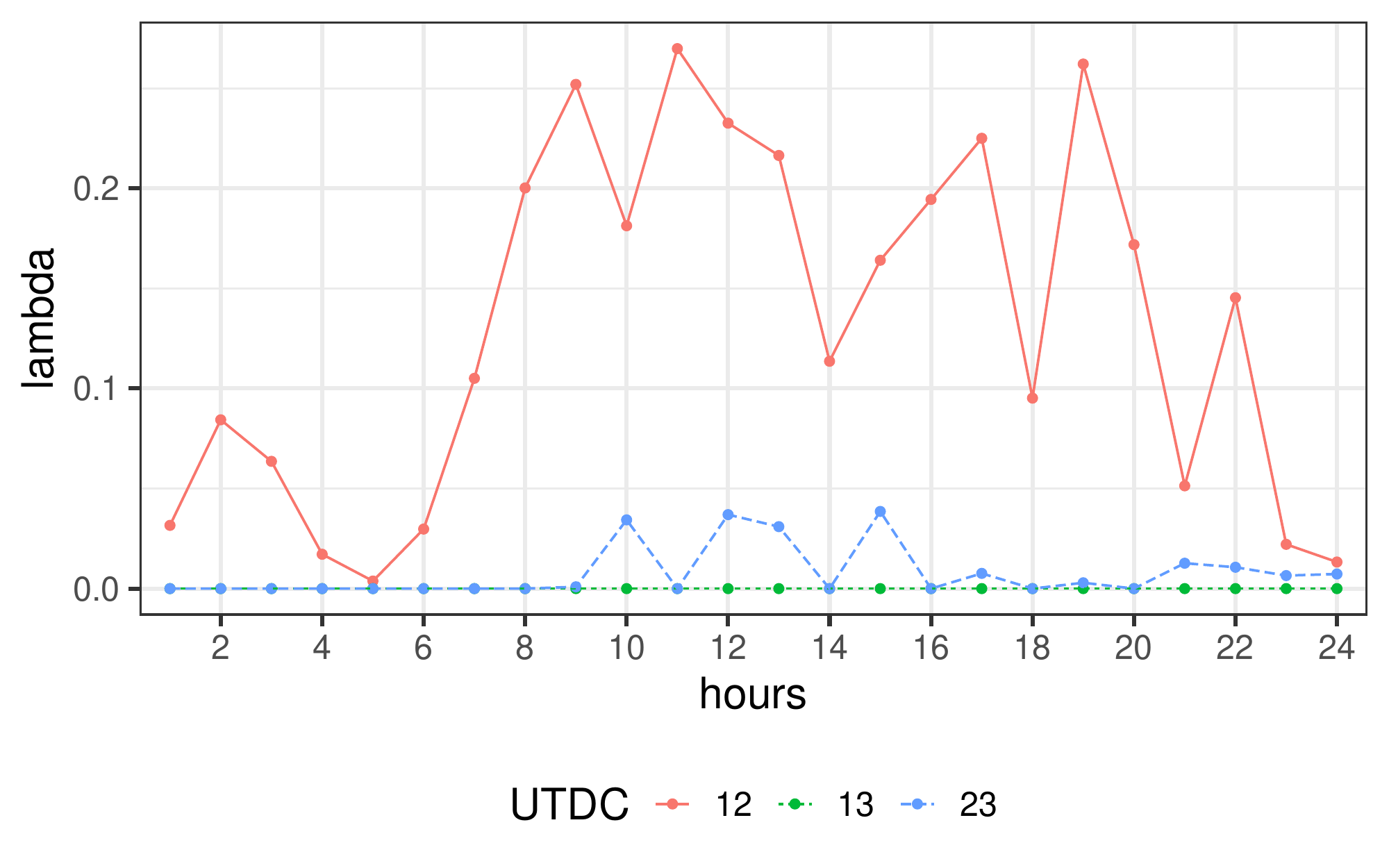}
			&
			\includegraphics[scale=0.48]{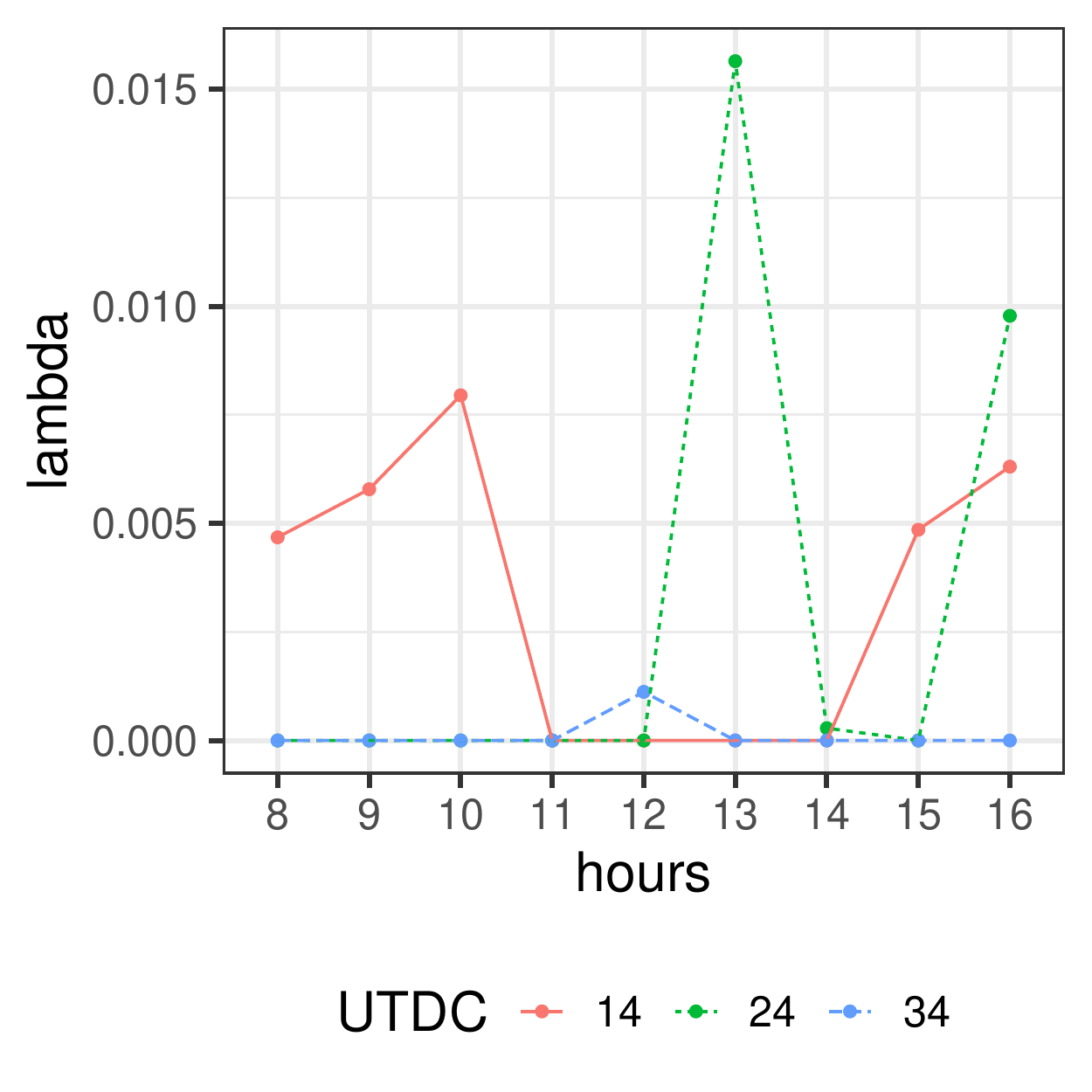}\\
			(c) & (d)\\
		\end{tabular}
		\caption{\footnotesize{Pairwise lower (a-b) and upper (c-d) tail dependence coefficients induced by the R-Vine Copula model specification over the $24$ hours between Electricity Prices (1), Forecasted Demand (2), Forecasted Wind (3), and Forecasted Solar PV (4).}}
		\label{Fig_2TDC_Insample}
	\end{figure}

	Apart from the global correlation, measures of tail dependence have been considered and computed, as described before, between the different variables during the whole $24$ hours. Figure \ref{Fig_2TDC_Insample} shows the model-based pairwise upper and lower tail dependence coefficients (UTDC and LTDC, in short) to capture the extra effect of one variable on the high/low values of the other variable in a pairwise tail dependence, resulting from the multivariate structure.
	
	It can be easily observed that independently from the tails, the coefficient of tail correlation between prices and demand is always positive and varying over the day with dynamics recalling the intra-daily profiles: lower correlations early in the morning and in the evening, higher ones during the middle of the day. This comes with no surprise apart the magnitudes expected to be higher over the right tail when demand pushes power plants under pressures, hence resulting in higher equilibrium prices. However, here the multivariate dependence detects also the interaction between demand and wind, which is indeed higher on the left tail during central hours, and thus resulting in a higher influence on prices. Instead, the most striking result is the asymptotic tail independence between prices and wind on both tails and across all hours, since previous studies have shown how wind instead does influence the left tail of prices  at finite, i.e. non-extreme, quantile levels.
	
	When also solar is included in the model, the tail dependence coefficients with prices are extremely low: around $0.005$ at $11$ for the LTDC and $0.0055$ at $10$ for the UTDC. In the former case, it may indicate some residual effect of high demand, whereas in the latter case it clearly shows the dependence exactly out of the solar peak generation, that is from $9$ to $11$ and from $14$ to $16$. Moreover, the correlation between demand and solar is at its maximum values at hours $13$ and $16$; again recalling the intra-daily profiles.
	
\begin{figure}[ht]
		\centering
		\includegraphics[scale=0.5]{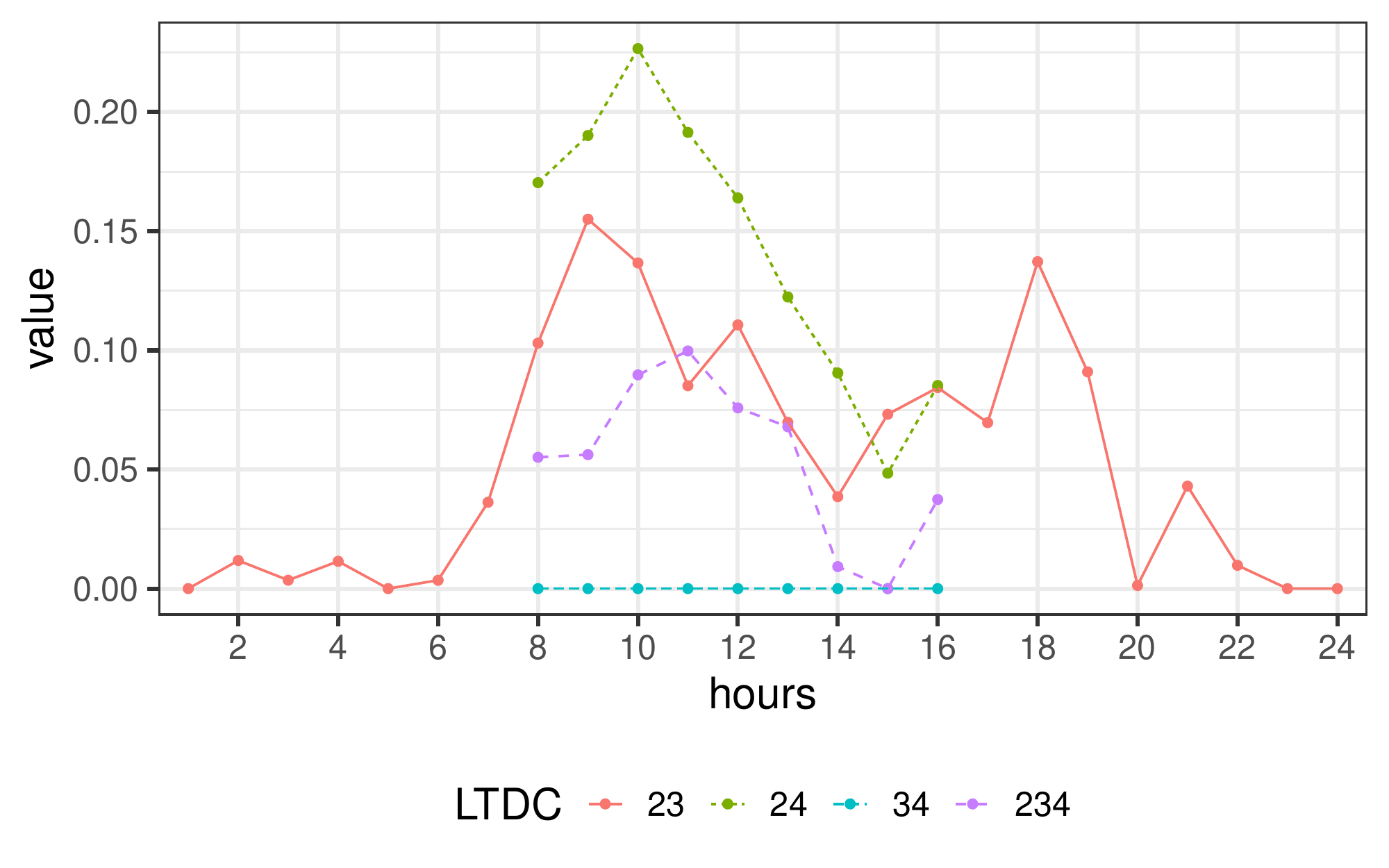}\\
		\includegraphics[scale=0.5]{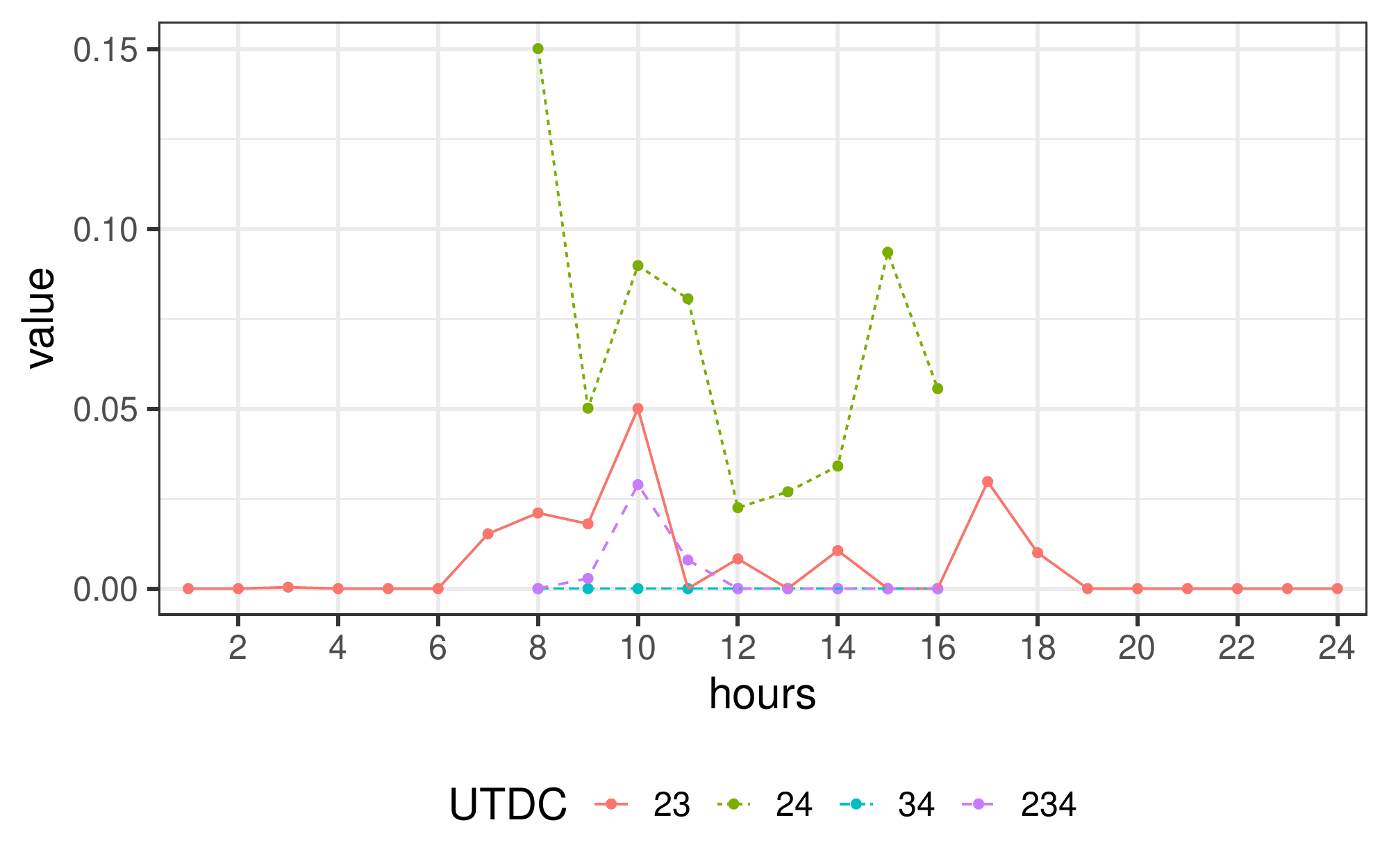}
		\caption{\footnotesize{Multivariate lower (top) and upper (bottom) tail dependence coefficients induced by the R-Vine Copula model specification related to Electricity Prices (1) given the other variables: Forecasted Demand (2), Forecasted Wind (3), and Forecasted Solar PV (4).}}
		\label{Fig_TDC_Price_Insample}
	\end{figure}

	Together with the pairwise analysis, it is relevant to visualize the joint effect of two or more variables on electricity prices. In particular, it is relevant to inspect whether high (respectively, low) values of electricity prices are influenced by extreme events occurring to: $a)$
	both forecasted demand and forecasted wind; $b)$ both forecasted demand and forecasted solar; $c)$ forecasted solar and forecasted wind; and finally, $d)$ all the three previous variables together. To this end, we use the indices $\lambda_L^K$ and $\lambda_U^K$ discussed in section \ref{sec:tail} to quantify how much high (respectively, small) values of one variable are influenced by simultaneous extreme values of two or more variables. To visualize the case when high (respectively, small) price values are influenced by extreme high (respectively, low) values for all the other three variables, the multivariate tail dependence coefficients are considered as depicted by the R-Vine Copula model related to prices. Results are shown in Figure \ref{Fig_TDC_Price_Insample}.
	
	According to the same methodology, we also describe how high electricity prices are linked with high demand, but low wind and solar power (this is identified as the HLL scenario); or with high demand and wind, but low solar (the HHL scenario); and finally, with high demand, low wind and high solar (HLH scenario). These different combinations of variables reflect the idea of looking at the dependence structure from many facets of the joint distribution; see for instance \cite{QueUbe12}.
	
\begin{figure}[ht]
		\centering
		\begin{tabular}{ccc}
			\includegraphics[scale=0.4]{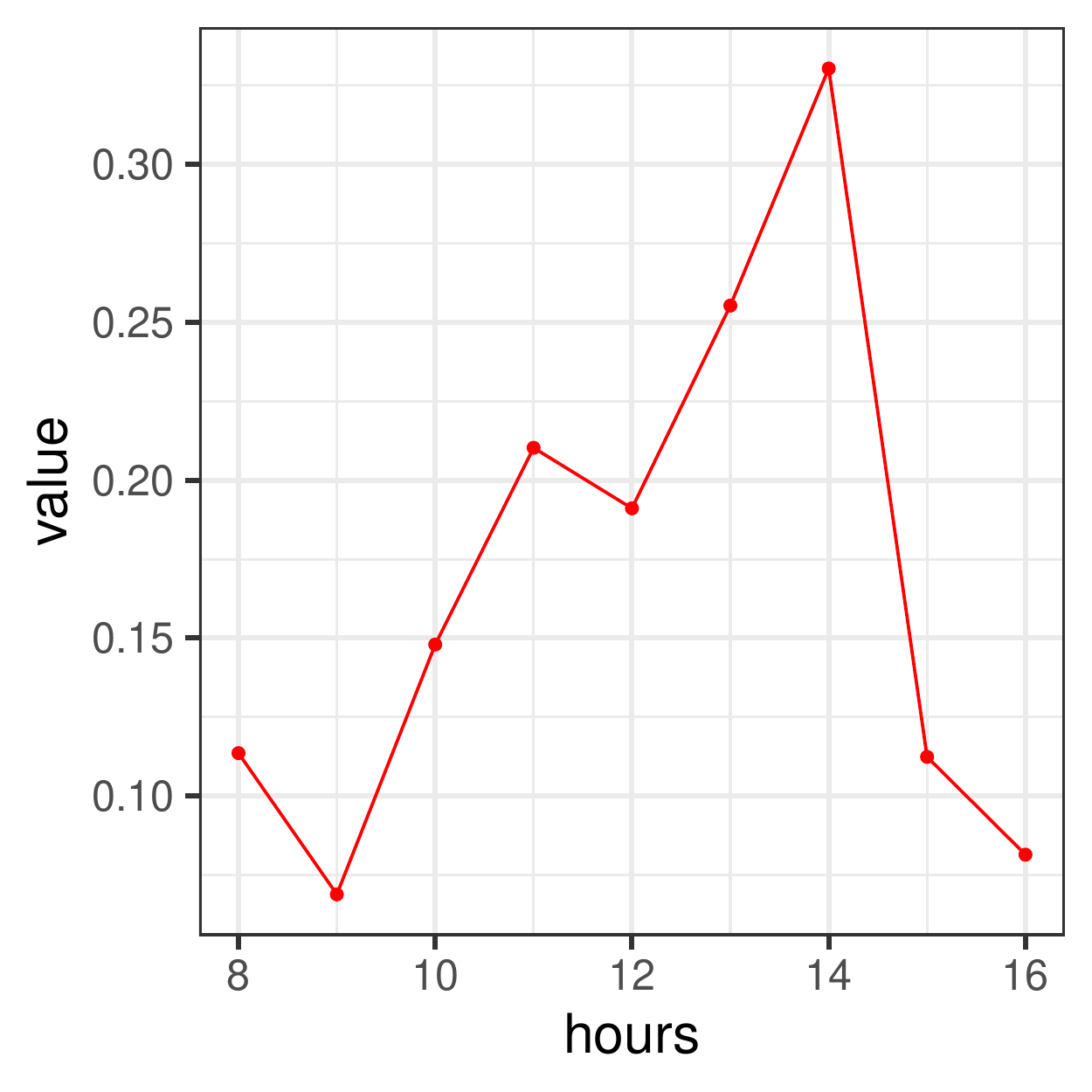}
			&
			\includegraphics[scale=0.4]{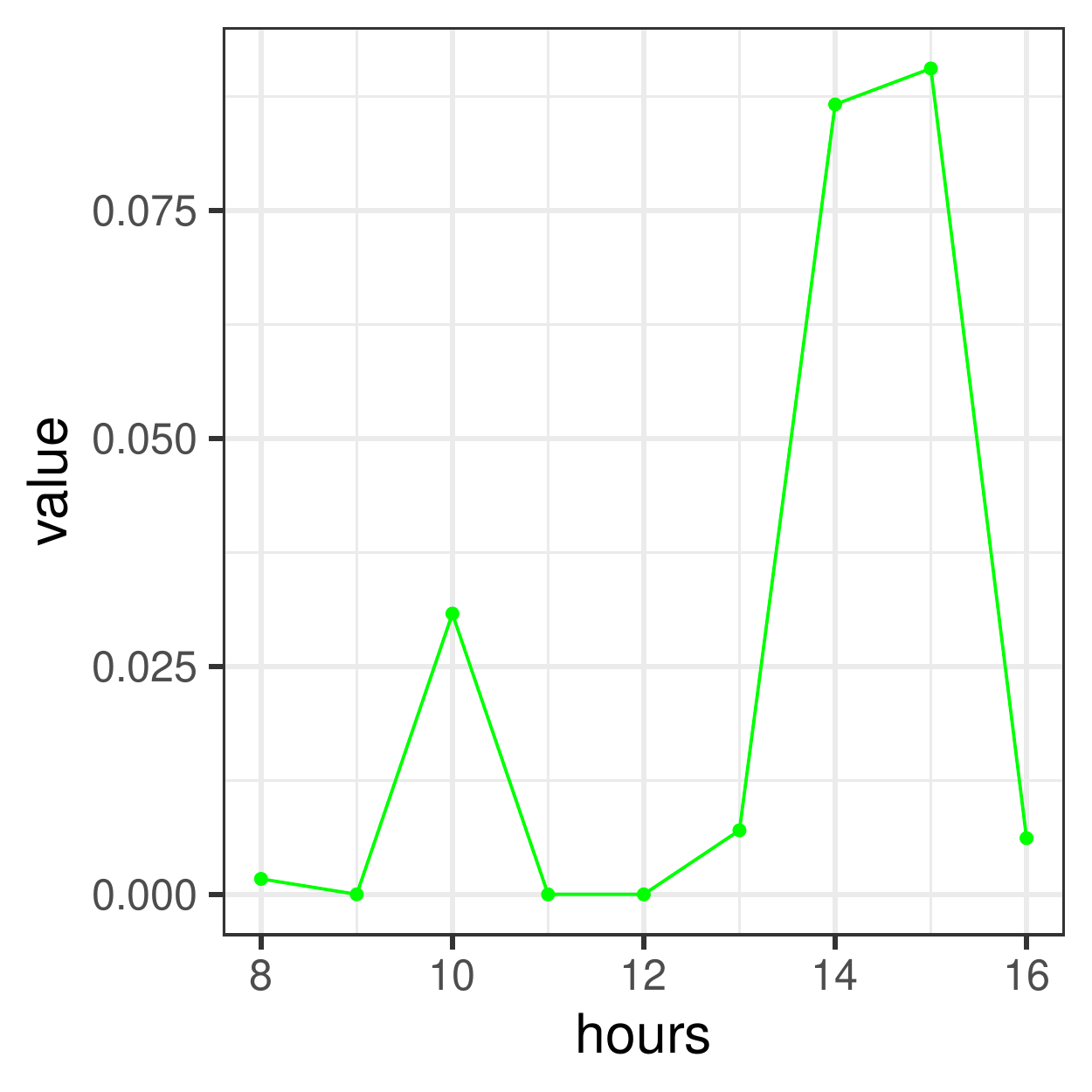}
			&
			\includegraphics[scale=0.4]{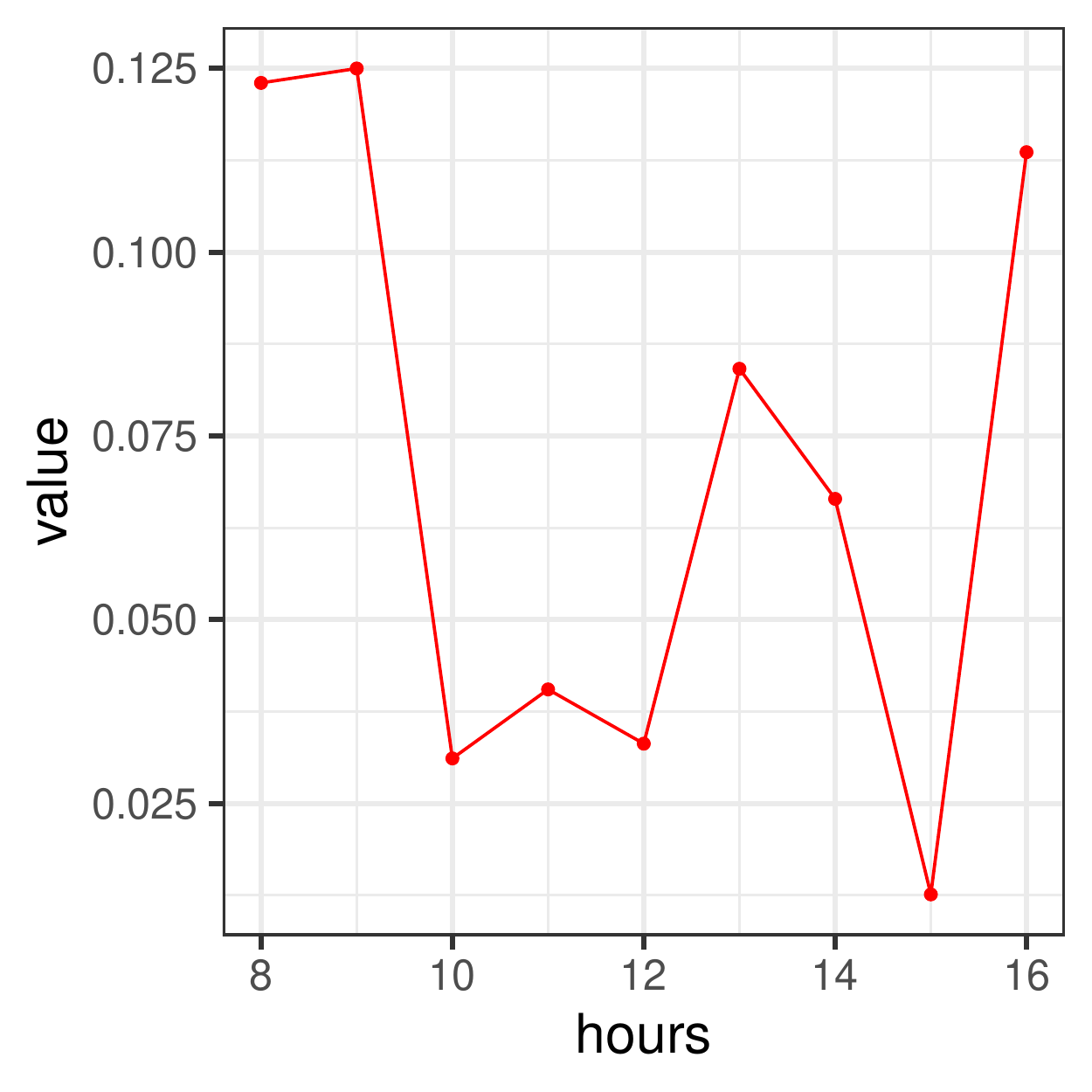}
		\end{tabular}
		\caption{\footnotesize{Multivariate upper tail dependence coefficients induced by the R-Vine Copula model specification related to Electricity Prices in the scenario HLL (left), HHL (middle), and HLH (right) related to Forecasted Demand, Forecasted Wind, and Forecasted Solar PV.}}
		\label{Fig_TDC_Strange_Insample}
	\end{figure}

	Figure \ref{Fig_TDC_Strange_Insample} shows the multivariate upper tail dependence coefficients induced by the R-Vine Copula model related to prices. In the HLL scenario, high prices confirm a clear positive dependence from demand, especially at $11$ and $14$. In the HHL scenario, prices exhibit a positive dependence but much lower than the previous situation, with more remarkable reductions especially at hour $11$ (from $0.21$ to $0$) and at hour $14$ (from around $0.35$ to $0.09$), as an effect of high wind. In the HLH scenario, instead, it is possible to detect the effect of solar peaking reducing the multivariate dependence, for instance at hour $10$ from $0.15$ (in HLL) to $0.03$, at hour $13$ from $0.25$ (in HLL) to $0.08$, or at hour $14$ from around $0.35$ (in HLL) to $0.06$.
	
\begin{figure}[ht]
		\centering
		\begin{tabular}{ccc}
			\includegraphics[scale=0.5]{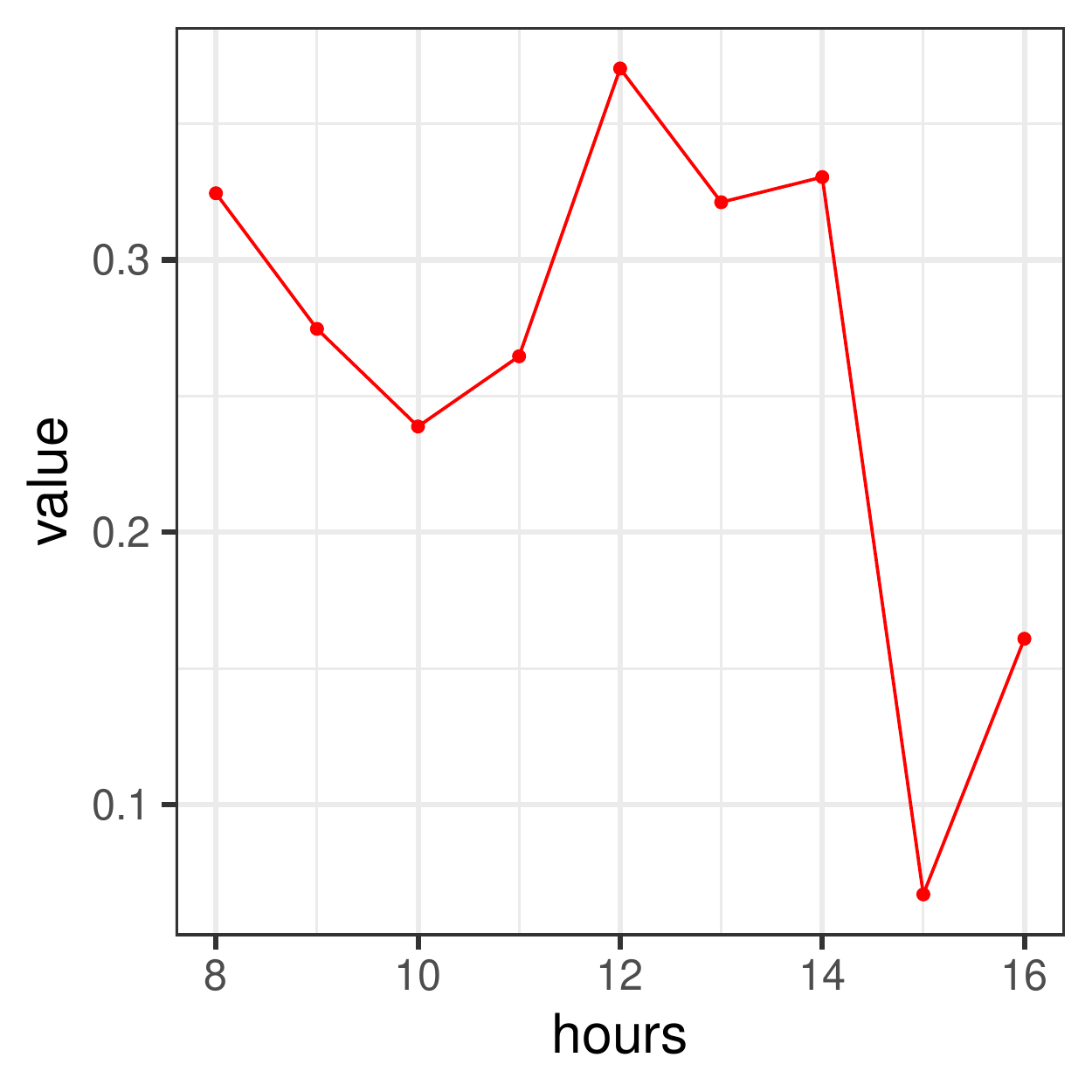}
			&
			\includegraphics[scale=0.5]{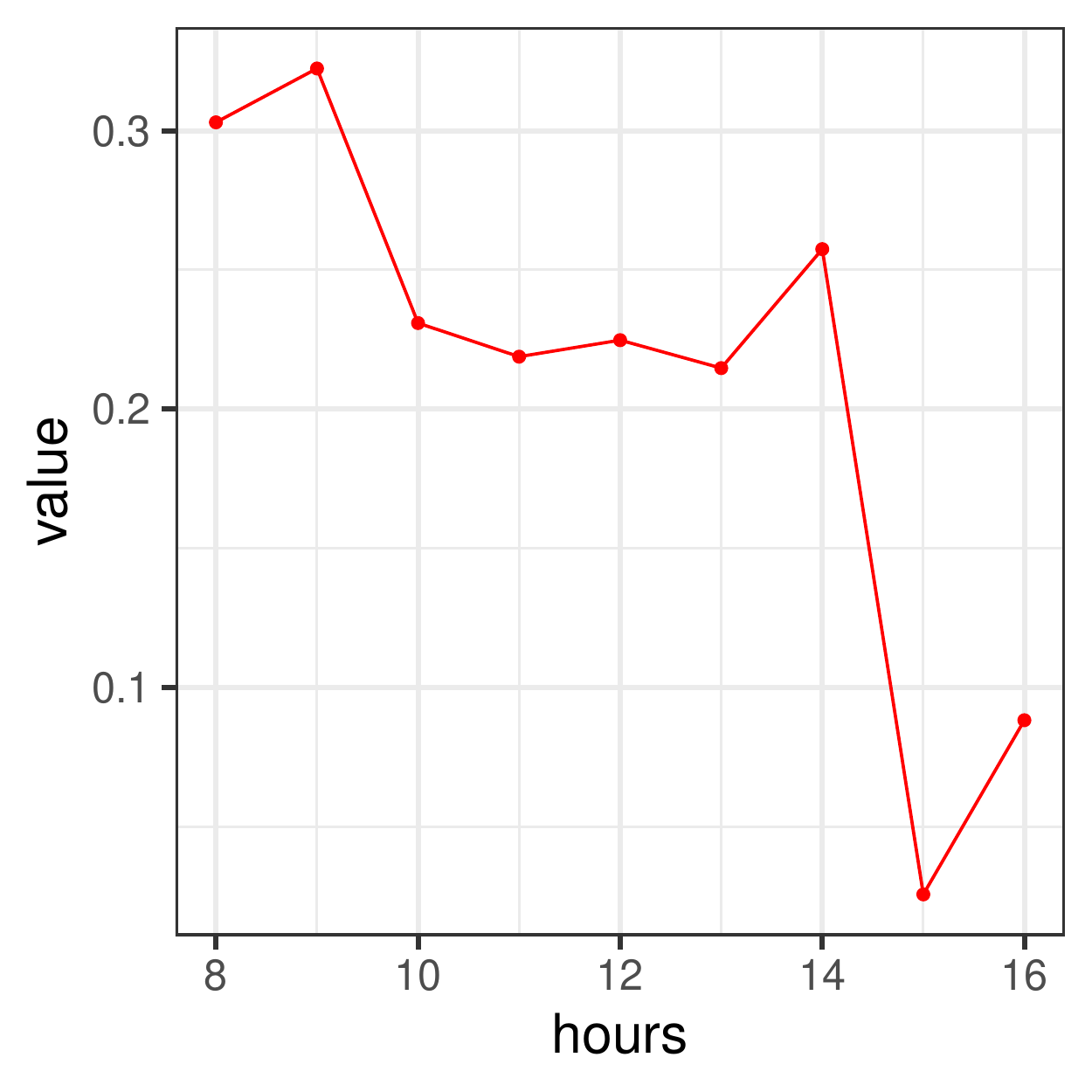}
		\end{tabular}
		\caption{\footnotesize{Multivariate lower tail dependence coefficients induced by the R-Vine Copula model specification related to Electricity Prices in the scenario LHH (left), and LHL (right) related to Forecasted Demand, Forecasted Wind, and Forecasted Solar PV.}}
		\label{Fig_TDC_Strange2_Insample}
	\end{figure}

	When the multivariate lower tail dependence is considered, the most interesting results refer to low prices in conjunction with low demand levels and high wind infeed. Then, results for the LH scenarios are considered with respect to the levels of solar power, reported in Figure \ref{Fig_TDC_Strange2_Insample}. All show positive dependence between low prices and demand but high wind, in both cases of high and low solar, but again with reduced magnitudes in the latter case.
	
	\subsection{{Time-varying Analysis with Rolling Windows}}
	
	Differently from what done previously, where the dependence parameters have been estimated using the whole time series as having \textit{static} trivariate and quadrivariate copulas, in what follows, instead, it is briefly inspected whether a time-varying dynamics of the involved variables can describe some additional features. Then the analysis starts by estimating the dependence model using a subset of the data and adopting a year rolling window approach. Using a window size of $2$ years, the first estimate of the dependence model is based on the window from 01 January 2011 to 31 December 2012; the second estimate is based on the window from 02 January 2011 to 01 January 2013 and so on until the last window is rolled to the end of the sample on December 31, 2019.
	
\begin{sidewaysfigure}
		\centering
		\begin{tabular}{ccccc}
			$8$ & $10$ & $12$ & $14$ & $16$ \\
			\includegraphics[width=4cm]{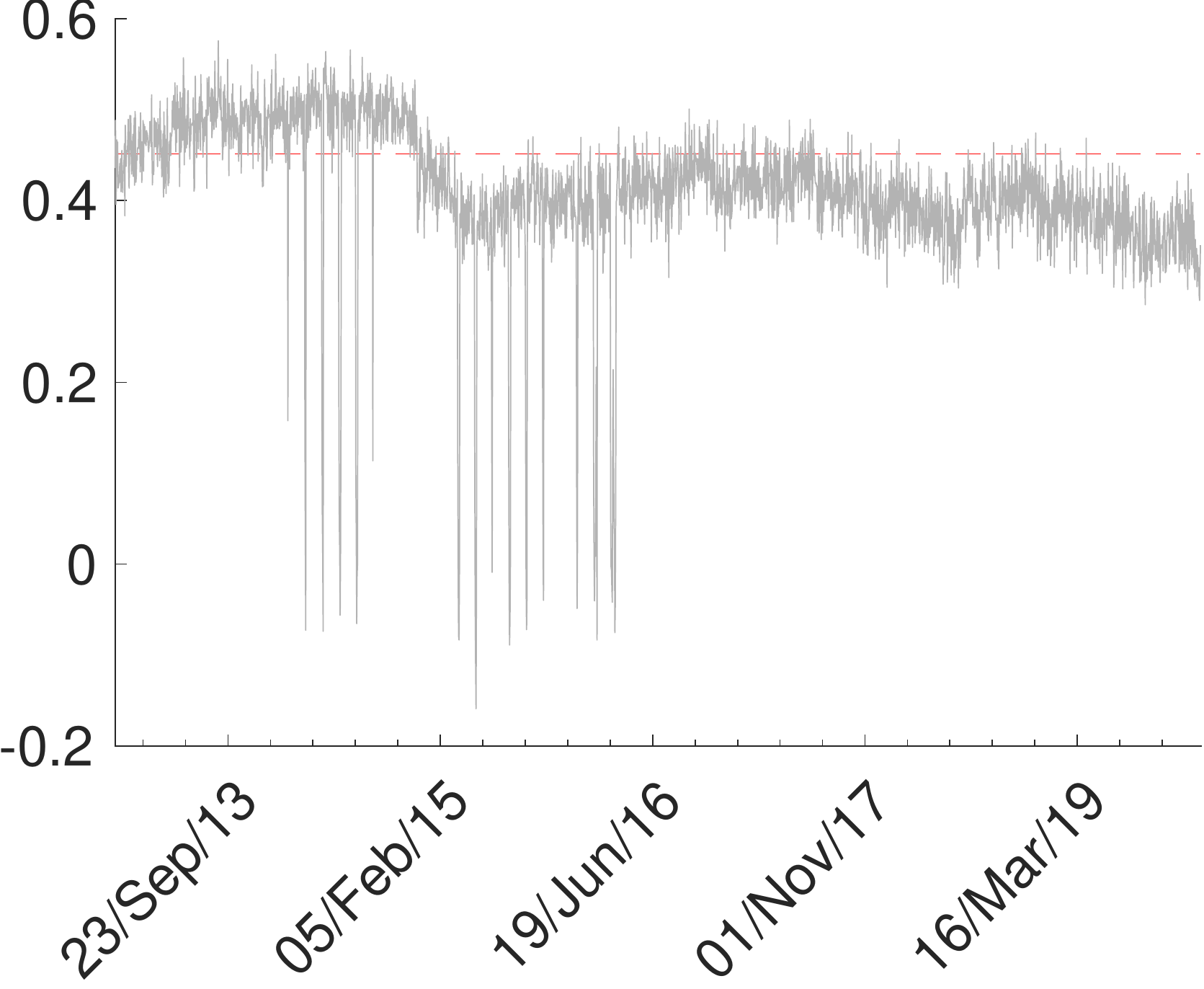}&
			\includegraphics[width=4cm]{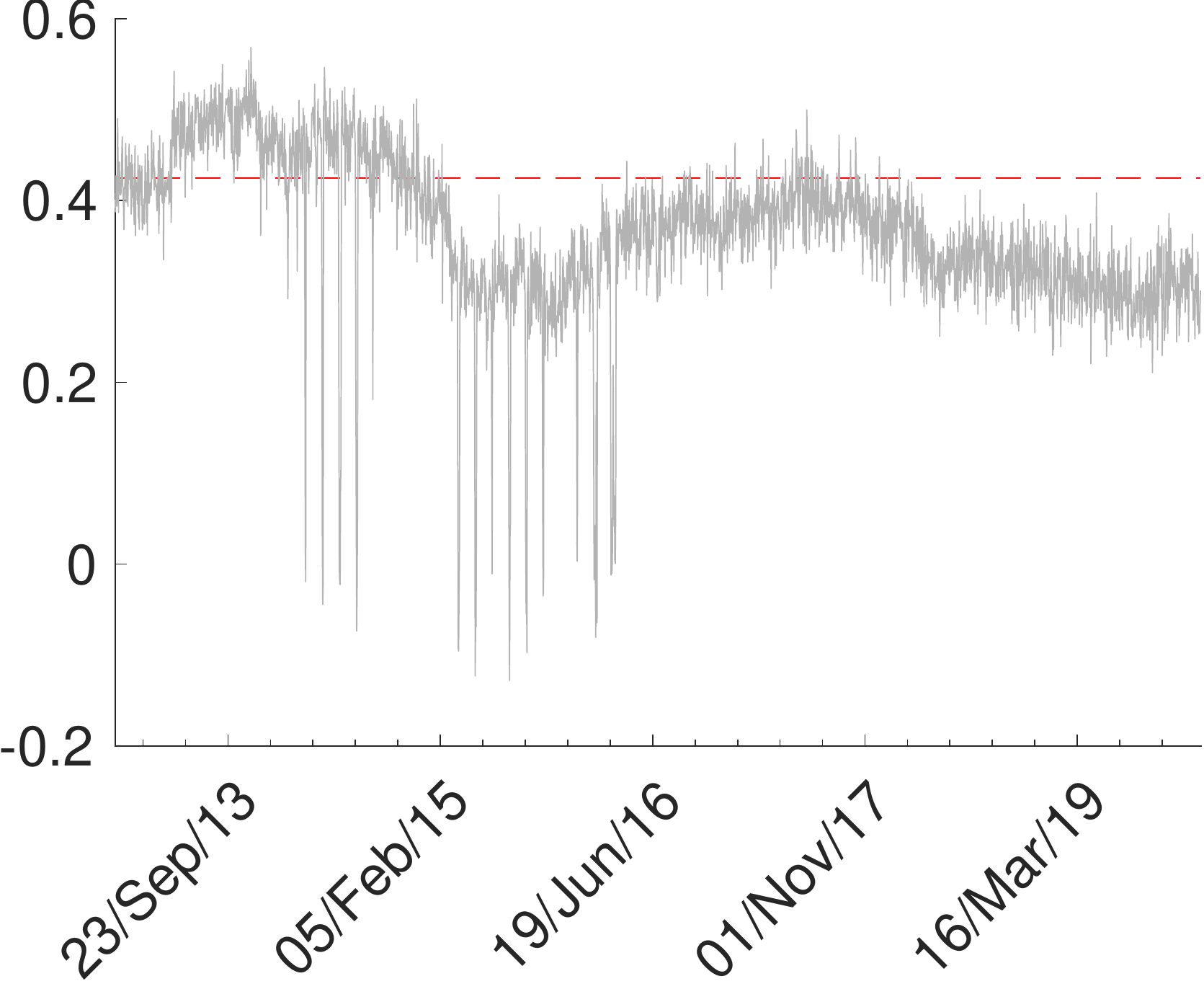}&
			\includegraphics[width=4cm]{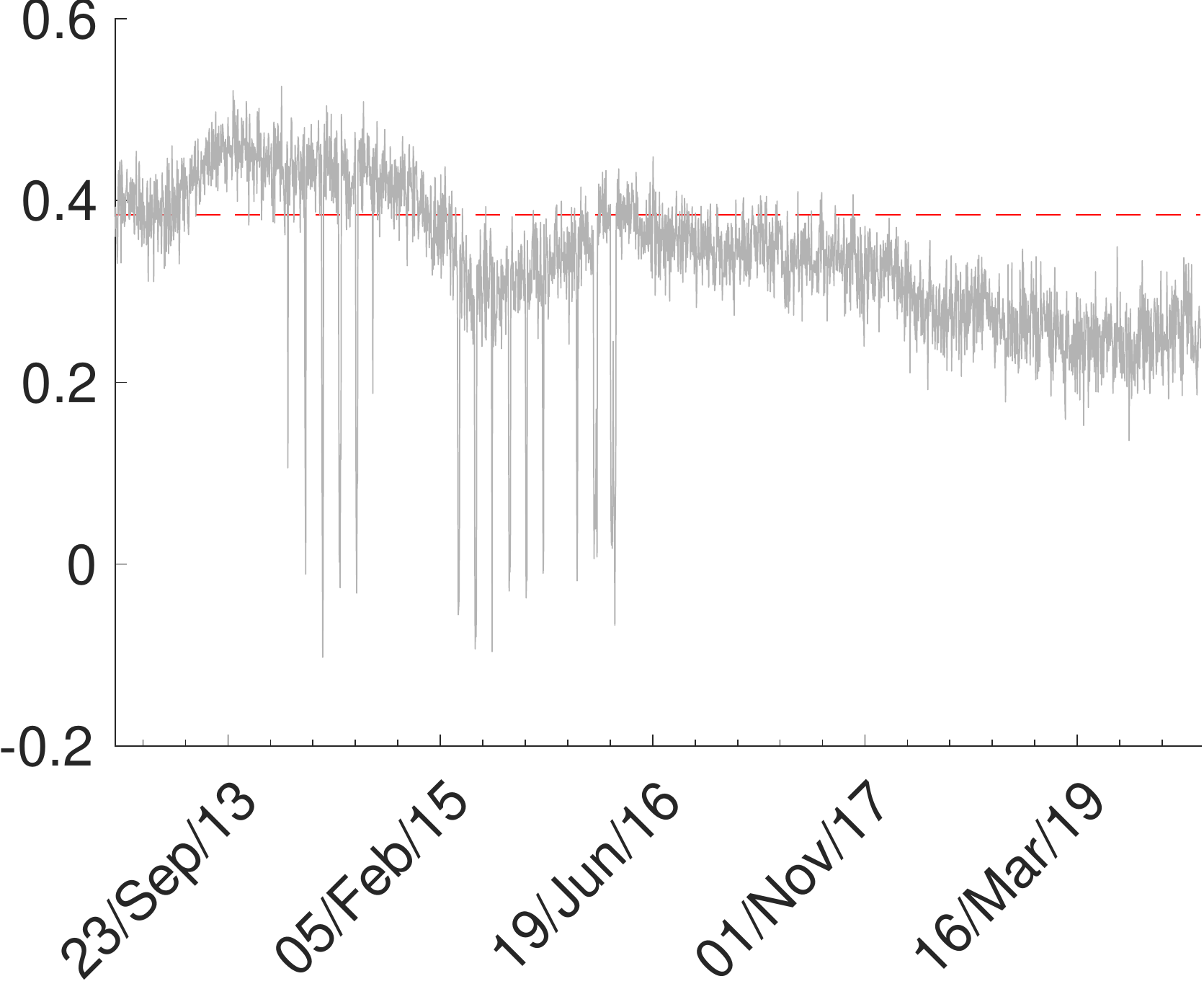}&
			\includegraphics[width=4cm]{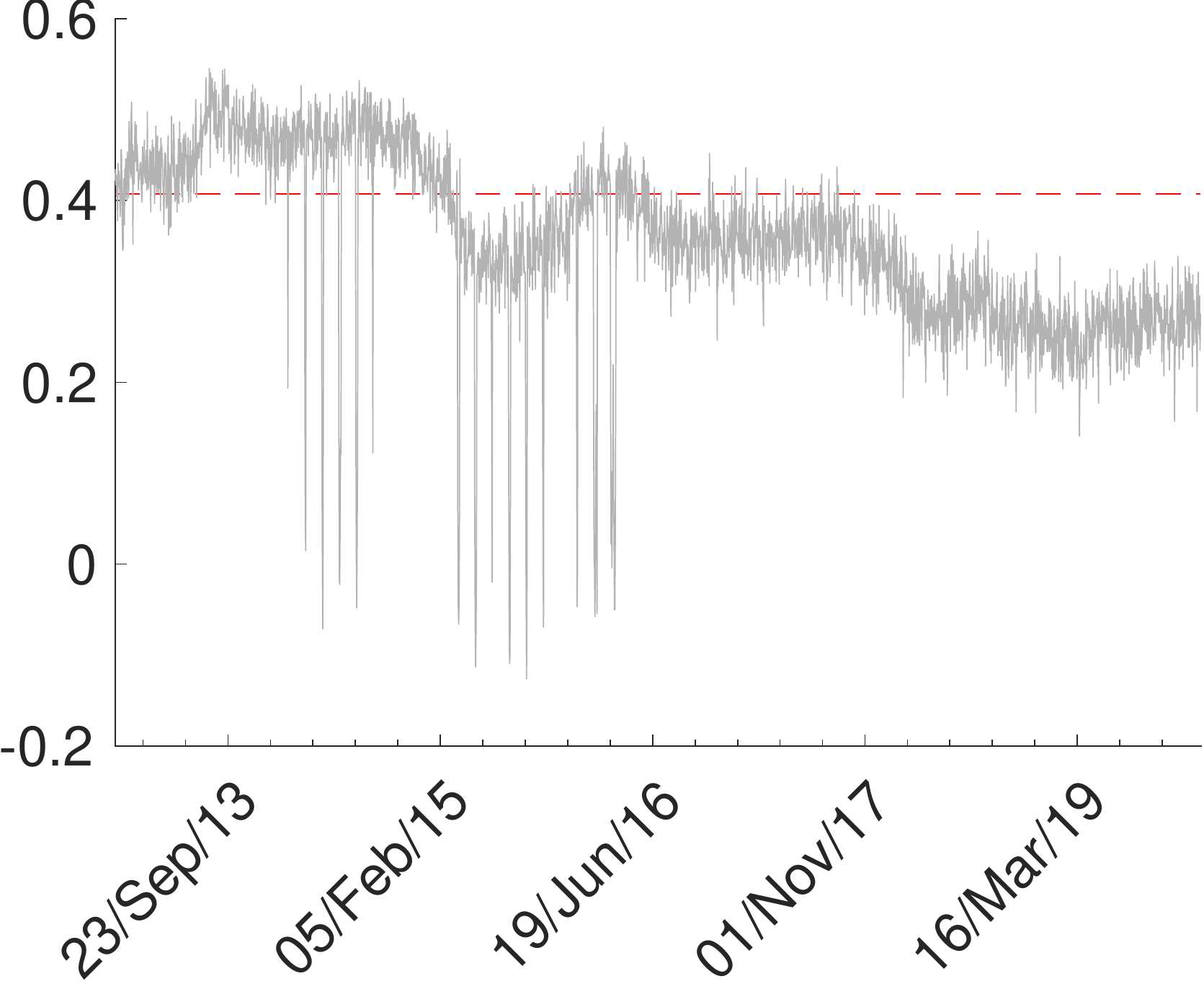}&
			\includegraphics[width=4cm]{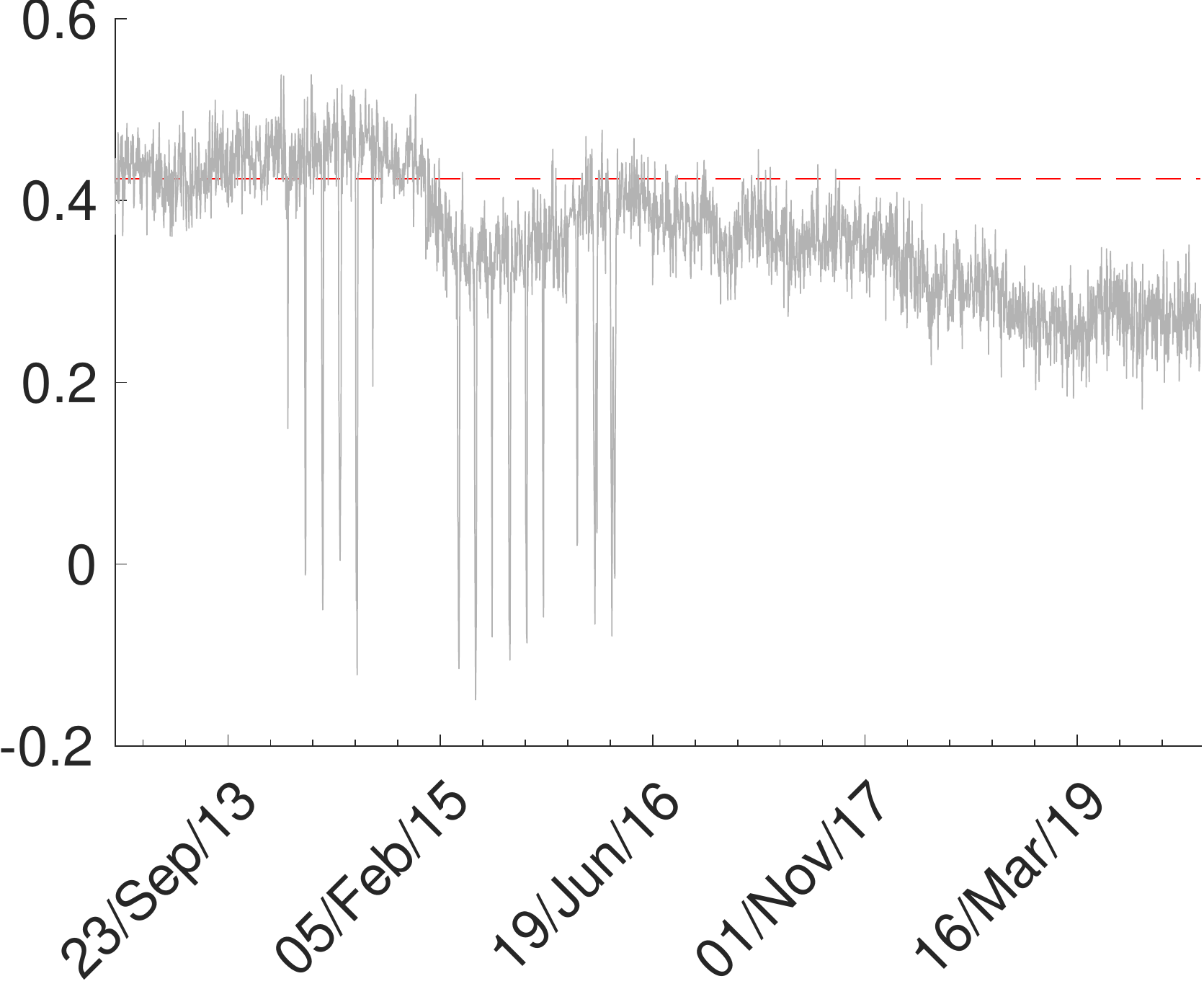}\\
			\includegraphics[width=4cm]{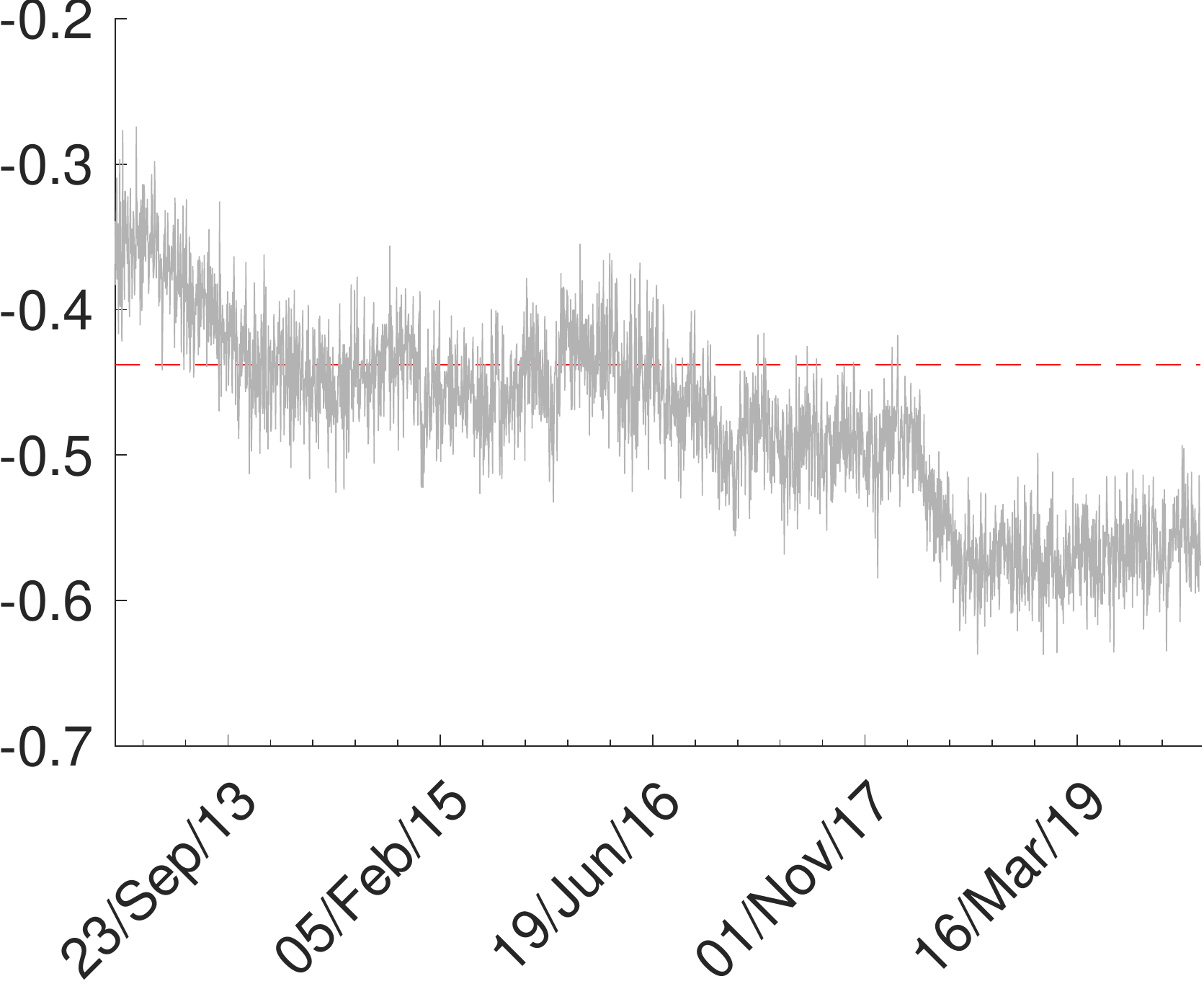}&
			\includegraphics[width=4cm]{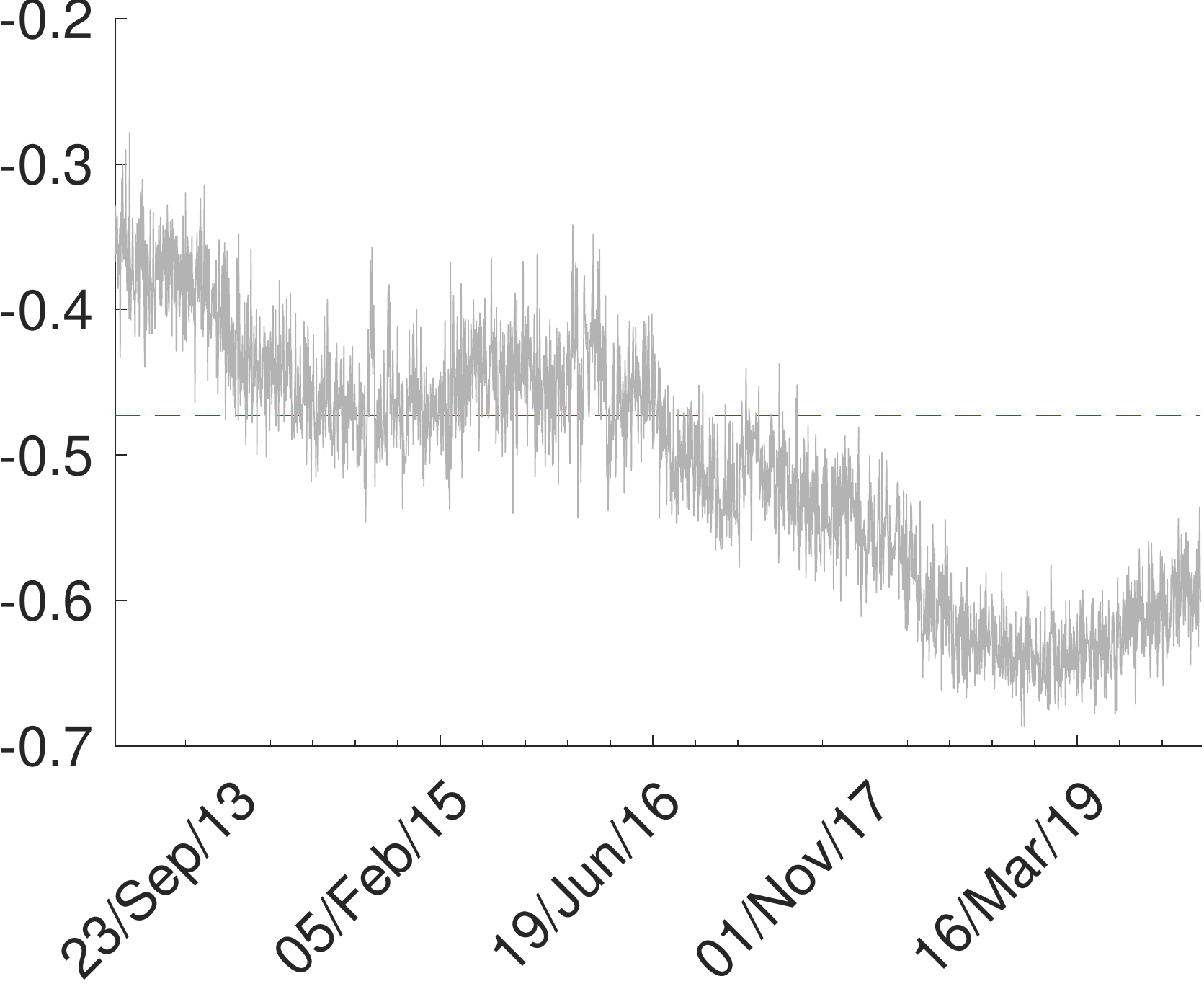}&
			\includegraphics[width=4cm]{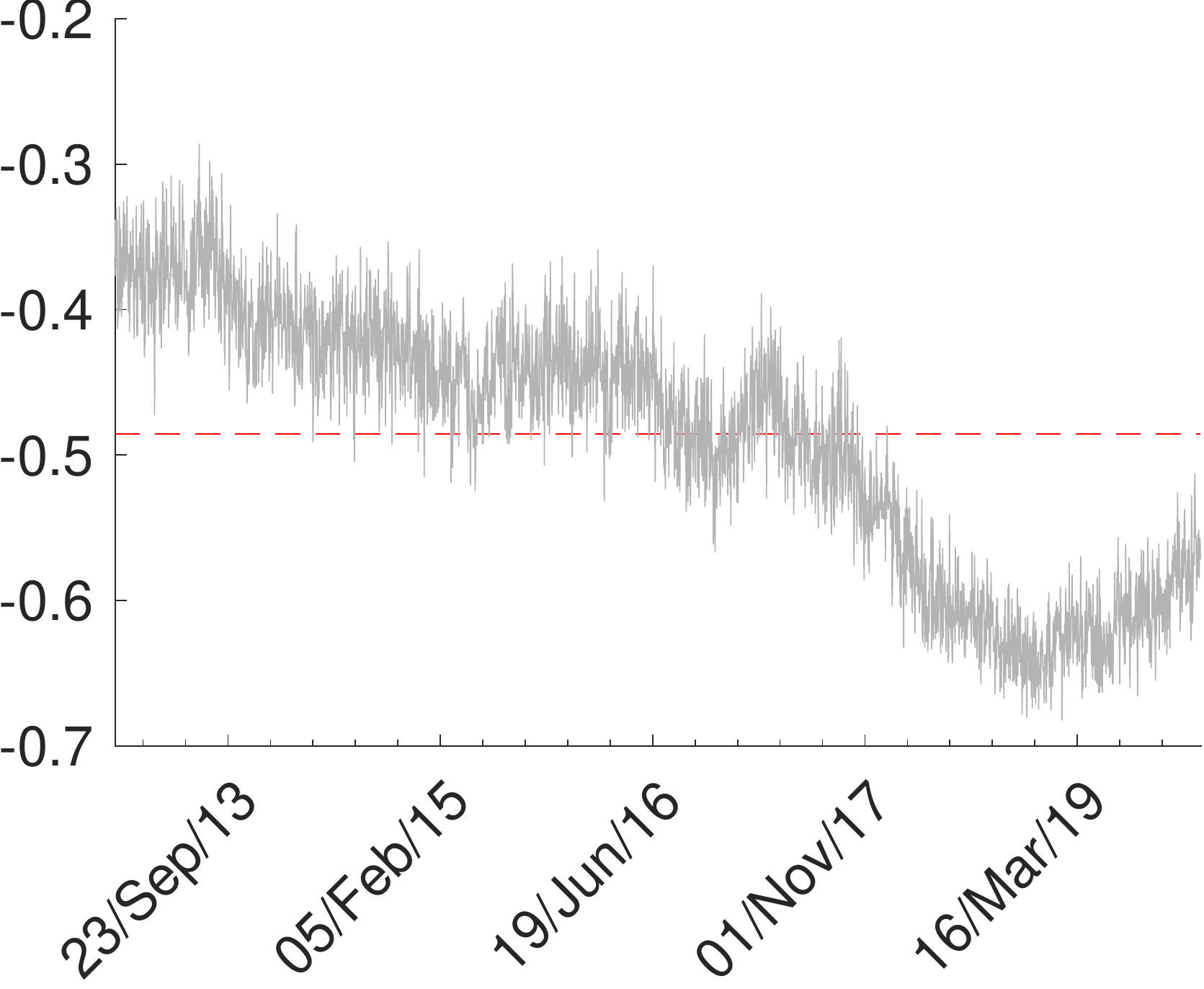}&
			\includegraphics[width=4cm]{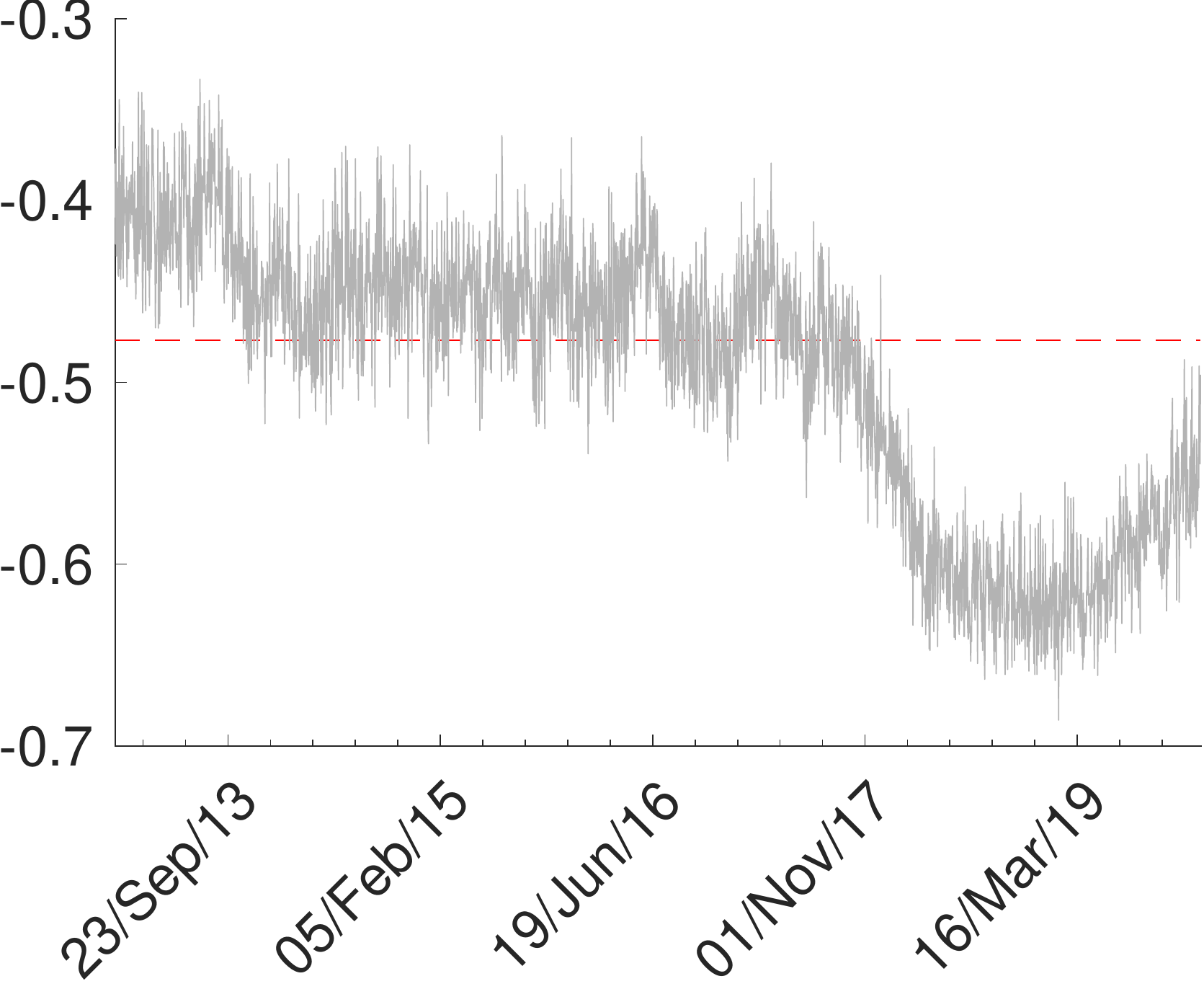}&
			\includegraphics[width=4cm]{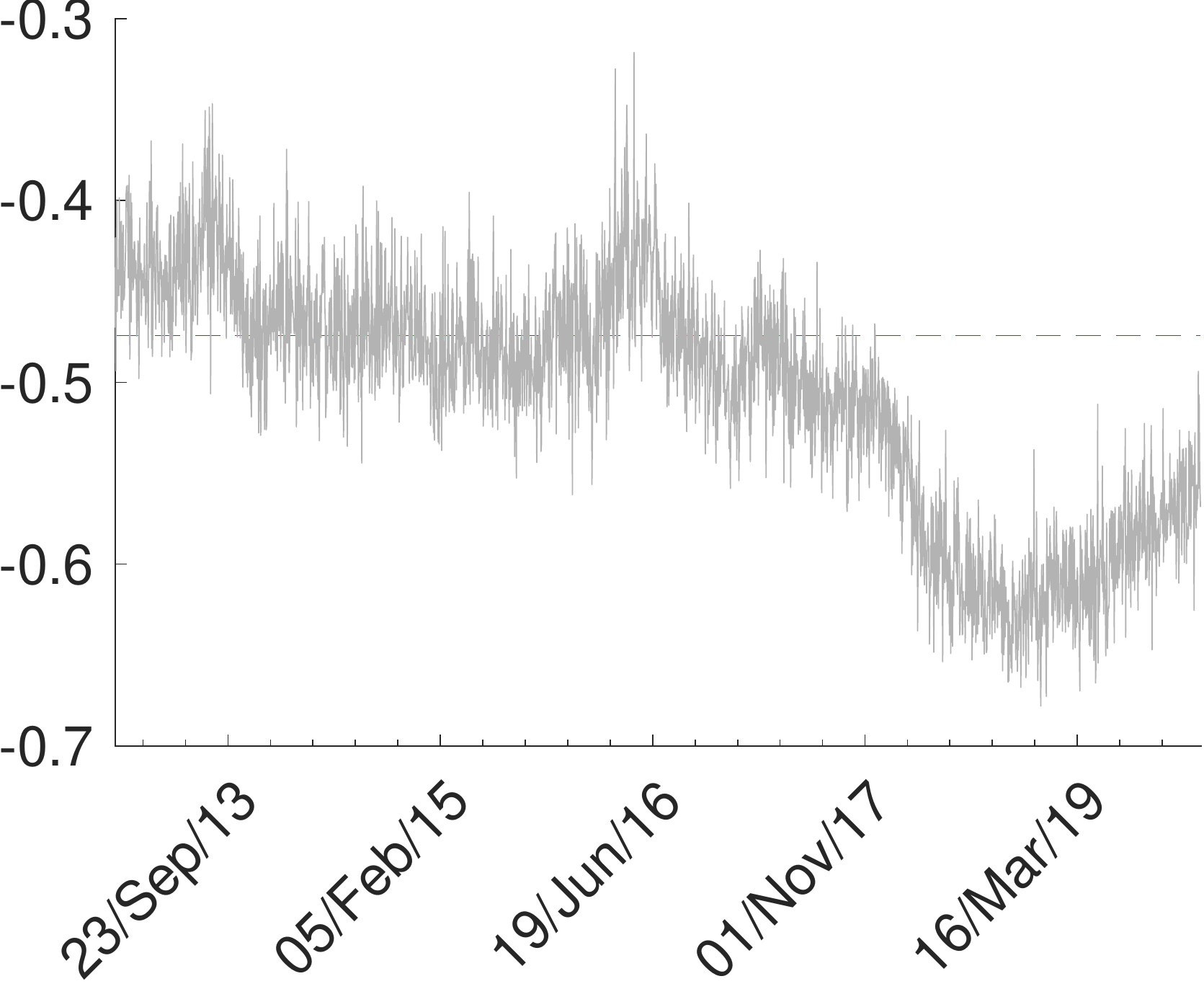} \\
			\includegraphics[width=4cm]{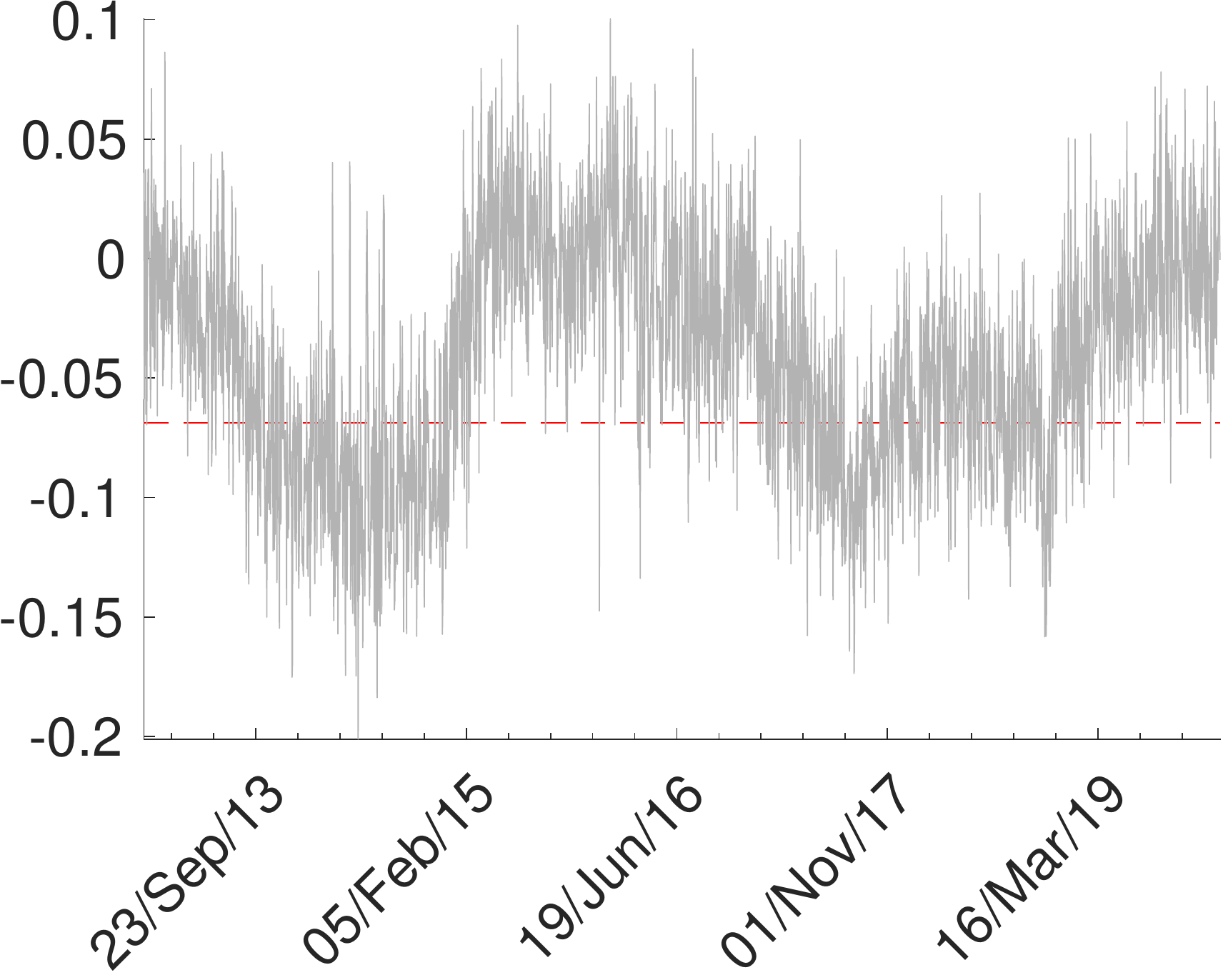}&
			\includegraphics[width=4cm]{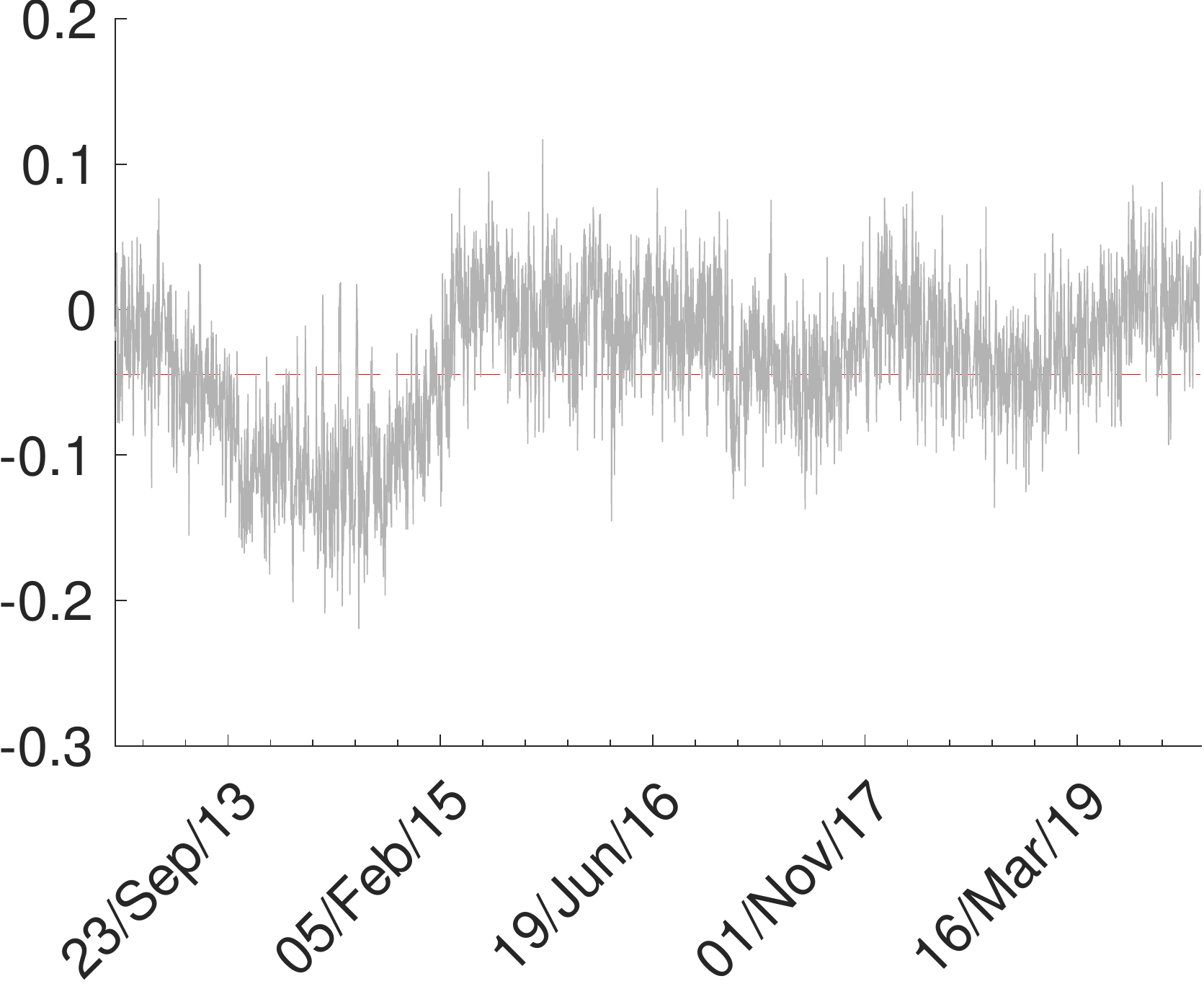}&
			\includegraphics[width=4cm]{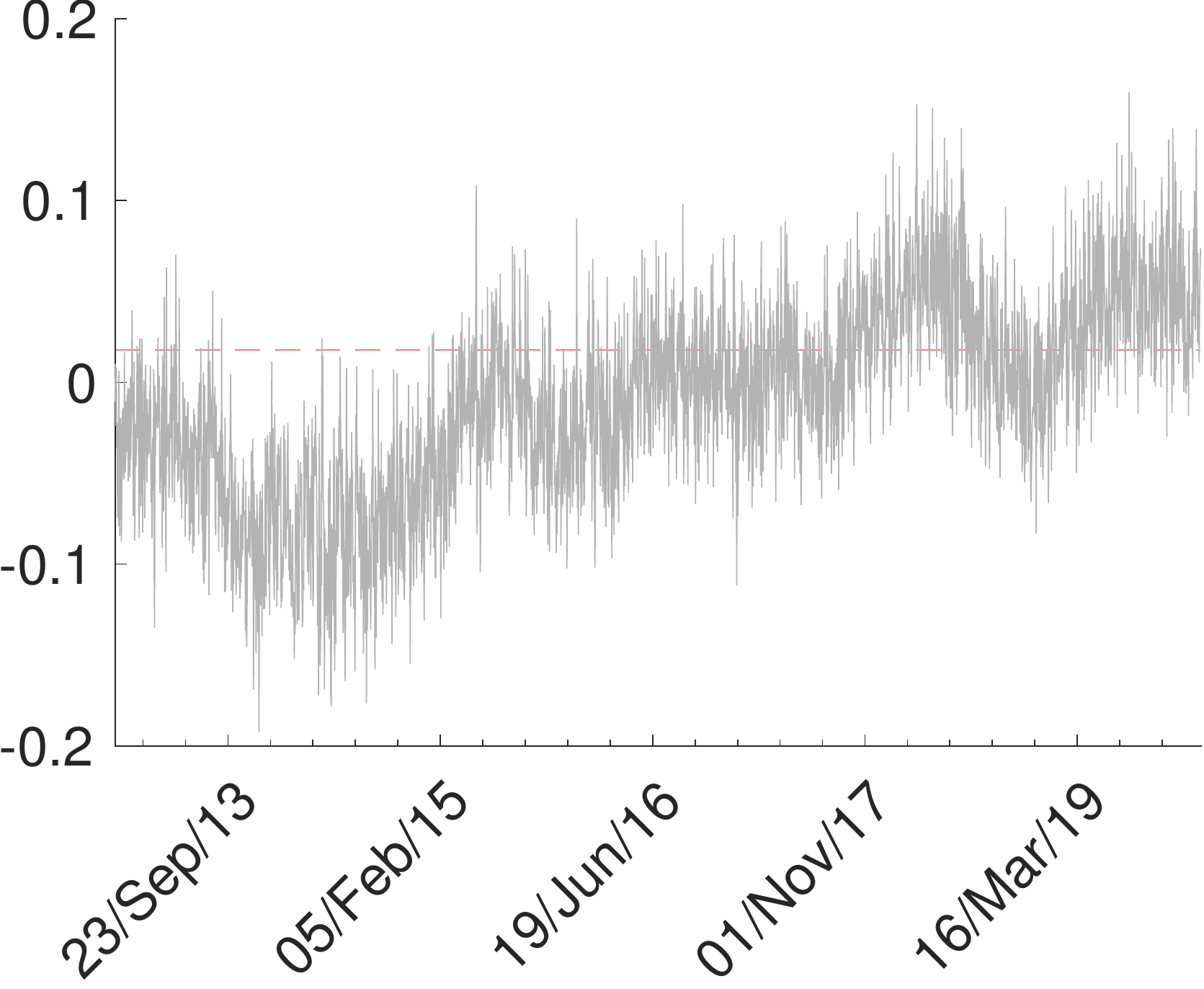}&
			\includegraphics[width=4cm]{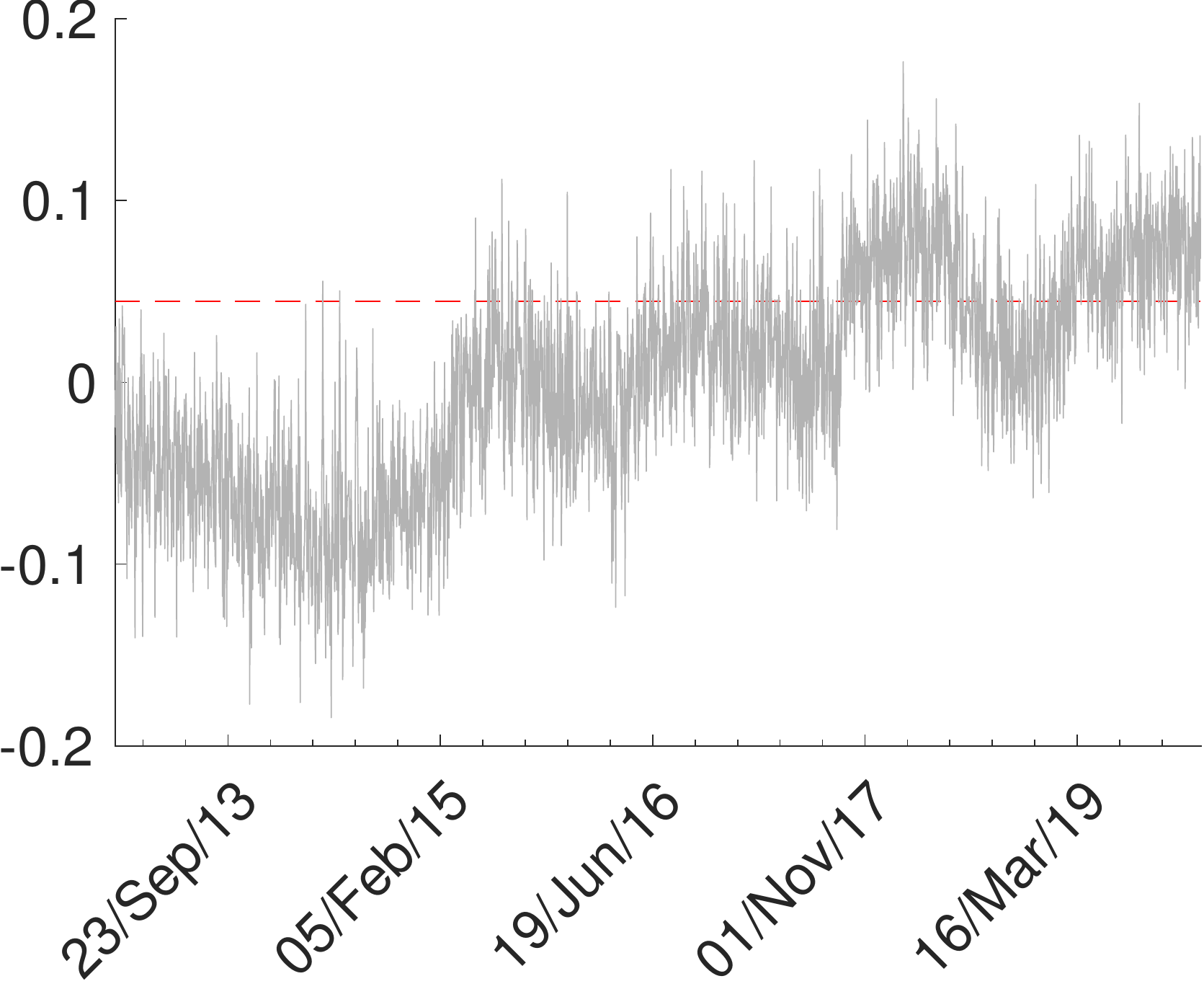}&
			\includegraphics[width=4cm]{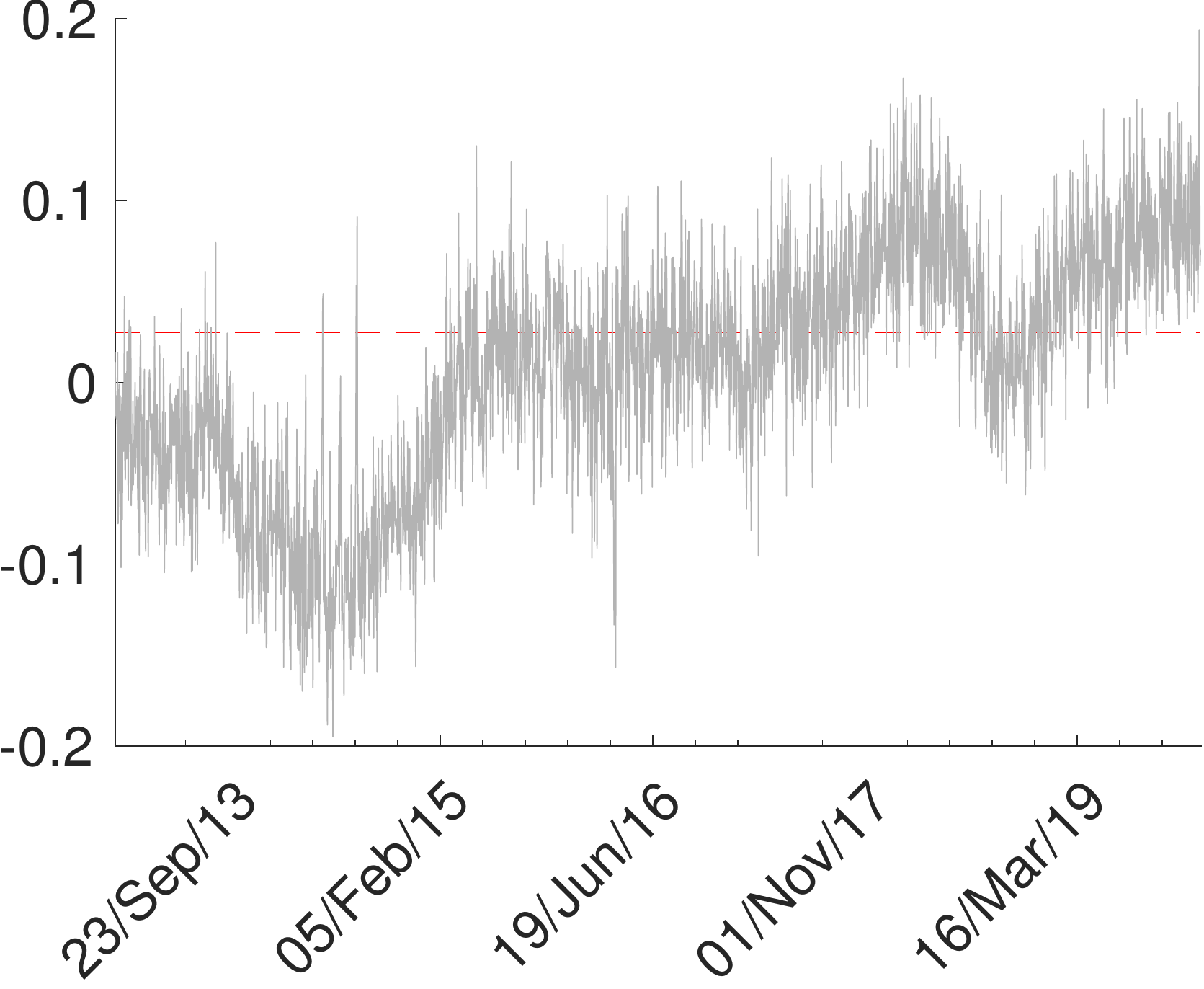}  \\
		\end{tabular}
		\caption{\footnotesize{Estimated Time-varying Correlation between Electricity Prices and Forecasted Demand (top row); Electricity Prices and Forecasted Wind (middle row) and Forecasted Demand and Forecasted Wind (bottom row) at five different hours for the trivariate estimations of the vine copula model specification (grey lines). The red dashed line is the dependence parameter for the vine copula model estimated on the whole sample.}}
		\label{Fig_Correl_Out_of_Sample_Price}
	\end{sidewaysfigure}

\begin{sidewaysfigure}
		\centering
		\begin{tabular}{ccccc}
			$8$ & $10$ & $12$ & $14$ & $16$ \\
			\includegraphics[width=4cm]{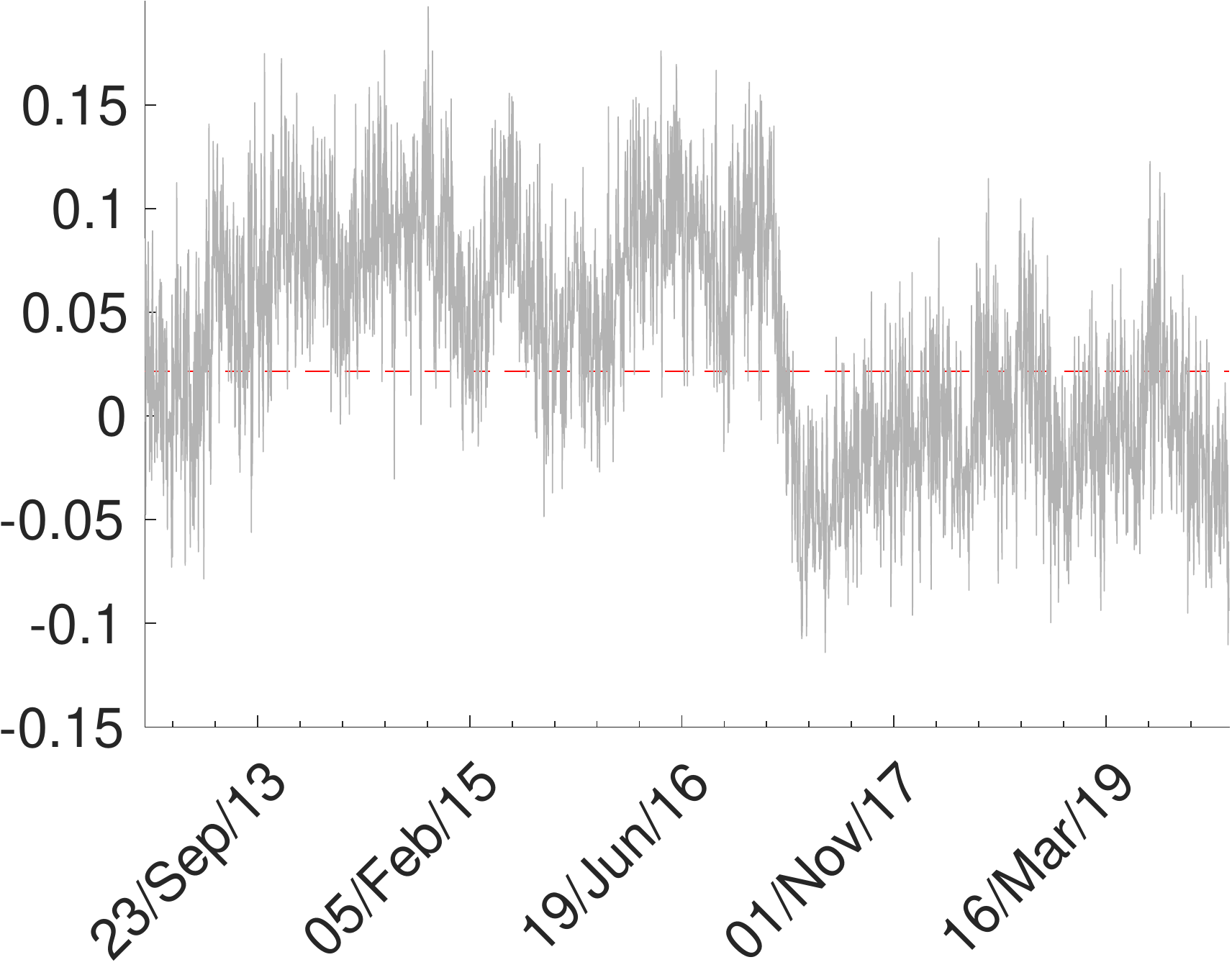}&
			\includegraphics[width=4cm]{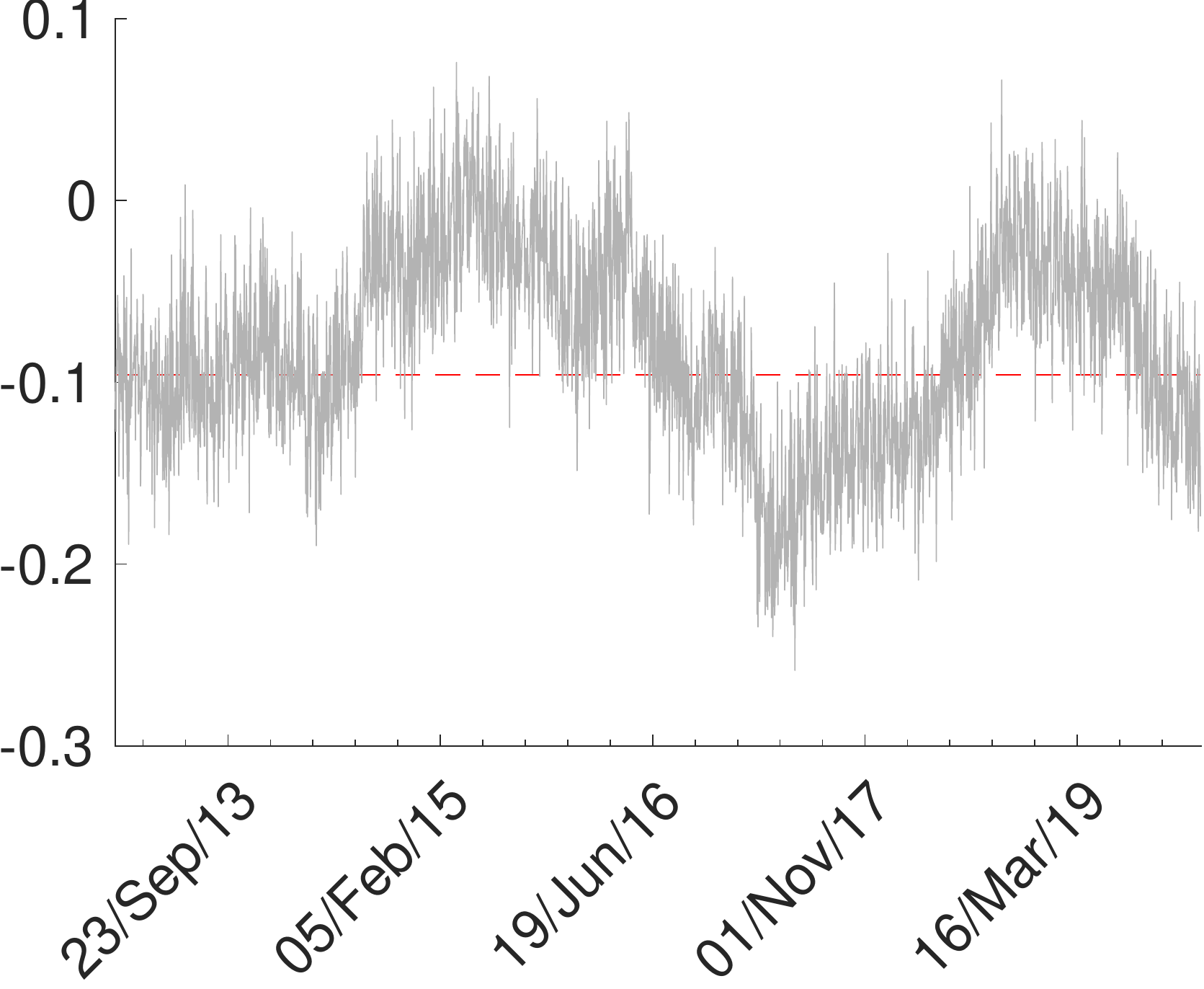}&
			\includegraphics[width=4cm]{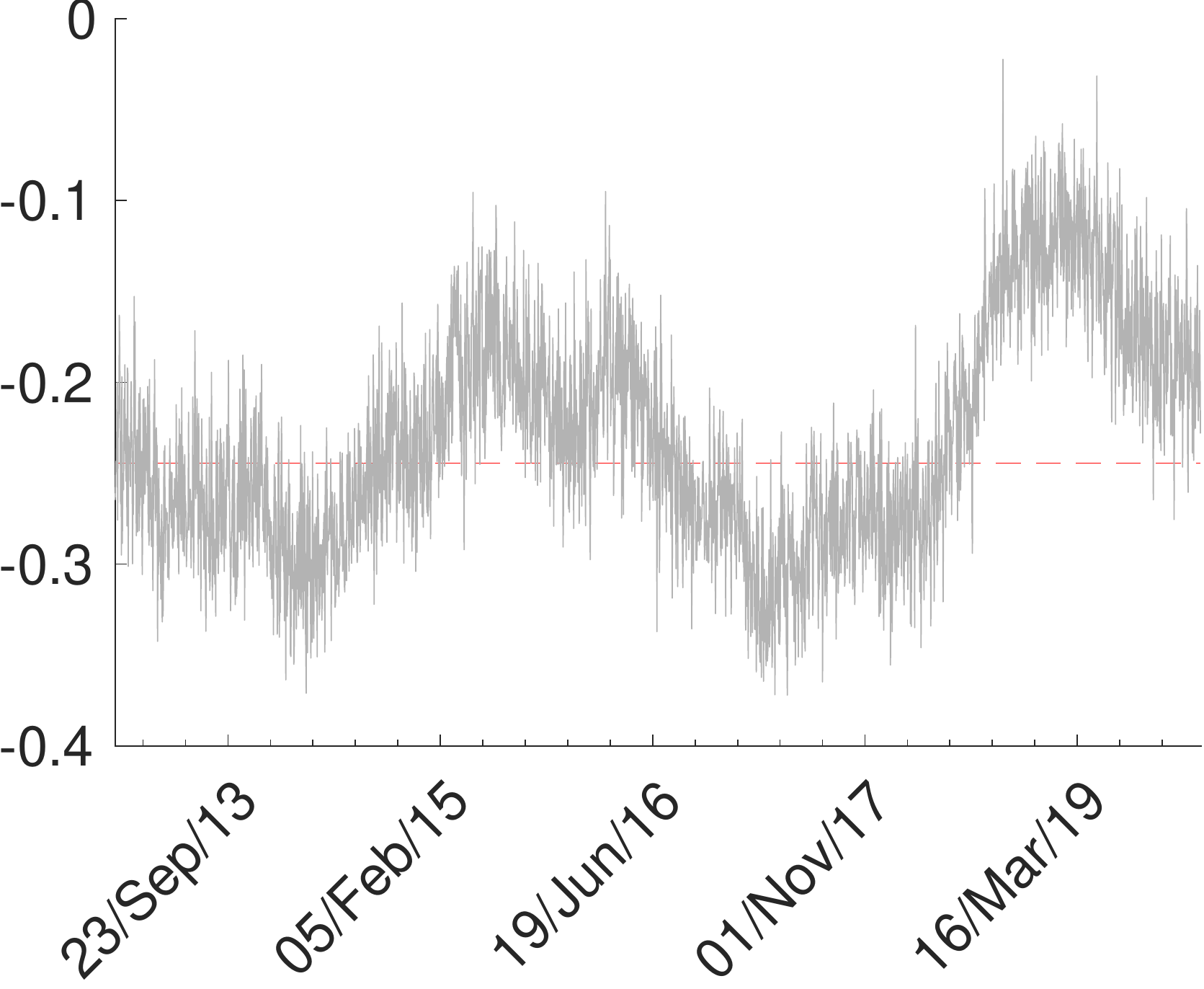}&
			\includegraphics[width=4cm]{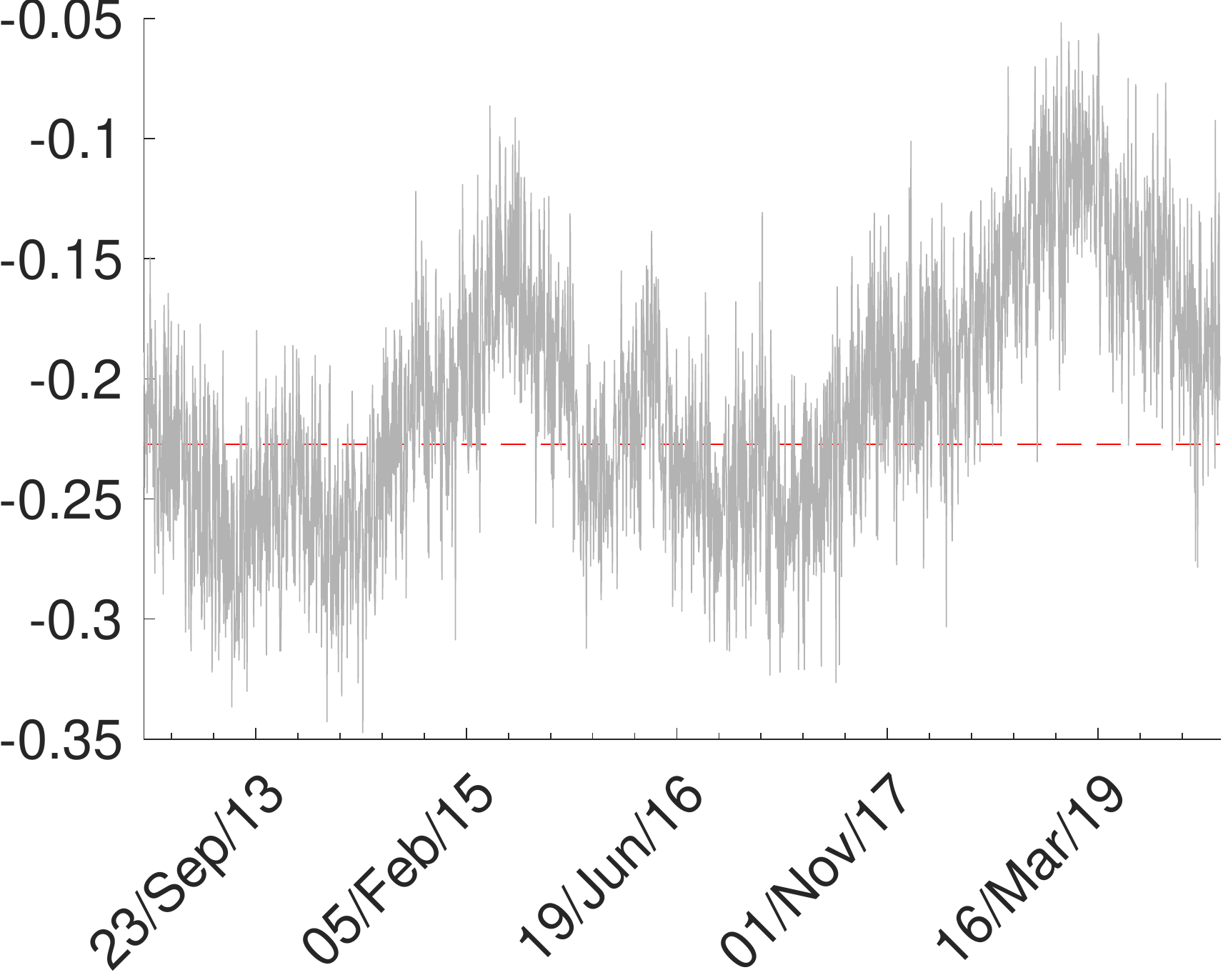}&
			\includegraphics[width=4cm]{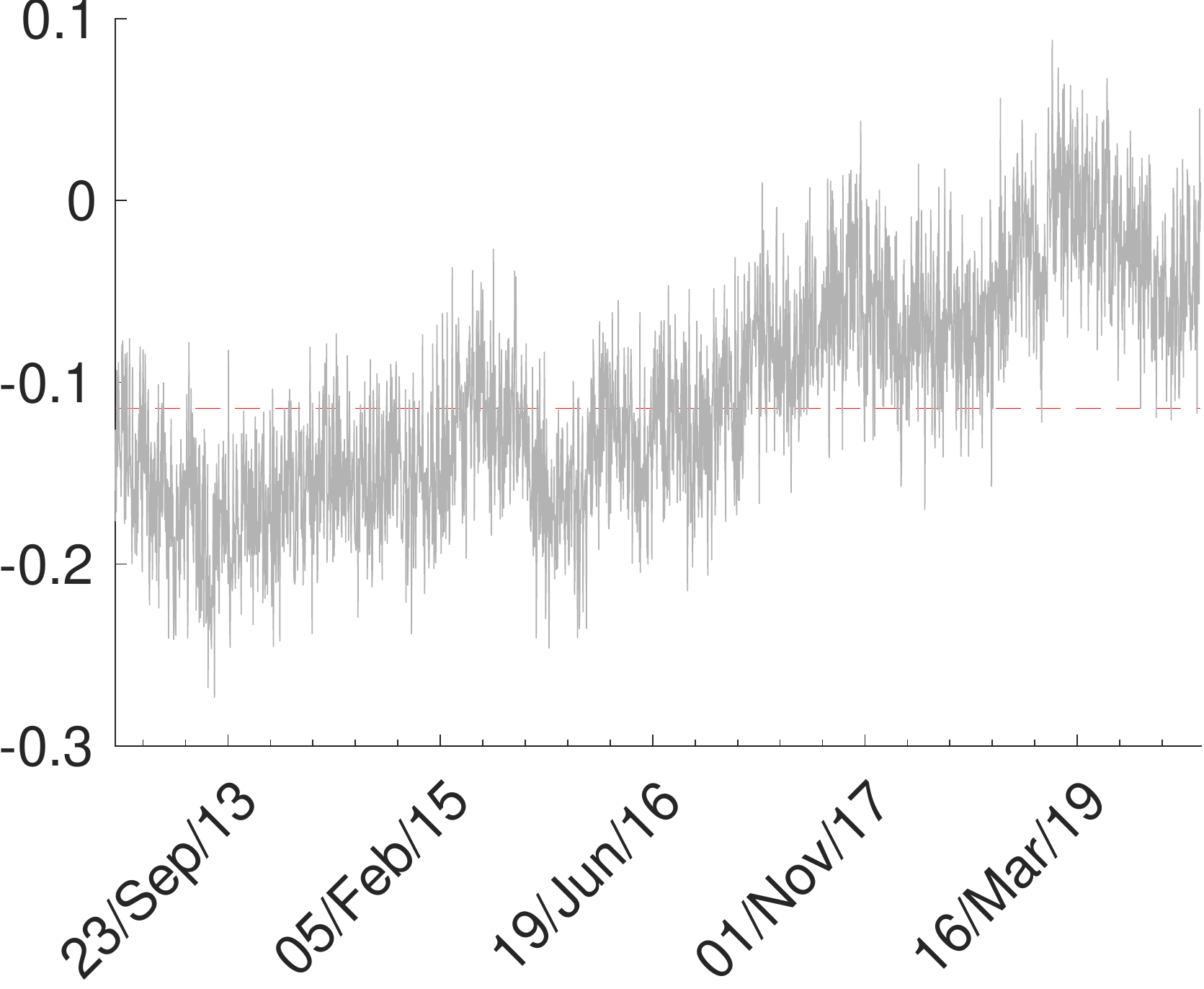}\\
			\includegraphics[width=4cm]{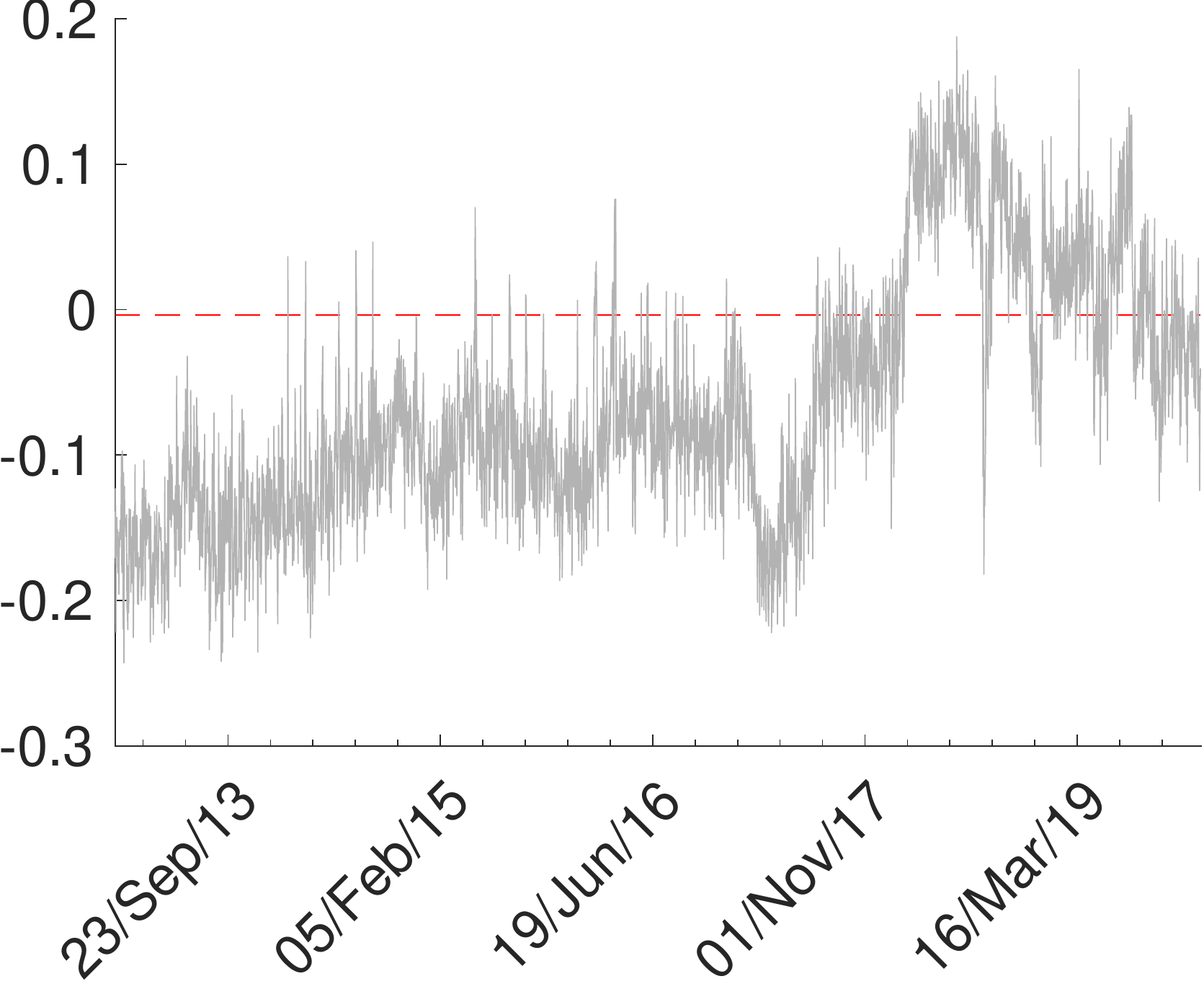}&
			\includegraphics[width=4cm]{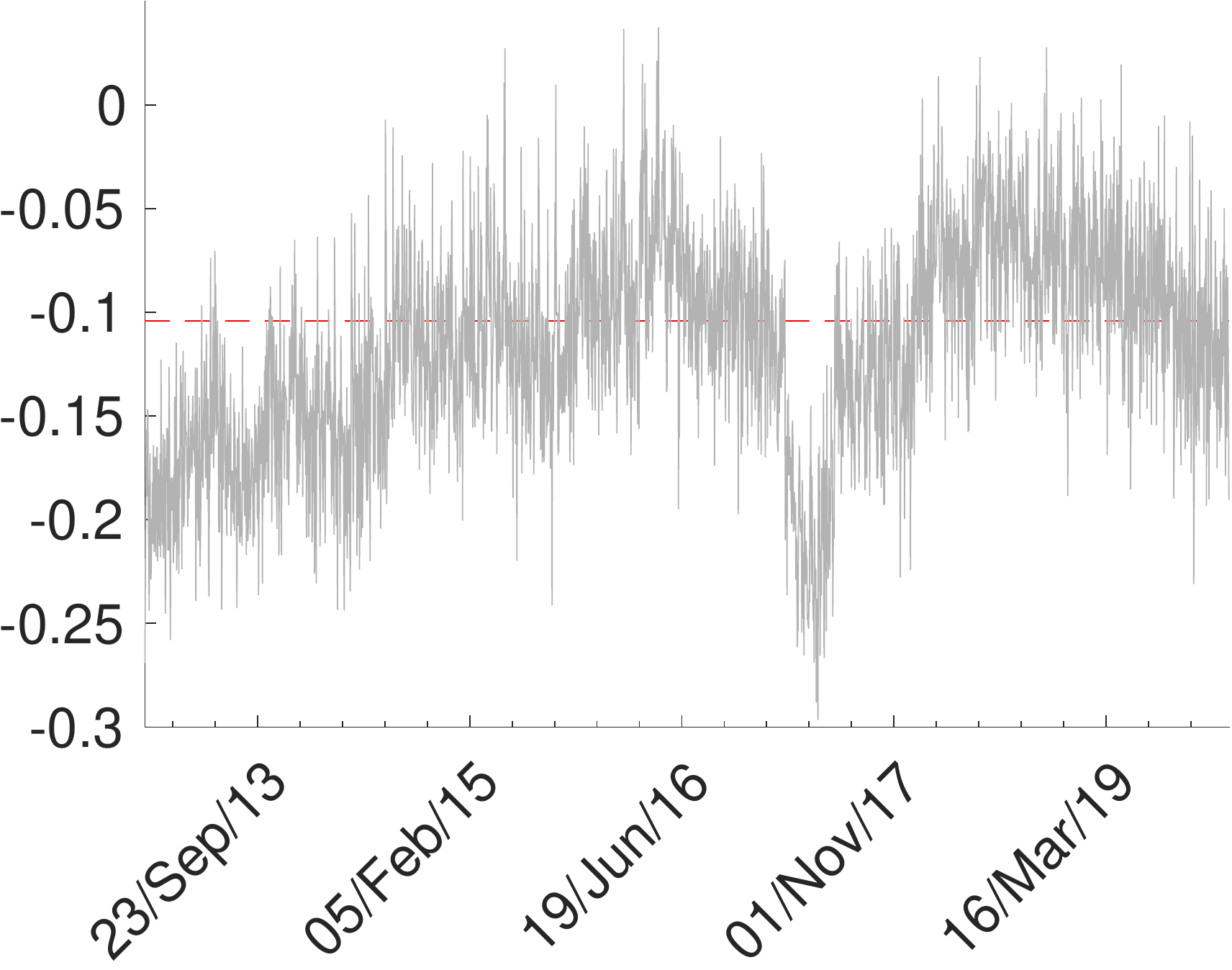}&
			\includegraphics[width=4cm]{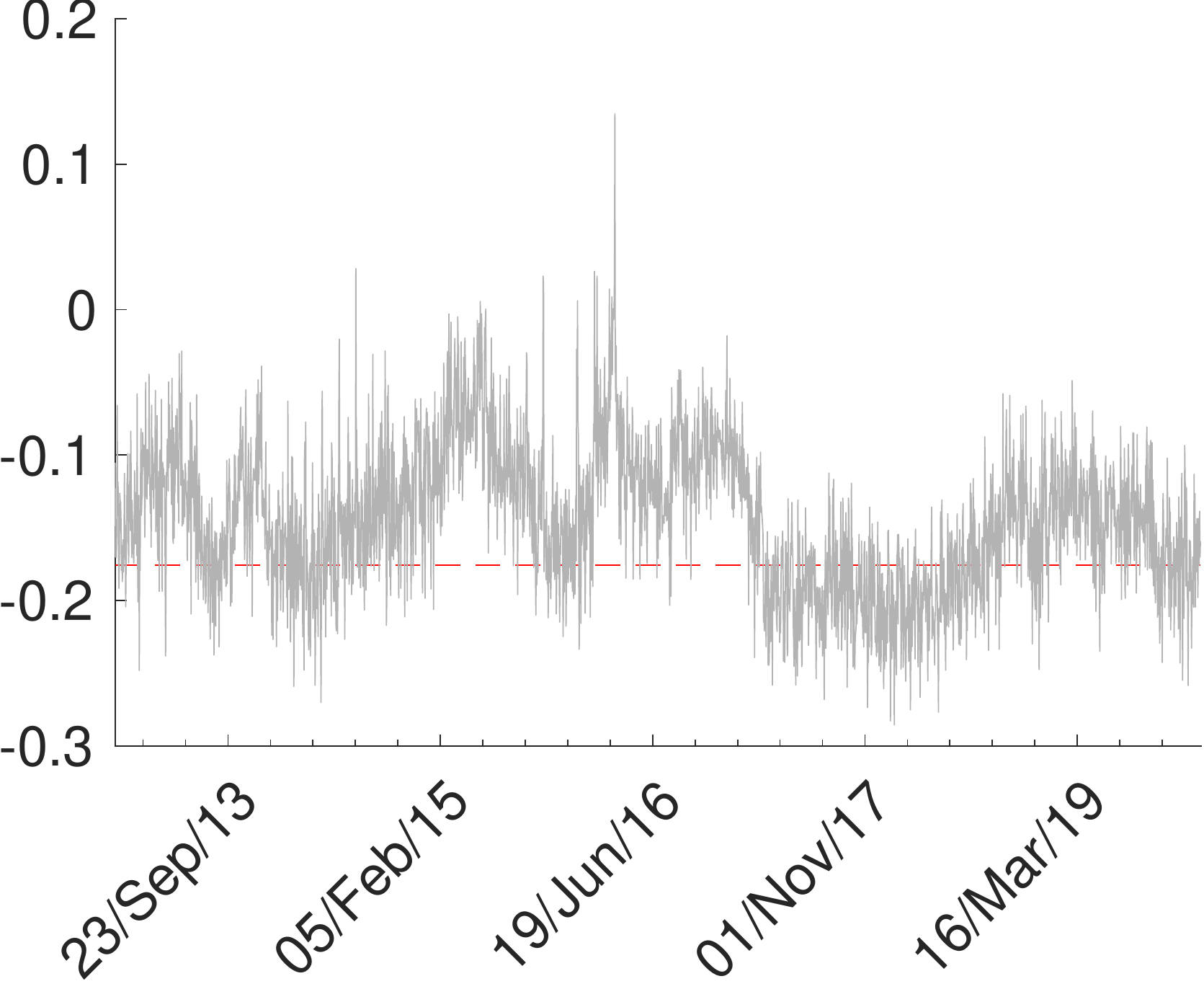}&
			\includegraphics[width=4cm]{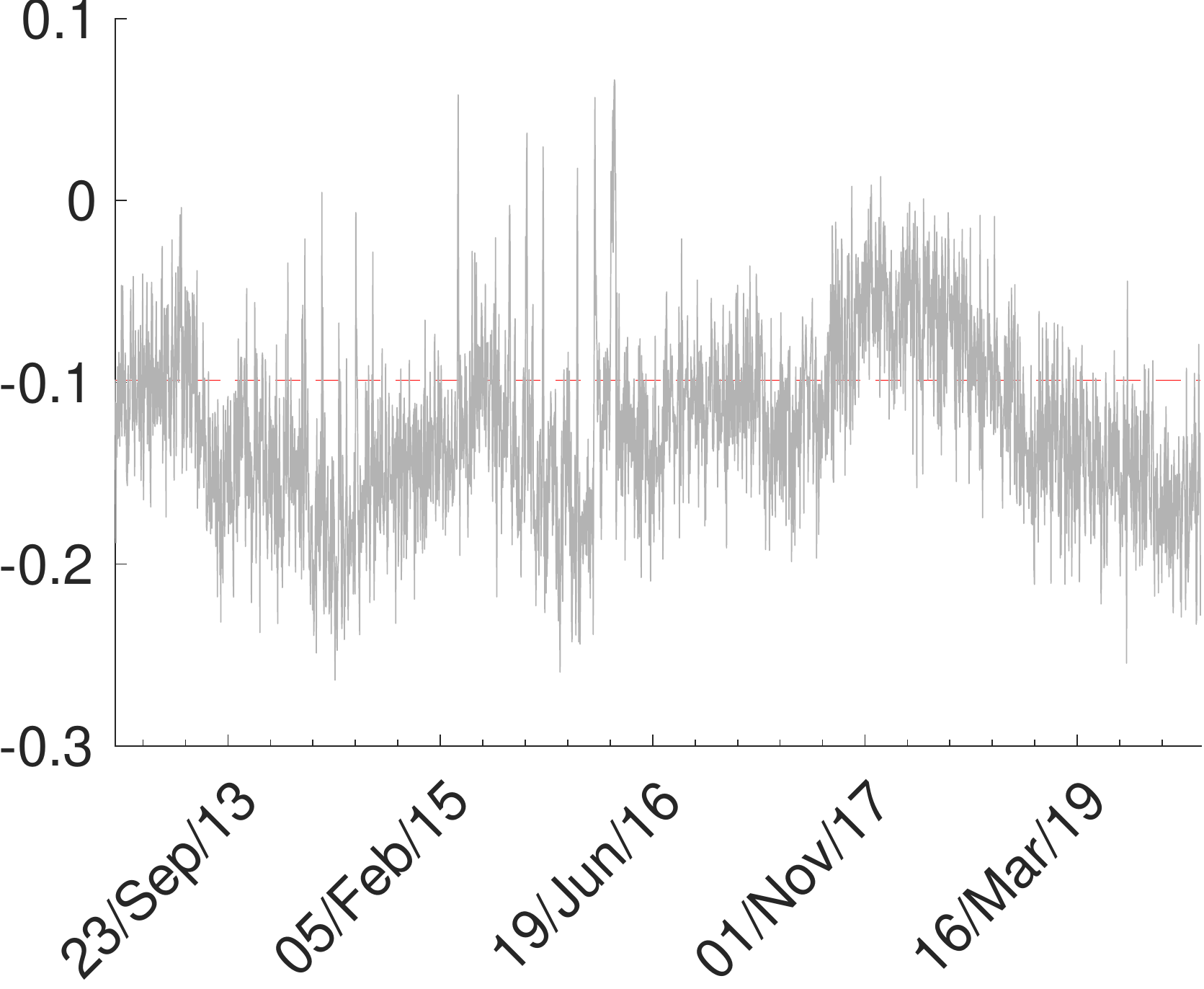}&
			\includegraphics[width=4cm]{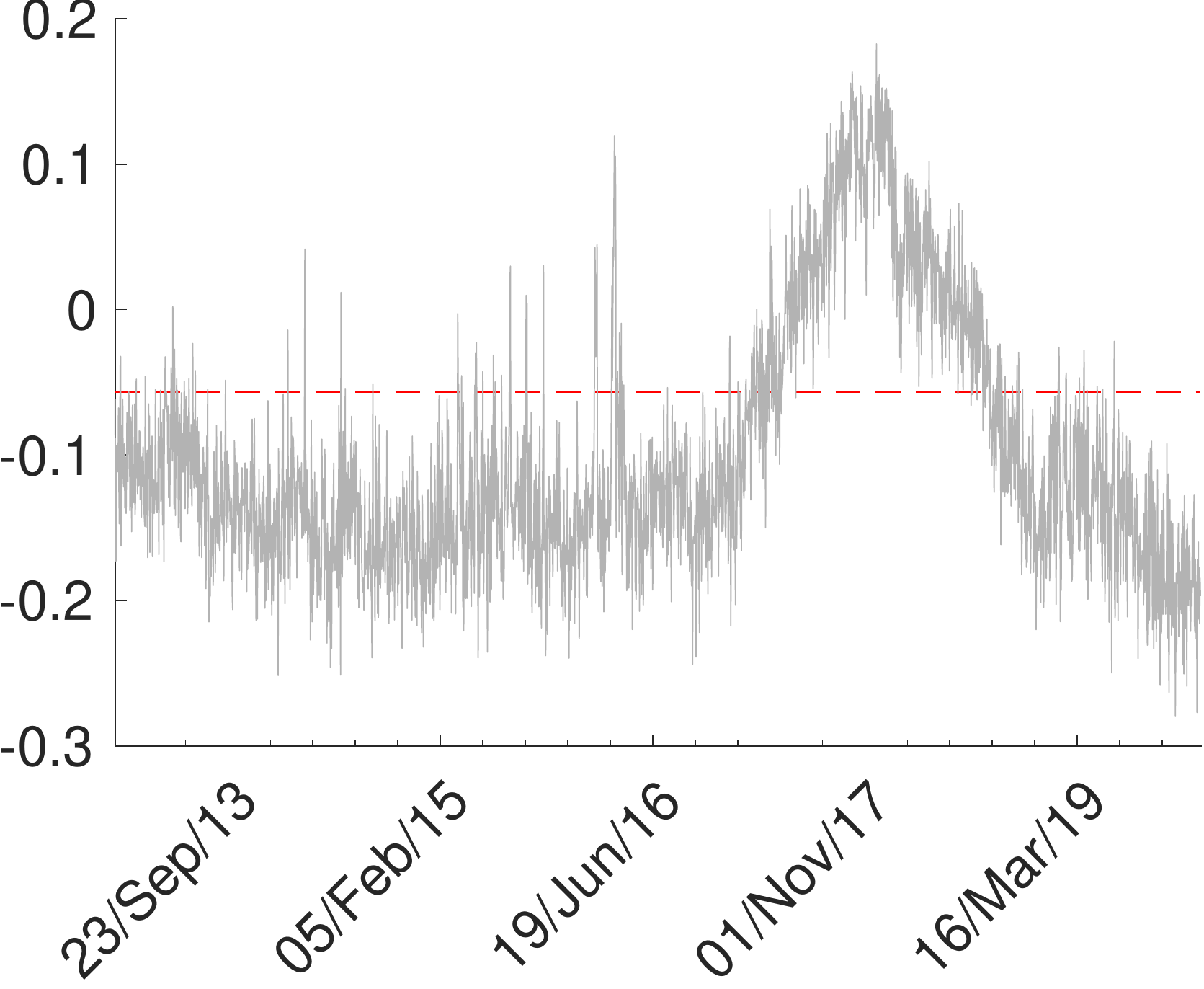}\\
			\includegraphics[width=4cm]{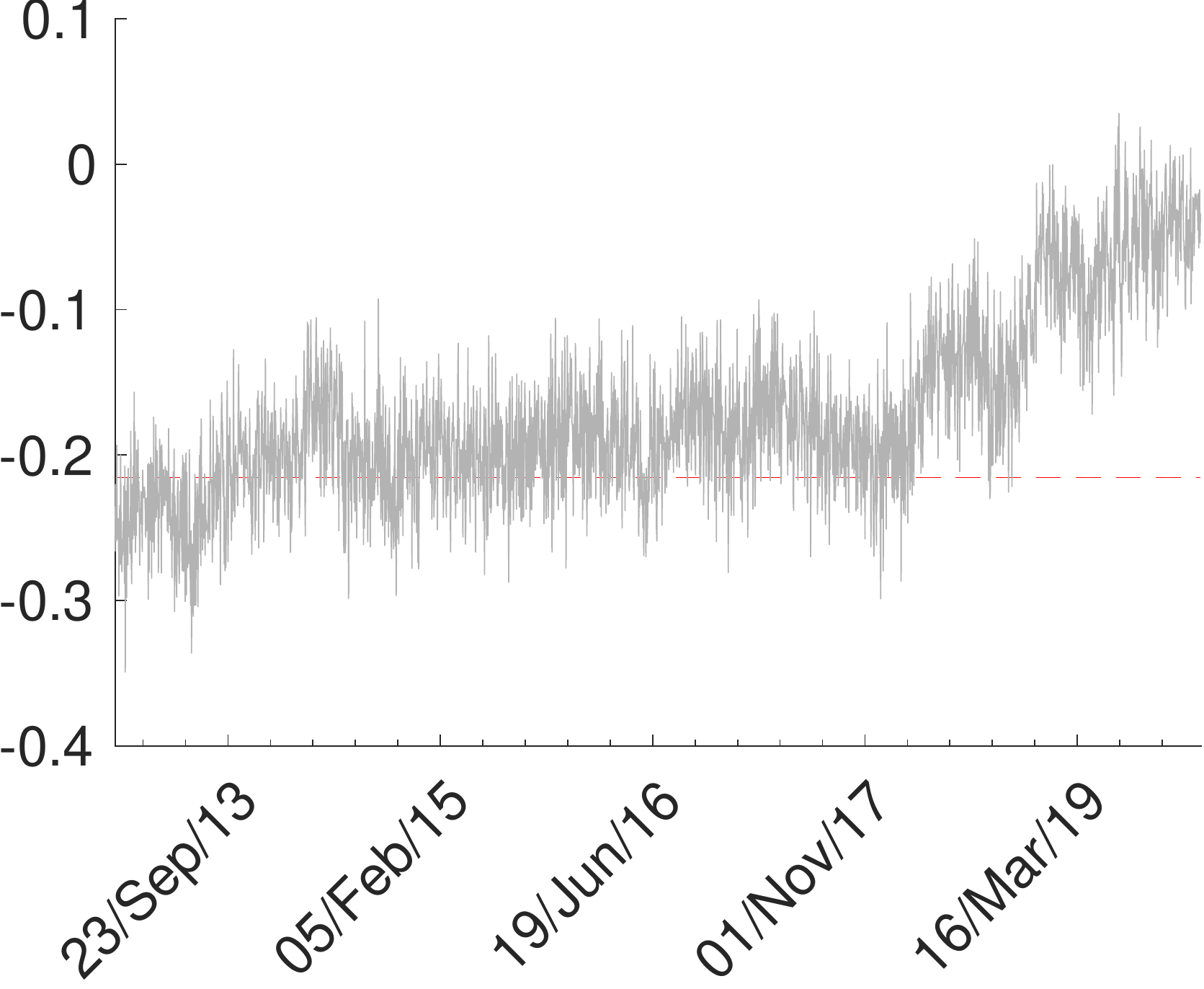}&
			\includegraphics[width=4cm]{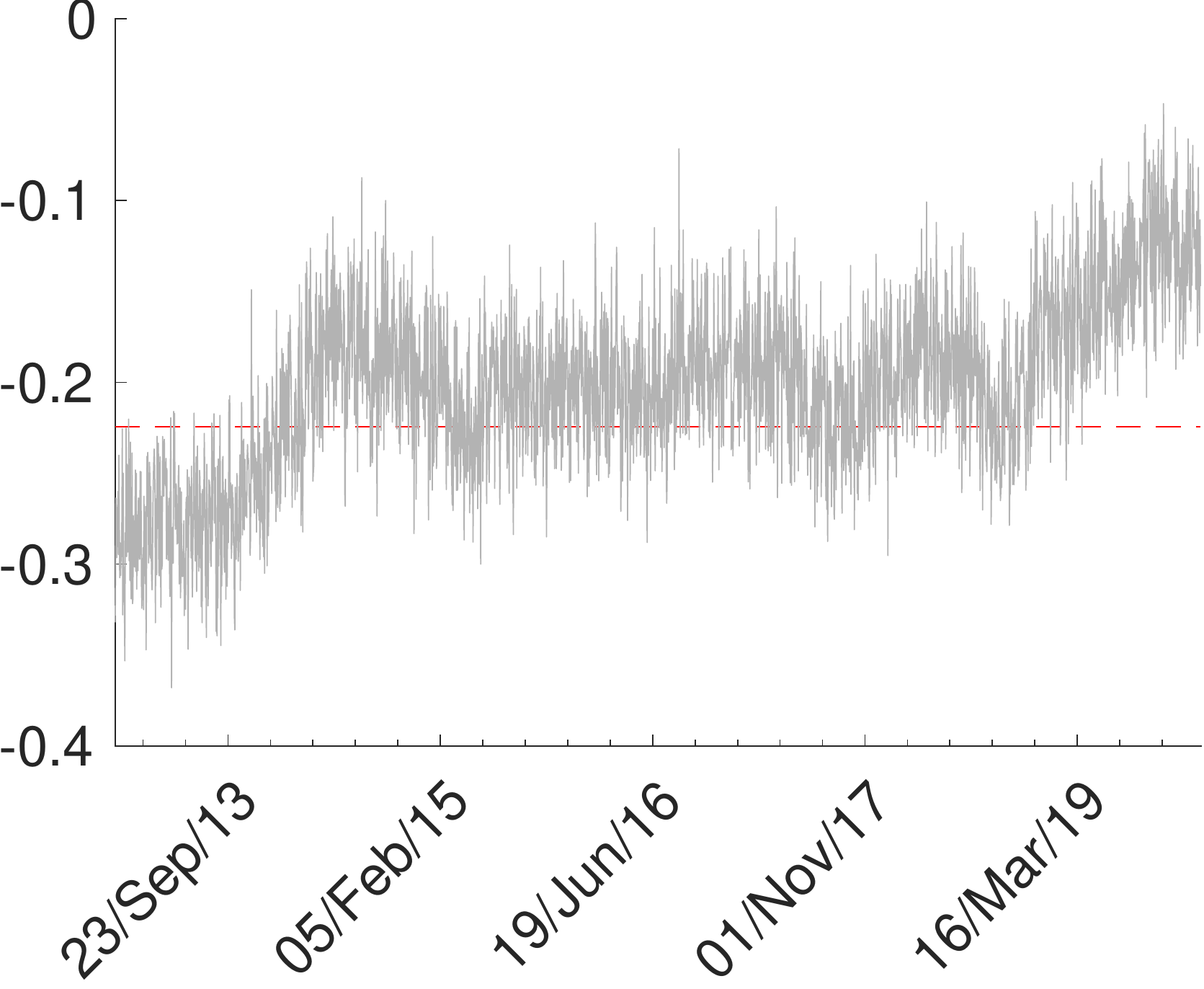}&
			\includegraphics[width=4cm]{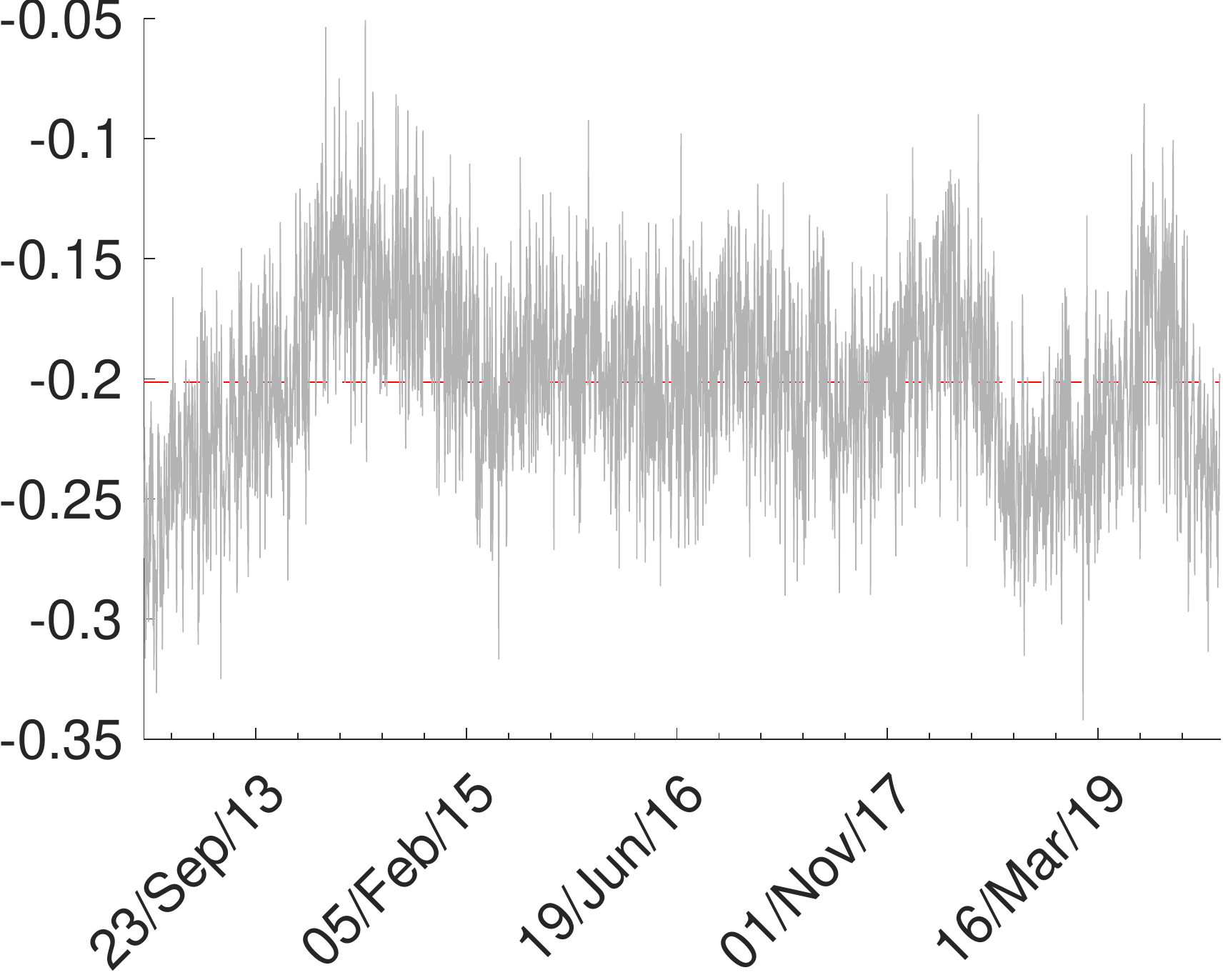}&
			\includegraphics[width=4cm]{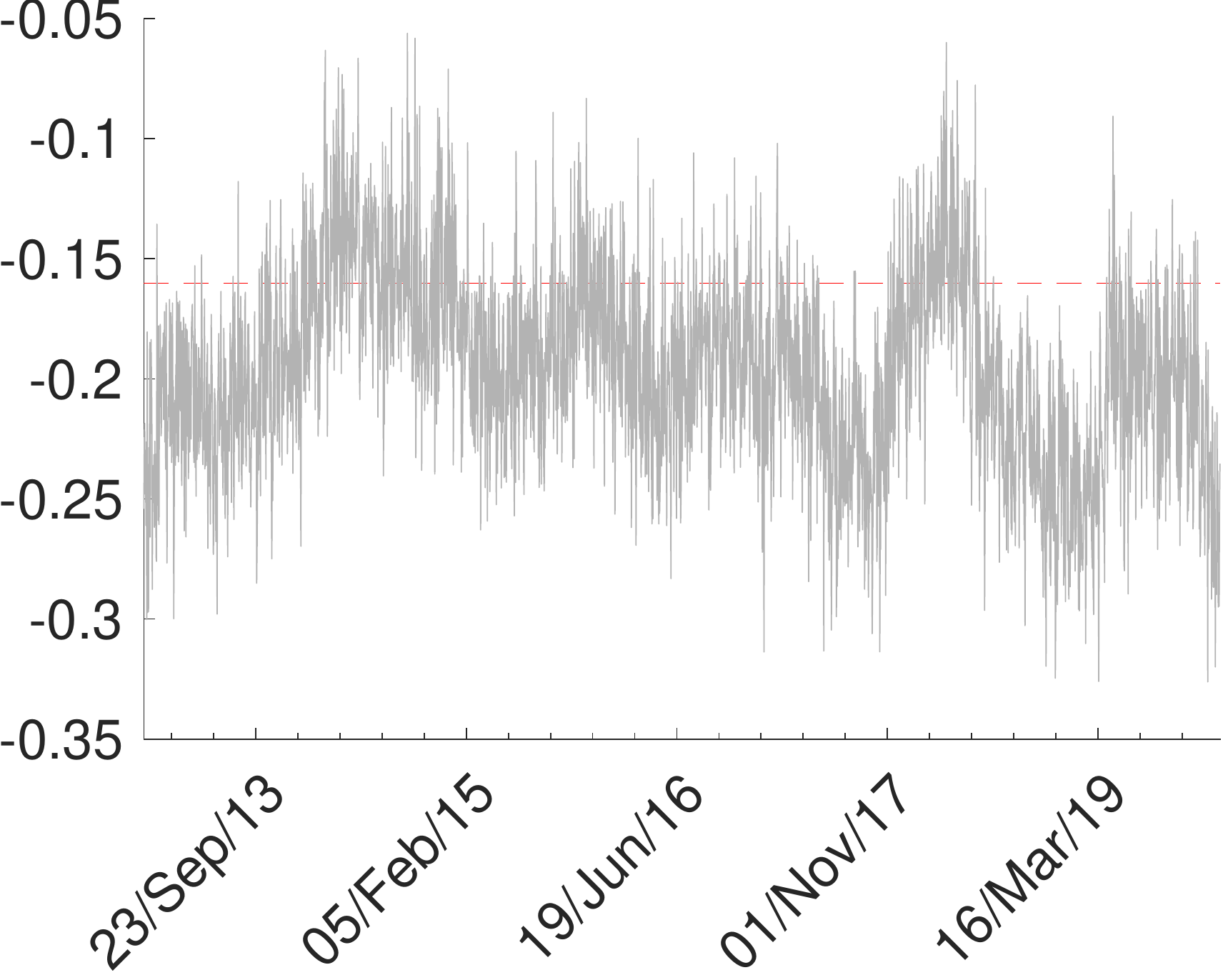}&
			\includegraphics[width=4cm]{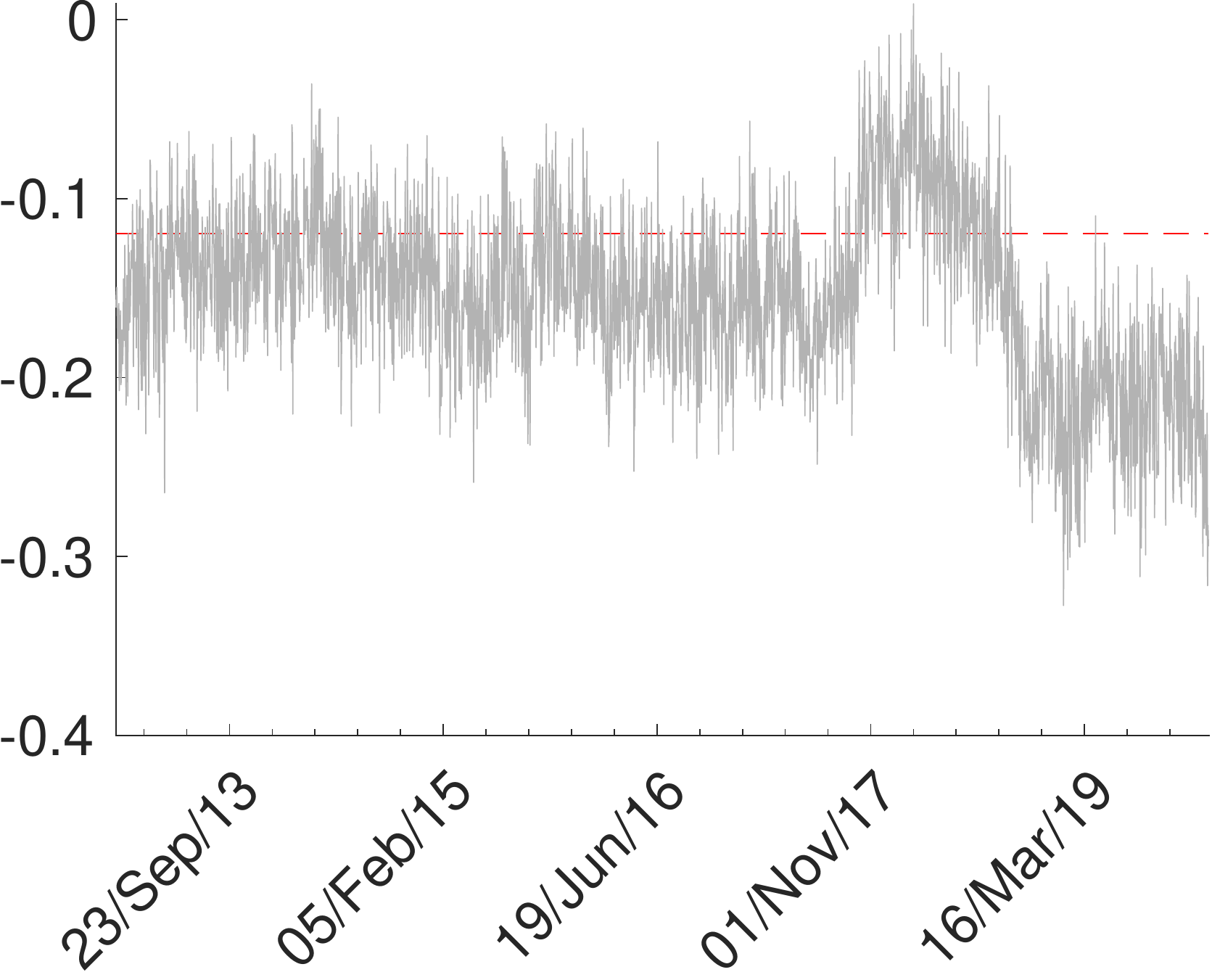}\\
		\end{tabular}
		\caption{\footnotesize{Estimated Time-varying Correlation between Electricity Prices and Forecasted Solar PV (top row), Forecasted Demand and Forecasted Solar PV (middle row), and Forecasted Wind and Forecasted Solar (bottom row) at five different hours for the quadrivariate estimations of the vine copula model specification (grey lines). The red dashed line is the dependence parameter for the vine copula model estimated on the whole sample.}}
		\label{Fig_Correl_Out_of_Sample_RES}
	\end{sidewaysfigure}

	Specifically, Figures \ref{Fig_Correl_Out_of_Sample_Price} and \ref{Fig_Correl_Out_of_Sample_RES} show the time-varying pairwise correlation induced by the trivariate and quadrivariate estimated vine copula models at five different hours ($8$, $10$, $12$, $14$, and $16$). For a matter of comparison, the horizontal line indicates the related correlation calculated on the whole sample. 
	
	Observing the first row in Figure \ref{Fig_Correl_Out_of_Sample_Price}, the dependence between electricity prices and forecasted demand changes slightly but consistently across the selected hours of the day and, more importantly, over the studied years. First, for every hour, the dependence seems to decrease through the sample, with a sharp decline around January 2015. Moreover, it can be observed that the static dependence parameter over the entire sample (represented with a red dashed line) seems to overestimate the dependence during the years 2015-2019, whereas it was underestimating the dependence at the beginning of the sample, that is over years 2013-2014.
	
	In the second row (of the same Figure), the dependence between prices and forecasted wind is depicted. Again, the dependence seems to act similarly across the hours of the day, with some differences in line with the amount of wind power produced, which differs across the hours of the day (as shown by its intra-daily profile). It is interesting to observe that the correlation was negative at the beginning of the sample and it has become progressively more negative through the years, which is consistent with the increasing generation of wind power.
	
	The rolling approach emphasizes the different behavior shown by the global dependence parameter, which underestimates at the beginning and overestimates at the end of the sample, corresponding to years in which wind had lower and then progressively higher levels of penetration.
	
	For completeness, the last row shows the dependence structure between the forecasted demand and forecast wind. As expected, this time-varying dependence does not seem to move strongly away from zero across hours and years. In other words, forecasted wind does not affect the demand, but only the supply curve and, through it, prices are consequently affected. However, both are influenced by weather conditions (even if with different magnitudes and together with other factors), therefore some correlation is observed.
	
	Moving to a quadrivariate dependence structure, Figure \ref{Fig_Correl_Out_of_Sample_RES} shows the time-varying correlations, at the five previously selected hours, for the dependencies between the forecasted solar PV and the remaining three variables (prices, demand, and wind). Results for the other dependencies are in line with results shown in Figure \ref{Fig_Correl_Out_of_Sample_Price} for the trivariate copula and have been omitted.
	
	The time-varying dependence between electricity prices and forecasted solar is shown in the first row. As anticipated by other studies, this relation is found to be marginal and negative across central hours, whereas it appears with slight different dynamics at hour $8$, when, however, solar production is limited. The negative time-varying dependence is decreasing and approaching null values over the more recent years. Notice that this comes with no surprise, since the main price reductions are induced by wind generation, and solar is expected to directly affect the level of demand. To this aim, the second row shows the dependence between forecasted demand and forecasted solar PV production, which is found to be strictly negative and erratic (especially at the central hours $10$, $12$, and $14$), thus reflecting the weather conditions for solar radiation.

	\section{Conclusions}
	\label{sec_Conclusions}
	
	Using a new compiled dataset, this paper investigates the multivariate dependence between hourly electricity prices, demand, and two different sources of renewable energy (wind and solar PV) in one of the largest producing countries of renewable energy in Europe, i.e., Germany.
	However, considering multivariate dependence structures is important in all countries for driving policy decisions, since increasing RES generation immediately affects both prices and demand. Therefore, identifying and adopting the appropriate methodology are two important tasks not only for the market studied in this analysis but also for all countries wishing to increase their green generation and reduce carbon emissions.
	
	By considering forecasted wind, solar PV generation, demand, and electricity prices, this work studies their joint dependence with a flexible copula approach. Moreover, the introduced multivariate tail dependence coefficients (depending on more than one variable) provide additional insights in the understanding of these relationships in the tail of their joint distribution. Indeed, applying suitable copula-based models for time series, a strong dependence is depicted and mapped between electricity prices, demand and RES during the day with important intra-daily and seasonal patterns.
	
	Apart from the methodological contribution related to the study of tail behavior in a multivariate setting, from an applied point of view, this paper contributes to the literature by filling the gap regarding the interrelationships between RES and demand and their combined effect on the electricity prices, given that there was no clear understanding of the effect of solar, especially its interactions with demand, and, eventually, with wind during central hours; however, here, this issue is addressed, and answers are provided. 
	
	\section*{Acknowledgments}
	
	The authors are grateful to the Editor and the Reviewers for their useful comments which significantly improved the quality of the paper.
	
	The authors thank the seminar and conference participants at the 12th International Conference on Computational and Financial Econometrics (CFE 2018) and 50th Meeting of the Italian Statistical Society (SIS).
	Europe Energy S.p.A. is acknowledged for funding this research project. In addition, Fabrizio Durante has been supported by the Italian Ministry MIUR under the PRIN project ``Stochastic Models for Complex Systems'' (grant no. 2017JFFHSH). Francesco Ravazzolo acknowledges financial support from Italian Ministry MIUR under the PRIN project ``Hi-Di NET - Econometric Analysis of High Dimensional Models with Network Structures in Macroeconomics and Finance'' (grant  no. 2017TA7TYC).
	
	\clearpage
	
	\newpage
	
	\clearpage
	{\footnotesize
		\bibliographystyle{chicago}
		\bibliography{DGRR_biblio}
		
	\end{document}